\documentclass[fleqn,usenatbib]{mnras}

\usepackage{newtxtext,newtxmath}

\usepackage[T1]{fontenc}
\usepackage{ae,aecompl}

\usepackage{graphicx}	
\usepackage{amsmath}	
\usepackage[dvipsnames]{xcolor}
\usepackage{multirow}
\usepackage{delarray}
\usepackage{color}
\usepackage{here}
\usepackage{float}
\usepackage{tensor}
\usepackage{xspace}
\usepackage{orcidlink}
\usepackage[multiple]{footmisc}

\newcommand{\mach}{\mathcal{M}}

\newcommand{\be}{\begin{equation}} \newcommand{\ee}{\end{equation}}

\newcommand{\SFE}{\mathrm{SFE}}
\newcommand{\MF}{\mathrm{MF}}
\newcommand{\CF}{\mathrm{CF}}
\newcommand{\solarmass}{\mathrm{M}_{\rm \sun}}
\newcommand{\msun}{\solarmass}

\newcommand{\tff}{t_\mathrm{ff}}

\newcommand{\Msonic}{M_{\rm sonic}}

\newcommand{\MJeans}{M_{\rm Jeans}}
\newcommand{\alphath}{\alpha_{\mathrm{th}}}
\newcommand{\alphaturb}{\alpha_{\mathrm{turb}}}
\newcommand{\alphaB}{\alpha_{\mathrm{B}}}

\newcommand{\pc}{\mathrm{pc}}
\newcommand{\AU}{\mathrm{AU}}

\newcommand{\dderiv}{\mathrm{d}}

\newcommand{\appropto}{\mathrel{\vcenter{
  \offinterlineskip\halign{\hfil$##$\cr
    \propto\cr\noalign{\kern2pt}\sim\cr\noalign{\kern-2pt}}}}}

\newcommand{\myquote}[1]{``#1''}

\usepackage[normalem]{ulem}

\defcitealias{grudic_starforge_methods}{Methods Paper}
\defcitealias{grudic_starforge_m2e4}{Paper I}
\defcitealias{guszejnov_starforge_clusters}{Paper II}
\defcitealias{guszejnov_starforge_imf}{Paper III}
\defcitealias{moe_distefano2016}{MDS17}

\title[STARFORGE: Binaries and multiples]{Effects of the environment on the multiplicity properties of stars in the STARFORGE simulations}

\author[]{
D\'avid Guszejnov\orcidlink{0000-0001-5541-3150}$^{1}$\thanks{guszejnov@utexas.edu},
Aman N. Raju\orcidlink{0000-0001-9339-0789}$^{1}$,
Stella S. R. Offner\orcidlink{0000-0003-1252-9916}$^{1}$,
Michael Y. Grudi\'{c}\orcidlink{0000-0002-1655-5604}$^{2}$,
\newauthor
Claude-Andr{\'e} Faucher-Gigu{\`e}re\orcidlink{0000-0002-4900-6628}$^{3}$
Philip F. Hopkins\orcidlink{0000-0003-3729-1684}$^{3}$,
Anna L. Rosen\orcidlink{0000-0003-4423-0660}$^{4}$
\\
$^{1}$Department of Astronomy, University of Texas at Austin, TX 78712, USA \\
$^{2}${CIERA and Department of Physics and Astronomy, Northwestern University, 2145 Sheridan Road, Evanston, IL 60208, USA}\\
$^{3}$TAPIR, Mailcode 350-17, California Institute of Technology, Pasadena, CA 91125, USA \\
$^{4}$Center for Astrophysics $|$ Harvard \& Smithsonian, 60 Garden St, Cambridge, MA 02138, USA \\
}

\date{\today \vspace{-0.6cm}}

\pubyear{2022}

\begin{document}
\label{firstpage}
\pagerange{\pageref{firstpage}--\pageref{lastpage}}
\maketitle

\begin{abstract}
Most observed stars are part of a multiple star system, but the formation of such systems and the role of environment and various physical processes is still poorly understood. We present a suite of radiation-magnetohydrodynamic simulations of star-forming molecular clouds from the STARFORGE project that include stellar feedback with varied initial surface density, magnetic fields, level of turbulence, metallicity, interstellar radiation field, simulation geometry and turbulent driving. In our fiducial cloud the raw simulation data reproduces the observed multiplicity fractions for Solar-type and higher mass stars, similar to previous works. However, after correcting for observational incompleteness the simulation under-predicts these values. The discrepancy is likely due to the lack of disk fragmentation, as the simulation only resolves multiples that form either through capture or core fragmentation. The raw mass distribution of companions is consistent with randomly drawing from the initial mass function for the companions of $>1\,\msun$ stars, however, accounting for observational incompleteness produces a flatter distribution similar to observations. We show that stellar multiplicity changes as the cloud evolves and anti-correlates with stellar density. This relationship also explains most multiplicity variations between runs, i.e., variations in the initial conditions that increase stellar density (increased surface density, reduced turbulence) decrease multiplicity. While other parameters, such as metallicity, interstellar radiation, and geometry significantly affect the star formation history or the IMF, varying them produces no clear trend in stellar multiplicity properties.
\end{abstract}



 \section{Introduction}\label{sec:intro}
 
Stars form in highly clustered environments \citep{clustering_lada}, and both young and older stellar populations have a significant fraction of multiples, which are defined as bound systems of two or more stars. 
The likelihood of a star being in a multiple is observed to increase monotonically with mass (see reviews of \citealt{duchene2013, Lee_2020_IMF_multiplicity_review} and references therein). It is generally understood that multiple systems form either during the star-forming phase of the parent cloud, where the dominant channels are the fragmentation of a protostellar core \citep{Goodwin_2004_turbulent_fragmentation} or disk \citep{Adams_1989_disk_fragmentation}, or through dynamical evolution during the dissolution of the cluster \citep{kouwenhoven_2010,Parker_Meyer_2014}.

The detailed multiplicity properties of a stellar system are characterized by several metrics, which are usually defined as a function of the mass of the most massive star, i.e., the \emph{primary}, in the system. 
 One commonly measured property is the mass ratio $q$ of the secondary to the primary mass. For Solar-type stars the mass ratio distribution is statistically consistent with a flat distribution \citep{raghavan2010} for most of the companion mass range, except for two features: a lack of brown dwarf-scale companions (\myquote{brown dwarf desert}, see e.g., \citealt{Kraus2008b}) and an excess of companions at near-unity mass ratio (\myquote{twins}, see \citealt{El_Badry_2019_binary_twins}). Other metrics concern the orbits of the companions, which can be characterized with the orbital period/semi-major axis and orbital eccentricity distributions. The semi-major axis distribution of Solar-type stars is well-described by a lognormal distribution that peaks around 100 AU \citep{raghavan2010}. The eccentricity distribution $f(e)$ for companions of Solar-type stars with separations $>50\,\AU$ follows $f(e)\approx 1.2e+0.4$ and in general eccentricity increases with orbital period \citep{Tokovinin_2016_solar_eccentricity_dist}. 

There has been significant theoretical effort to explain these observations, mainly through detailed hydrodynamical simulations. Simulations of star cluster formation show good agreement with observed multiplicity statistics \citep{Bate09a,bate12a,krumholz_2012_orion_sims, Lee_2019_MHD_binary_separations, li_2018_sf_mhd_jets}) using different combinations of physical processes. Unfortunately the dynamic range of simulations is unavoidably limited, leading to either insufficient resolution to resolve close binary formation \citep[e.g., ][]{Mathew_Federrath_2021_IMF_multiplicity_2021} or low number statistics due to the small cloud size \citep[e.g., ][]{Rohde_2021_outflows_IMF_multiplicities}. Thus pinpointing the key physics is challenging.

Note that most simulations only follow the star-forming phase of the cluster evolution, so they in fact predict the multiplicity of stars close to formation. Both observations \citep{Duchene1999, Kraus2008b, Kraus2011,Tobin_2016_protostellar_multiplicity_Perseus, Tobin_2021_protostellar_multiplicity} 
and simulations suggest that stars are born in complex, multiple systems that dynamically evolve (e.g., ejection of stars) causing multiplicity to drop (\citealt{Goodwin_Kroupa_2005_primordial_multiplicity, Goodwin2007, Kaczmarek2011}) and the period/separation distribution to shift to shorter periods/ closer separations (\citealt{Kroupa1995,Marks2011}). 
This can be understood as the result of wide-separation binaries becoming unbound due to either internal dynamical evolution or by interacting with external tidal fields, the latter of which also increases the average binding energy between the remaining stars. In general, dynamical interactions cause strongly bound binaries to be even more bound (i.e., \myquote{harden}), while the separation of weakly bound companions increases  \citep{Heggie_1975_binary_dynamics}. 

To date most simulations have attempted to recover the observed multiplicity properties without conducting a detailed parameter study on how their initial conditions might affect multiplicity (see \citealt{Lee_2019_MHD_binary_separations} for an exception). In this work we present the first comprehensive analysis of how cloud properties affect stellar multiplicity properties. We use simulations from the STAR FORmation in Gaseous Environments (STARFORGE) project\footnote{\url{http://www.starforge.space}} that include all relevant physical processes of star formation. These radiation-magnetohydrodynamic (RMHD) simulations achieve a dynamic range in mass resolution that allow us to simulate the detailed evolution of molecular clouds while following the formation of individual low-mass stars (see \citealt{grudic_starforge_methods}, henceforth referred to as the \citetalias{grudic_starforge_methods}). In this study we analyze a set of runs with varied initial cloud surface density, level of turbulence, magnetic field strength, metallicity and interstellar radiation field and compare them to a fiducial run with parameters representing a typical Milky Way molecular cloud (similar to \cite{grudic_starforge_m2e4}, henceforth referred to as \citetalias{grudic_starforge_m2e4}). We focus on the evolution of multiplicity properties from the onset of star formation until cloud disruption.

The paper is structured as follows: \S\ref{sec:methods} provides a brief overview of the code (for details on numerical methods as well as tests see the \citetalias{grudic_starforge_methods}) and the initial conditions of the runs. We present our results for the fiducial run in \S\ref{sec:results_fiducial} and compare them with observations. In \S\ref{sec:results_var} we explore how multiplicity properties change in response to variations in the initial parameters. An analysis of the clustering properties, the star formation history and the initial mass functions of these runs are presented in \citet{guszejnov_starforge_clusters} and \citet{guszejnov_starforge_imf}, henceforth referred to as \citetalias{guszejnov_starforge_clusters} and \citetalias{guszejnov_starforge_imf} respectively. We discuss the implications of our results to observations and future work in \S\ref{sec:discussion}. Finally, we present our conclusions in \S\ref{sec:conclusions}.


 \section{Numerical Methods}\label{sec:methods}
 
  \subsection{The STARFORGE simulations}\label{sec:starforge}
  
For this work we utilize simulations from the STARFORGE project, which are run with the {\small GIZMO} code\footnote{\url{http://www.tapir.caltech.edu/~phopkins/Site/GIZMO.html}}. A full description and presentation of the STARFORGE methods including a variety of tests and algorithm details are given in the \citetalias{grudic_starforge_methods}. We only briefly summarize the key points here. Note that in this work we use the same physics modules as \citetalias{grudic_starforge_m2e4} and our fiducial run uses identical initial conditions as the run presented there. Readers familiar with the STARFORGE simulation methods should skip ahead to \S\ref{sec:metrics} where we define the various metrics used in this study.

\subsubsection{Physics}\label{sec:physics}

  
We simulate star-forming clouds with the {\small GIZMO} code \citep{hopkins2015_gizmo} using the Lagrangian meshless finite-mass (MFM) method for magnetohydrodynamics \citep{hopkins_gizmo_mhd}, assuming ideal MHD. 
Individual stars in the simulations are represented by sink particles . Once they form they follow the protostellar evolution model from \citet{Offner_2009_radiative_sim}, extended past the main sequence by the mass-loss and stellar lifetime prescriptions presented in \citetalias{grudic_starforge_methods}. 

The presented STARFORGE runs utilize the radiative cooling and thermochemistry module from \citet{fire3} that contains detailed metallicity-dependent cooling and heating physics, including recombination, thermal bremsstrahlung, metal lines, molecular lines, fine structure and dust collisional processes. The cooling module self-consistently solves for the internal energy and ionization state of the gas. The simulations co-evolve the gas, dust, and radiation temperature self-consistently, including the stellar luminosity in various bands accounting for photon transport, absorption and emission using dust opacity. In addition to local sources (i.e. stars) we include an external heating source at the boundary of the simulation domain that represents the interstellar radiation field (ISRF).

The simulations account for the dominant stellar feedback processes, including protostellar jets, radiative feedback from both protostars and main sequence stars, stellar winds and supernovae. See \citetalias{grudic_starforge_m2e4} and the \citetalias{grudic_starforge_methods} for details on the numerical implementations.

\subsubsection{Initial Conditions \& Parameters of Clouds}\label{sec:initial_conditions}


We use cloud initial conditions (ICs) identical to those presented in \citetalias{guszejnov_starforge_imf}, so we only give a short summary here.

We generate our initial conditions using {\small MakeCloud} \citep{Grudic_MakeCloud_2021}. Our default IC geometry is the \emph{\myquote{Sphere}} where we initialize the cloud as a homogeneous sphere near thermal pressure equilibrium with a low density ambient medium. We apply an initial random velocity field with a power spectrum of $E_k\propto k^{-2}$ with amplitude set by the $\alpha_{\rm turb}\equiv 5 \sigma^2 R_\mathrm{cloud}/(3 G M_0)$ turbulent virial parameter. The cloud is initially threaded by a uniform magnetic field $B_z$ whose strength is set by the normalized mass-to-flux ratio $\mu$. 
There is no external driving in \emph{\myquote{Sphere}} simulations. 

We also run a simulations using the \emph{\myquote{Box}} geometry, a periodic box with externally driven turbulence whose side length $L_\mathrm{box}$ gives the box a volume equal to that of a {\it Sphere} cloud model of similar mass. The box is initialized with a uniform density and stationary gas and then \myquote{stirred} for five global freefall times $\left(t_{\mathrm{ff}}\equiv\sqrt{\frac{3 \pi}{32 G \rho_0}}\right)$, to achieve saturated MHD turbulence. An important difference between the \textit{Sphere} and \textit{Box} runs is that in the latter case the magnetic field is enhanced by dynamo action during the stirring phase (e.g., \citealt{federrath_2011_dynamo,2016MNRAS.461.1260T}).

\begin{figure*}
\begin {center}
\includegraphics[width=0.33\linewidth]{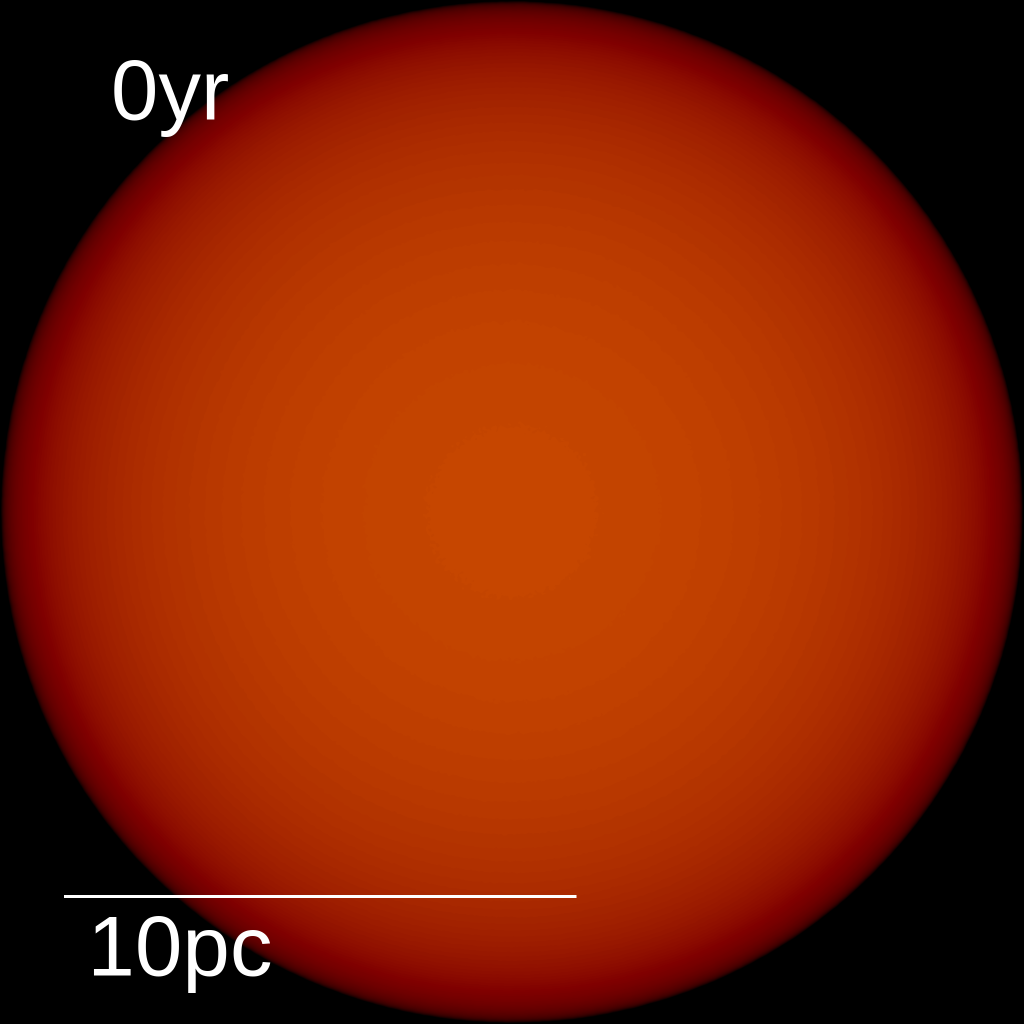}
\includegraphics[width=0.33\linewidth]{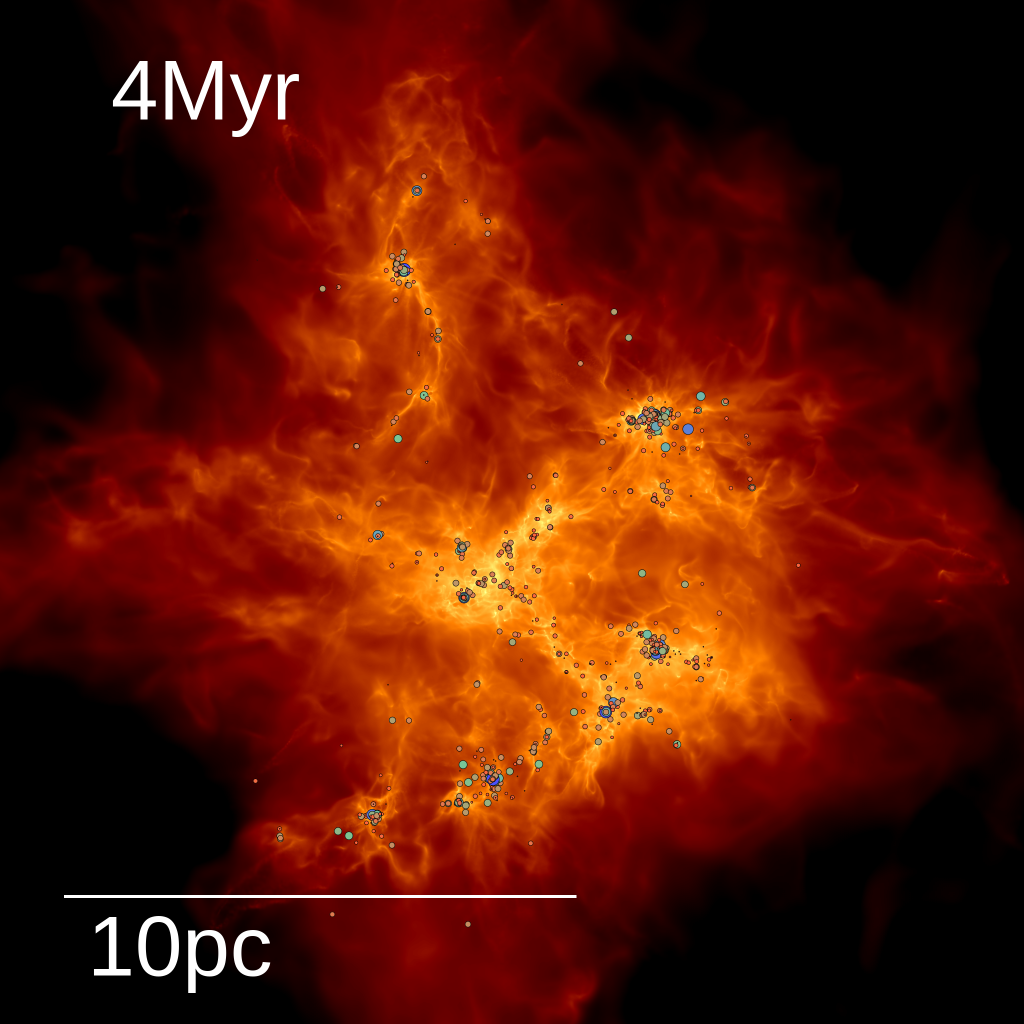}
\includegraphics[width=0.33\linewidth]{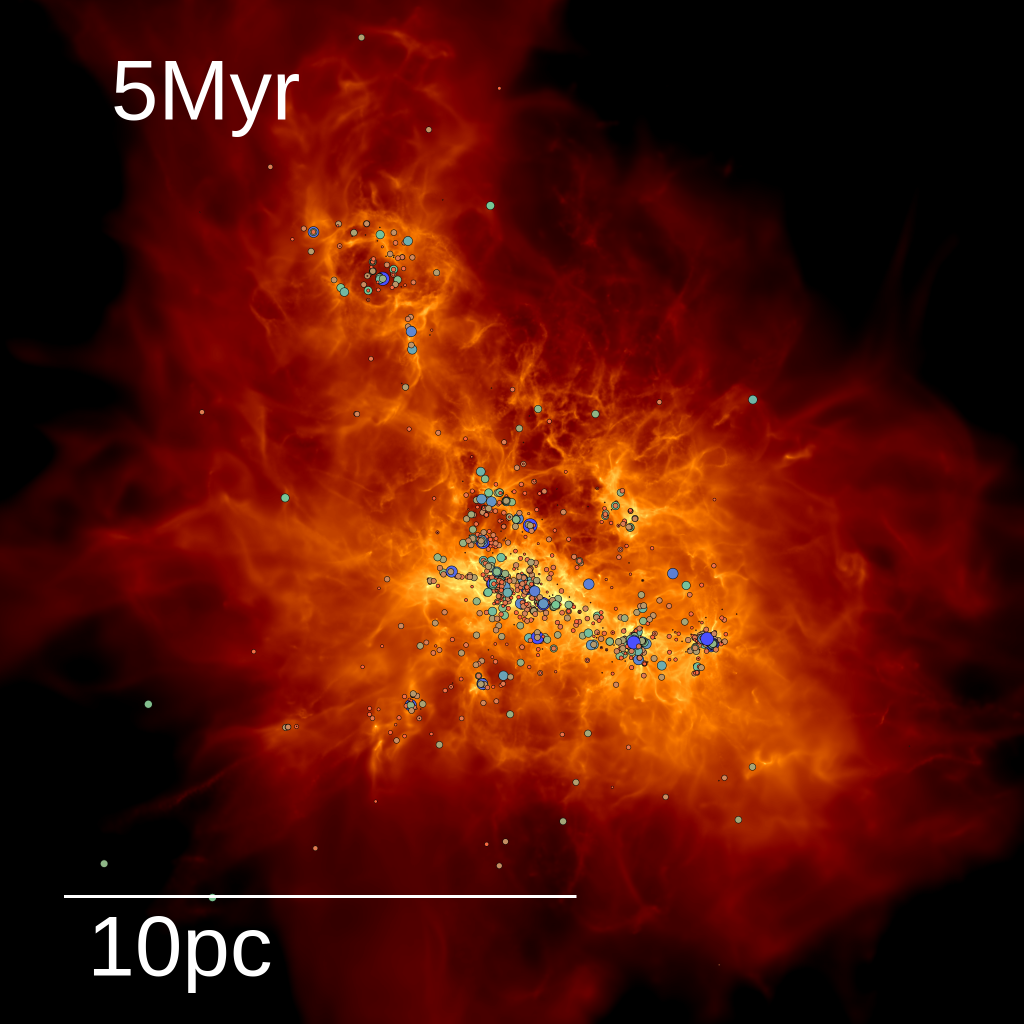}\\
\includegraphics[width=0.33\linewidth]{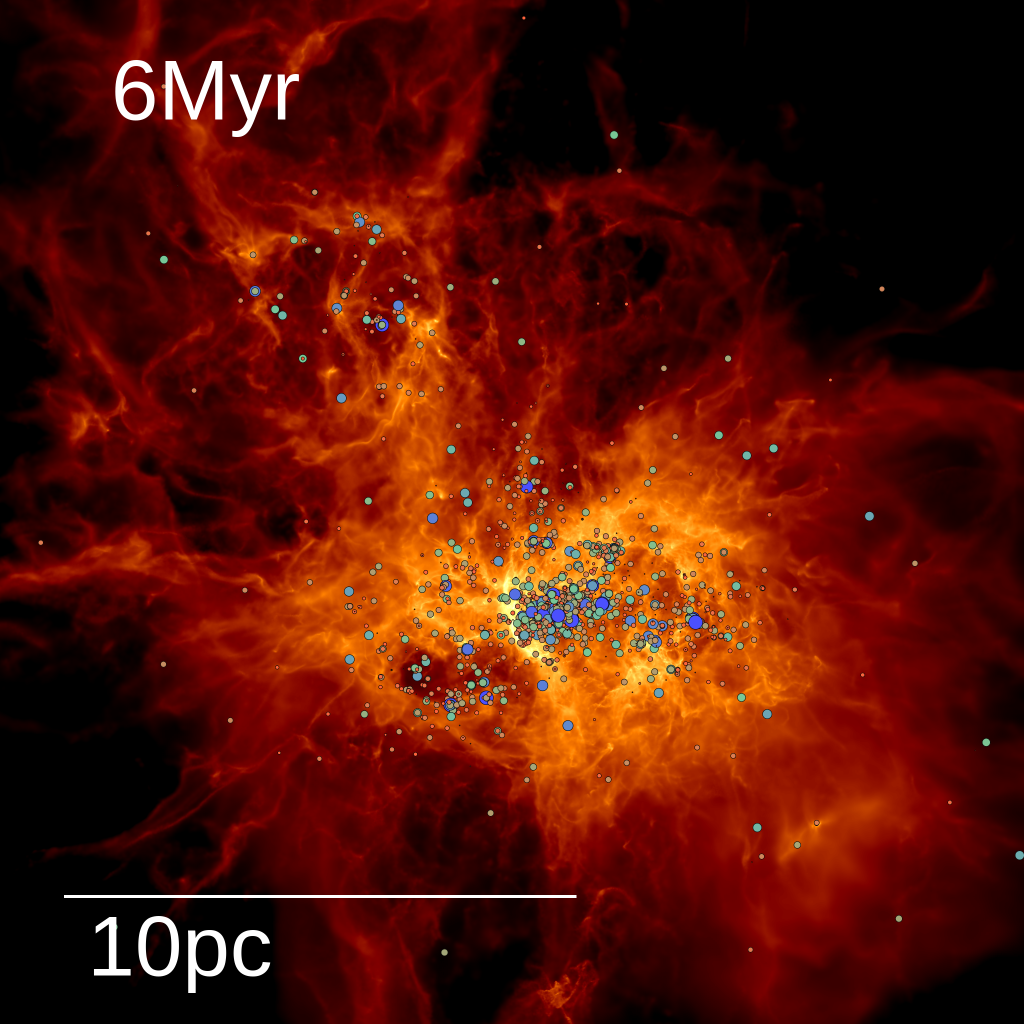}
\includegraphics[width=0.33\linewidth]{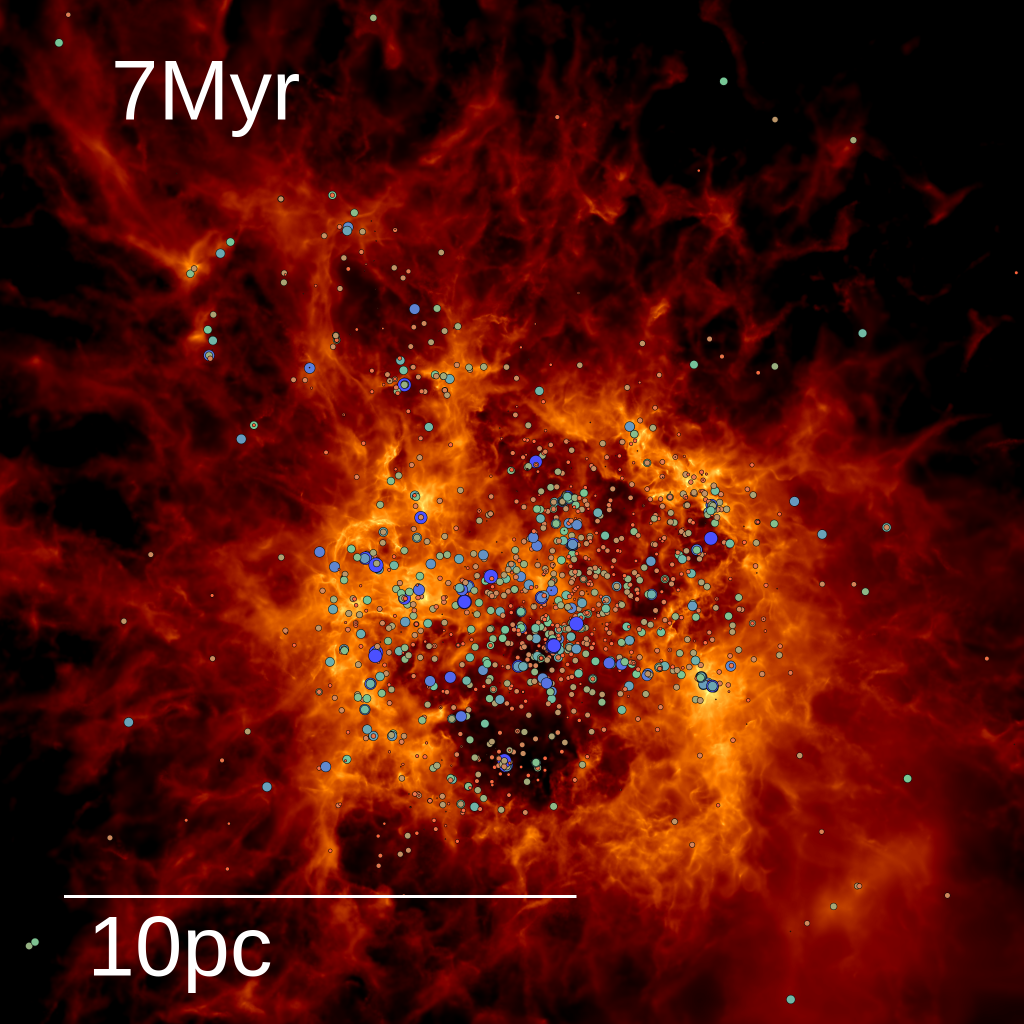}
\includegraphics[width=0.33\linewidth]{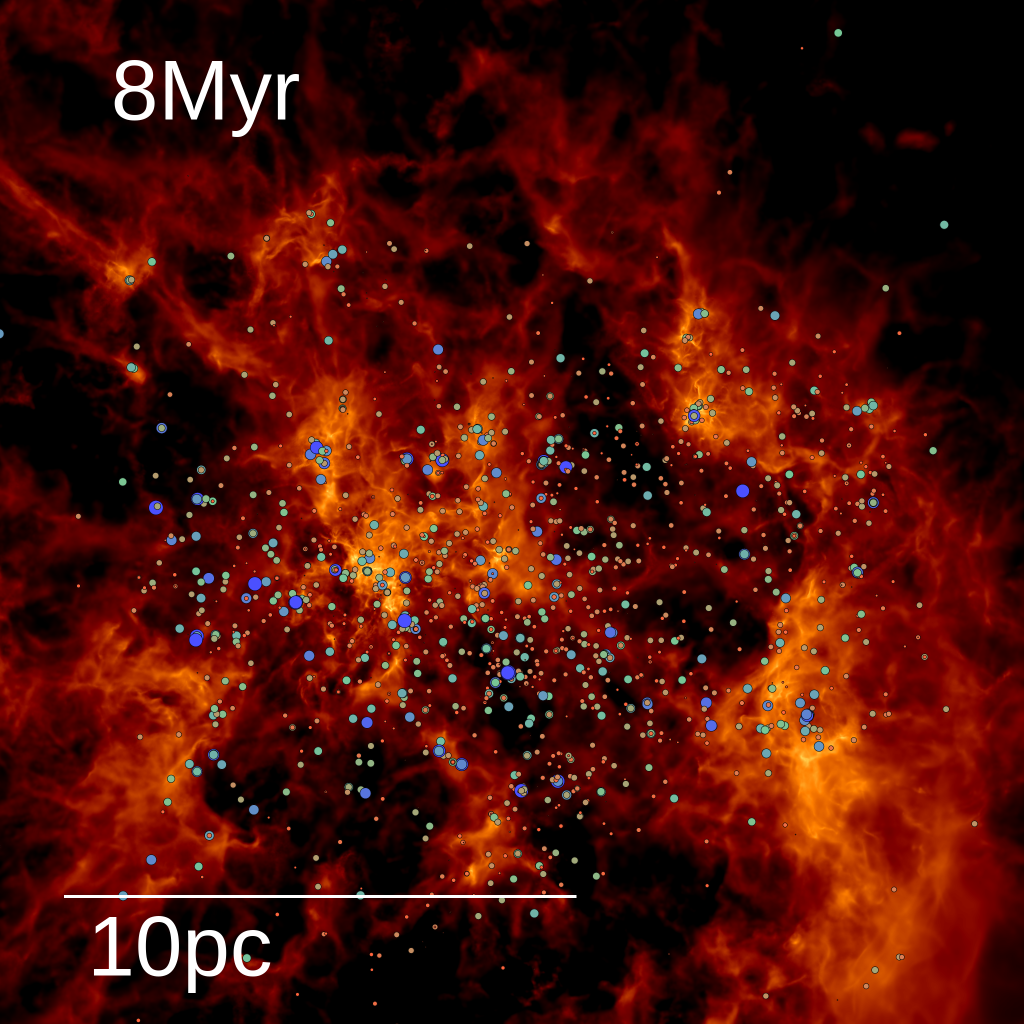}\\
\caption{Surface density maps for \textbf{M2e4} (our fiducial run), which is an $M_0=2\times 10^4\,\msun$ mass cloud resolved with $M_0/\Delta m=2\times 10^7$ initial gas cells (see Table \ref{tab:IC_phys}) at different times, from the beginning of the simulation until cloud disruption. The color scale is logarithmic, and the circles represent sink particles (stars) that form in high-density regions where fragmentation can no longer be resolved. The size of the circles  increases with mass, and their color changes from red ($M\sim0.1\,\msun$), green ($M\sim1\,\msun$) to blue ($M\sim10\,\msun$). This simulation resolves a dynamic range from $\sim\!\mathrm{20\,pc}$ down to $\sim\!\mathrm{30\,AU}$ and evolves until stellar feedback quenches star formation and disrupts the cloud.}
\label{fig:M2e4_series}
\vspace{-0.5cm}
\end {center}
\end{figure*}

Table \ref{tab:IC_phys} shows the target parameters for the runs we present in this paper. The input parameters are the cloud mass $M_0$, size $R_0$, turbulent virial parameter $\alphaturb$, normalized magnetic mass-to-flux ratio $\mu$, metallicity $Z$ and the energy density of the ISRF $e_\mathrm{ISRF}$ (note that initial temperature is set by the ISRF). Similar to \citetalias{grudic_starforge_m2e4} our fiducial cloud satisfies the observed Milky Way cloud mass-size relation (e.g. \citealt{larson_law,lada:2020.sigma.gmc}, specifically assuming $\Sigma\equiv M_\mathrm{0}/ \uppi R_\mathrm{cloud}^2 = 63 \msun\,\mathrm{pc}^{-2}$). The cloud is marginally bound ($\alphaturb=2$) and begins in thermal equilibrium with the ISRF. The initial gas metallicity is assumed to be equal to the solar value. For the initial magnetization we assume $-E_\mathrm{mag}/E_\mathrm{grav}=0.01$, which translates to $\mu=4.2$. The STARFORGE simulations we use have a mass resolution of $\Delta m=10^{-3}\,\msun$, making the mass function incomplete stars with masses below $0.1\,\msun$, which we omit from our analysis (see \citetalias{guszejnov_starforge_imf} for a convergence test). {\it Sphere} runs continue until stellar feedback quenches star formation and subsequently disrupts the cloud (see Figure \ref{fig:M2e4_series}). In the case of the Box runs the periodic boundary conditions trap both radiation and cloud material, so we terminate the run when the box becomes saturated by stellar radiation. 

\begin{table*}
    \setlength\tabcolsep{2.0pt} 
	\centering
	\begin{tabular}{ | c | c | c | c | c | c |c |c | }
	\hline
	\textbf{Physics label} & Thermodynamics & MHD & Protostellar Jets & Stellar Radiation & Stellar Winds \& SNe \\
	\hline
	\textbf{C\_M\_J\_RT\_W} & Non-isothermal, RHD (C) & Ideal (M) & Included (J) & Included (RT) & Included (W)   \\ 
	\hline
    \end{tabular}
	\begin{tabular}{|cccccccc|ccccccccc|cc|}
	     \multicolumn{17}{c}{}\\ 
		 \multicolumn{1}{c}{}&
		 \multicolumn{8}{c}{\bf Input Parameters} &
		 \multicolumn{8}{c}{\bf Derived Parameters}&
		 \multicolumn{2}{c}{\bf Results} \\
		\hline
		\bf Cloud label & $\frac{M_0}{\msun}$ & $\frac{R_{\mathrm{cloud}}}{\pc}$ & $\frac{L_{\mathrm{box}}}{\pc}$ & $\alphaturb$ & $\mu$ & $\frac{Z}{Z_\mathrm{\odot}}$ & $\frac{e_\mathrm{ISRF}}{e_\mathrm{ISRF,solar}}$ & $\frac{\sigma}{\mathrm{km/s}}$  &  $\alphath$ & $\alpha$ & $\mach_{\rm A} $ & $\beta$ & $\alphaB$ & $\frac{\MJeans}{M_0}$ & $\frac{\Msonic}{M_0}$ & $\frac{M_{\Phi}}{M_0}$ &  SFE [\%] &  $t_\mathrm{disrupt}/\tff$  \\
		\hline
		\textbf{M2e4} (fiducial) & $2\times 10^4$ & 10 & & 2 & 4.2 & 1 & 1 
		& 3.2 & 0.008 & 2.03 & 10 &  0.78 & 0.02 & 
		$3\times 10^{-3}$ & $7 \times 10^{-5}$ & 0.1 &  $\mathrm{9\pm 0.3}$ & $\mathrm{1.6\pm 0.2}$ \\
		\hline
		\textbf{M2e4} (Box) & $2\times 10^4$ &  & 16 & 2 & 4.2 & 1 & 1 
		& 3.2 & 0.008 & 2.03 & 10 &  0.78 & 0.02 & 
		$3\times 10^{-3}$ & $7 \times 10^{-5}$ & 0.1 &   &  \\
		\hline
		\bf M2e4\_R3 & $2\times 10^4$ & 3 &  & 2 & 4.2 & 1 & 1 
		& 5.8 & 0.008 & 2.02 & 10 &  0.23 & 0.02 & 
		$5\times 10^{-4}$ & $7 \times 10^{-6}$ & 0.1 &  14 & 2.0 \\
		\hline
		\bf M2e4\_R30 & $2\times 10^4$ & 30 &  & 2 & 4.2 & 1 & 1 
		& 1.9 & 0.02 & 2.04 & 10 &  2.3 & 0.02 & 
		$1\times 10^{-2}$ & $6 \times 10^{-4}$ & 0.1 &  1 & 1.7 \\
		\hline
		\bf M2e4\_a1 & $2\times 10^4$ & 10 &  & 1 & 4.2 & 1 & 1 
		& 2.3 & 0.008 & 1.03 & 10 &  0.78 & 0.02 & 
		$3\times 10^{-3}$ & $4 \times 10^{-5}$ & 0.1 &  11 & 1.2 \\
		\hline
		\bf M2e4\_a4 & $2\times 10^4$ & 10 &  & 4 & 4.2 & 1 & 1 
		& 4.5 & 0.008 & 4.03 & 10 &  0.78 & 0.02 & 
		$3\times 10^{-3}$ & $1 \times 10^{-4}$ & 0.1 &  4 & 2.1 \\
		\hline
		\bf M2e4\_mu1.3 & $2\times 10^4$ & 10 &  & 2 & 1.3  & 1 & 1 
		& 3.2 & 0.008 & 2.21 & 3.1 &  0.078 & 0.2 & 
		$3\times 10^{-3}$ & $7 \times 10^{-5}$ & 0.4 &  7 & 2.0 \\
		\hline
		\bf M2e4\_mu0.4 & $2\times 10^4$ & 10 &  & 2 & 0.42  & 1 & 1 
		& 3.2 & 0.008 & 4.01 & 3.1 &  0.0078 & 2 & 
		$3\times 10^{-3}$ & $7 \times 10^{-5}$ & 4 &  5 & 2.2 \\
		\hline
		\bf M2e4\_ISRF10 & $2\times 10^4$ & 10 &  & 2 & 4.2 & 1 & 10 
		& 3.2 & 0.008 & 2.03 & 10 &  0.78 & 0.02 & 
		$3\times 10^{-3}$ & $7 \times 10^{-5}$ & 0.1 &  10 & 1.6 \\
		\hline
		\bf M2e4\_ISRF100 & $2\times 10^4$ & 10 &  & 2 & 4.2 & 1 & 100 
		& 3.2 & 0.008 & 2.03 & 10 &  0.78 & 0.02 & 
		$3\times 10^{-3}$ & $7 \times 10^{-5}$ & 0.1 &  11 & 1.3 \\
		\hline
		\bf M2e4\_Z01 & $2\times 10^4$ & 10 &  & 2 & 4.2 & 0.1 & 1 
		& 3.2 & 0.008 & 2.03 & 10 &  0.78 & 0.02 & 
		$3\times 10^{-3}$ & $7 \times 10^{-5}$ & 0.1 &  7 & 1.1 \\
		\hline
		\bf M2e4\_Z001 & $2\times 10^4$ & 10 &  & 2 & 4.2 & 0.01 & 1 
		& 3.2 & 0.008 & 2.03 & 10 &  0.78 & 0.02 & 
		$3\times 10^{-3}$ & $7 \times 10^{-5}$ & 0.1 &  4 & 1.6 \\
		\hline
	\end{tabular}
        \vspace{-0.1cm}
 \caption{Simulations used in this paper described with STARFORGE label conventions. \textit{Top}: Physics modules included, see \citetalias{grudic_starforge_methods} for details on the individual physics modules. 
 \textit{Bottom}: Initial conditions of our simulated clouds, where $M_0$, $R_{\mathrm{cloud}}$, $\alphaturb$. $\mu$, $Z$ and $e_\mathrm{ISRF}$ are the initial cloud mass, size, virial parameter, mass to magnetic flux ratio, metallicity and the energy density of the ISRF, respectively. Note these runs  explicitly evolve the radiation field so the initial gas-dust temperature is set by the ISRF. We also report the initial 3D turbulent velocity dispersion $\sigma$, thermal virial parameter $\alphath$, total virial parameter $\alpha$, Alfv\'{e}n Mach number $\mach_{\rm A}$, plasma $\beta$, magnetic virial parameter $\alphaB$, as well as the relative Jeans, sonic and magnetic mass scales (note that these are all defined assuming as 10 K gas temperature, see \S2 in \citealt{Guszejnov_isoT_MHD} for definitions). 
 Note that \textit{Box} runs have slightly different initial parameters (e.g., Mach number, virial parameter) due to the non-exact scaling of the turbulent driving, so the values shown here are the target values that are to be reached at the end of the initial turbulent driving phase of 5 crossing times. In the last two column we show the final star formation efficiency ($\SFE=M_*/M_0$) and the disruption time for the Sphere runs, see \citetalias{guszejnov_starforge_imf} for detailed star formation histories. For the fiducial run these columns also show the standard variations between the three runs that were run with different initial turbulent realizations.}
 \label{tab:IC_phys}\vspace{-0.5cm}
\end{table*}

\subsection{Multiplicity calculation and metrics}\label{sec:metrics}

To derive multiplicity statistics in our simulation snapshots we first need to identify bound systems. We do so by using the hierarchical algorithm introduced by \cite{bate12a}, which has the following steps:
\begin{enumerate}
\item Calculate the binding energy between all pairs of stars.
\item Find the most bound pair (i.e., having the lowest total energy) and replace it with a single point mass with the same total mass and momentum, located at the center of mass of the removed pair.
\item Recursively repeat steps 1 and 2 until no more bound stars are left, with the exception that we do not combine pairs if the resulting bound aggregate would consist of more than 4 individual stars. If such an aggregate is the most bound pair at any point, we proceed to the next most bound pair, terminating if no other bound pair exists. 
\end{enumerate}

Using the above algorithm we produce a list of bound systems. For each star in these systems we assign one of the following labels:
\begin{enumerate}
	\item \textit{Unbound}: The star is not bound to any other stars.
	\item \textit{Primary}: The star is the most massive (primary) star of a multiple star system.
	\item \textit{Non-primary}: The star is part of a multiple star system, but it is not the primary star.
\end{enumerate}
Following the definitions from the literature (e.g., \citealt{duchene2013}) we introduce a set of multiplicity metrics (summarized in Table \ref{tab:mutip_param_guide}), starting with the \emph{multiplicity fraction} $\MF$:
\be
\MF \equiv \frac{B+T+Q}{S+B+T+Q},
\label{eq:multiplicity_fraction}
\ee
where $S,B,T,Q$ are the number of single, binary, triple, quadruple systems whose primary star is in mass bin $M$. Similarly we introduce the \emph{companion frequency} $\CF$:
\be
\CF \equiv \frac{B+2T+3Q}{S+B+T+Q},
\label{eq:companion_frequency}
\ee
which is the average number of companions in systems with primary mass of $M$. 
Due to the relatively small number of high mass stars in the simulations ($\sim 30$ have $>10\,\msun$ out of $\sim 2000$ stars), the uncertainty of $\MF$ and $\CF$ can be significant in high mass bins. We estimate the errors using a Bayesian method where we assume the number of multiples and companions follow a binomial and Poisson distribution respectively, see Appendix \ref{app:CF_MF_error} for details.

\begin{figure}
\begin {center}
\includegraphics[width=\linewidth]{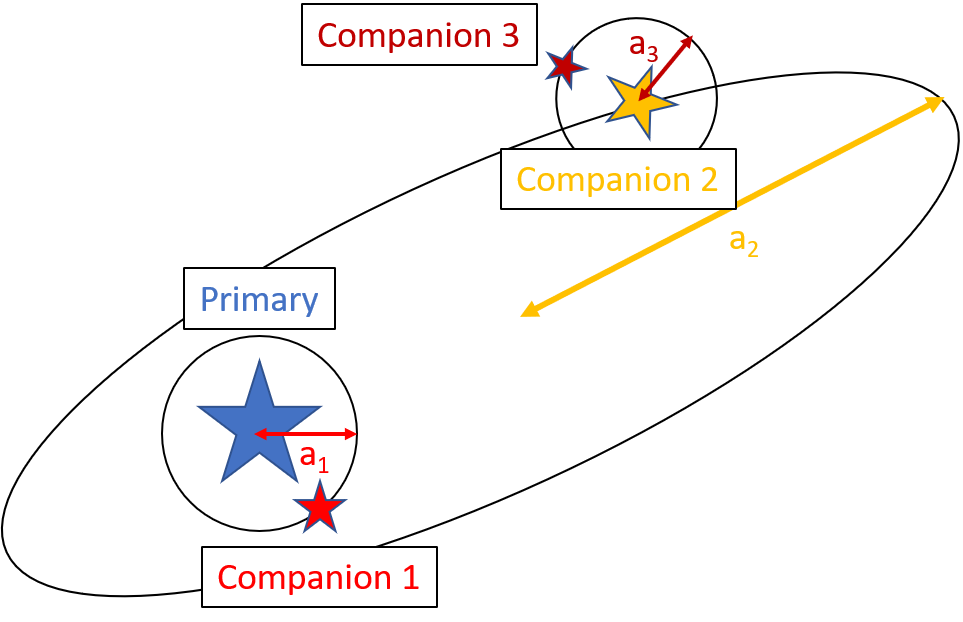}
\vspace{-0.4cm}
\caption{Cartoon illustration of the definition of orbits for each companion in a quadruple system. When calculating the semi-major axes or orbital periods for a companion, we take the orbit of the highest level subsystem that contains the chosen star but not its companion. We substitute the properties of these subsystems into Eqs \ref{eq:semi_major_def}-\ref{eq:period_def}. To calculate the semi-major axis or eccentricity distribution in \S\ref{sec:results_fiducial}-\ref{sec:results_var} we count the semi-major axis between all subsystems. Here, this would mean the orbits with semi-major axes of $a_1$, $a_2$ and $a_3$.}
\label{fig:semi_major_illustration}
\vspace{-0.5cm}
\end {center}
\end{figure}

For companions we characterize the companion separation as the semi-major axis $a$ with respect to the primary using the well-known two-body solution to the Kepler problem that yields 
\be
a = \frac{G M_1 M_2} {2 E_\mathrm{total}},
\label{eq:semi_major_def}
\ee
where $G$ is the gravitational constant, $M_1$ and $M_2$ are the masses of the primary and companion stars, and $E$ is the kinetic+gravitational energy of the system. Similarly, we calculate the orbital period $P$ as
\be
P = 2 \mathrm{\pi} \sqrt{\frac{a^3}{G (M_1+M_2)}}.
\label{eq:period_def}
\ee
To compute these quantities for higher order systems we adopt the orbit and mass of the highest level subsystem that contains the chosen star but not the primary, see Figure \ref{fig:semi_major_illustration} for an illustration. For each orbit we also calculate the corresponding eccentricity
\be
e = \sqrt{\frac{h^2}{a G (M_1+M_2)}-1},
\label{eq:eccentricity}
\ee
where $h$ is the specific relative angular momentum of the two-body system.

\begin{table}
    \setlength\tabcolsep{2.0pt} 
	\centering
	\begin{tabular}{ | c | c | }
	\hline
	Property & Definition \\
	\hline
	$M_p$ & Mass of the primary (i.e., most massive) star in the system \\
	\hline
	$q\equiv M/M_p$ & Mass ratio of a companion to the primary \\
	\hline
	$\MF$ & Multiplicity fraction for stars of $M_p$ primary mass, see Eq. \ref{eq:multiplicity_fraction} \\
    \hline
	$\CF$ & Companion frequency for stars of $M_p$ primary mass, see Eq. \ref{eq:companion_frequency} \\
    \hline
	$a$ & Semi-major axis of companion's orbit, see Eq. \ref{eq:semi_major_def} \\
	\hline
	$P$ & Orbital period of the companion, see Eq. \ref{eq:period_def} \\
    \hline
	$e$ & Eccentricity of companion's orbit, see Eq. \ref{eq:eccentricity} \\
    \hline
    \end{tabular}
    \vspace{-0.1cm}
 \caption{List of multiplicity properties used throughout the paper along with their definition.}
 \label{tab:mutip_param_guide}\vspace{-0.5cm}
\end{table}

To compare these metrics with observations we apply two corrections to the \myquote{raw} multiplicity properties of the simulations. First, all companions with  mass ratio $q<0.1$ are ignored, as most observations are incomplete in this regime, see  \citet{moe_distefano2016} henceforth referred to as \citetalias{moe_distefano2016}. We note that the exact choice of the cut-off $q$ value can significantly affect the companion frequency at the high mass end of the IMF, since there are many Solar-type companions around $>10\,\msun$ stars in our simulations. When specifically comparing the properties of simulated and observed Solar-type stars, we account for observational incompleteness by discarding shorter period companions ($\mathrm{Log} P<4.5$, or $a<30\,\AU$) for which observations are incomplete for $q<0.5$ and low-$q$ longer period binaries ($5.9<\mathrm{Log} P<6.7$, or $150<a/\AU<400$), which are only detectable for $q>0.2$ (see Figure 28 in \citetalias{moe_distefano2016}). The second correction we apply removes all short-lived companions from the distribution, i.e., stars that have only been companions to their primaries for $t_{\rm comp}<100\,\mathrm{kyr}$ or have not yet completed two orbits. This correction removes binary assignments that are the result of chance alignments between stars (i.e., cases where the pairwise comparison considers two stars bound but they are not when accounting for all interactions), see \S\ref{sec:results_fiducial} for details. 
Source confusion is likely only important in crowded regions, and indeed, we find that this correction has a relatively minor effect on the statistics. Note that we report the \myquote{raw} simulation values unless specified otherwise. 

Finally, we also examine the multiplicity properties of young stellar objects (YSOs). Observed YSOs are classified according to their spectral energy distributions \citep{Dunham_2014_protostar_evol_review}, which requires radiative transfer post-processing \citep[e.g.,][]{Offner_2012_protostar_SED_synthetic_obs}. Instead, we take a simpler approach and define YSOs in the simulation as stars (sink particles) that are younger than 0.5 Myr, as 0.5 Myr is approximately the Class 0 + Class I lifetime \citep{Dunham_2014_protostar_evol_review}.

  \section{Multiplicity properties for the fiducial cloud}\label{sec:results_fiducial}
  
We run our fiducial cloud (\textbf{M2e4}) until star formation is quenched, and the cloud is fully disrupted by stellar feedback (see Figure \ref{fig:M2e4_series}). We identify star systems in all snapshots of the run using the method outlined in \S\ref{sec:metrics}. Unless stated otherwise we show multiplicity properties at the end of the simulation, when star formation has quenched and the cloud has been fully disrupted. 

\begin{figure*}
\begin {center}
\includegraphics[width=0.33\linewidth]{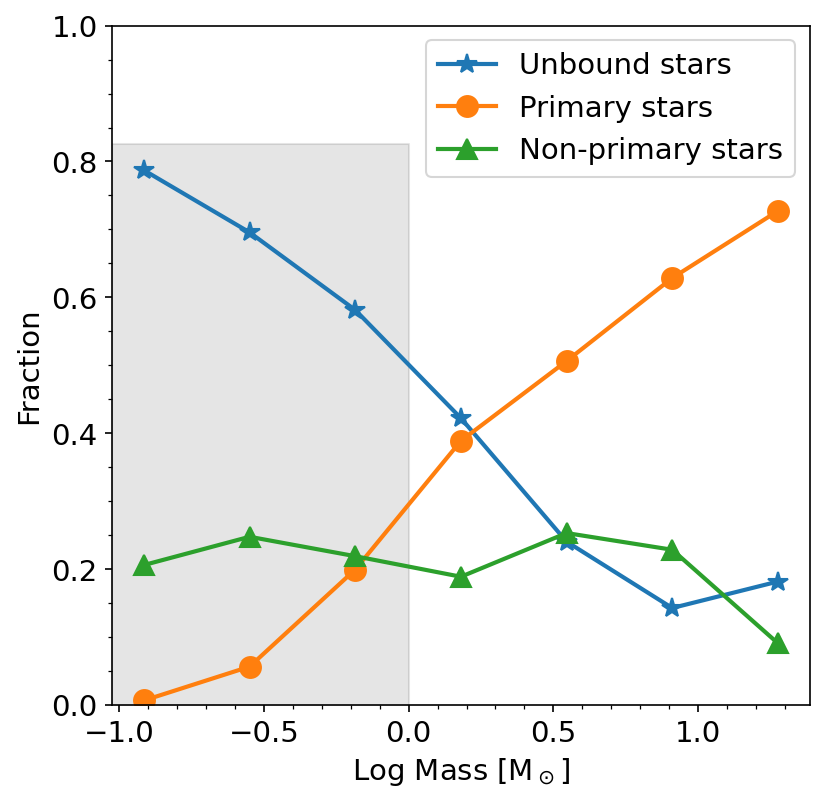}
\includegraphics[width=0.33\linewidth]{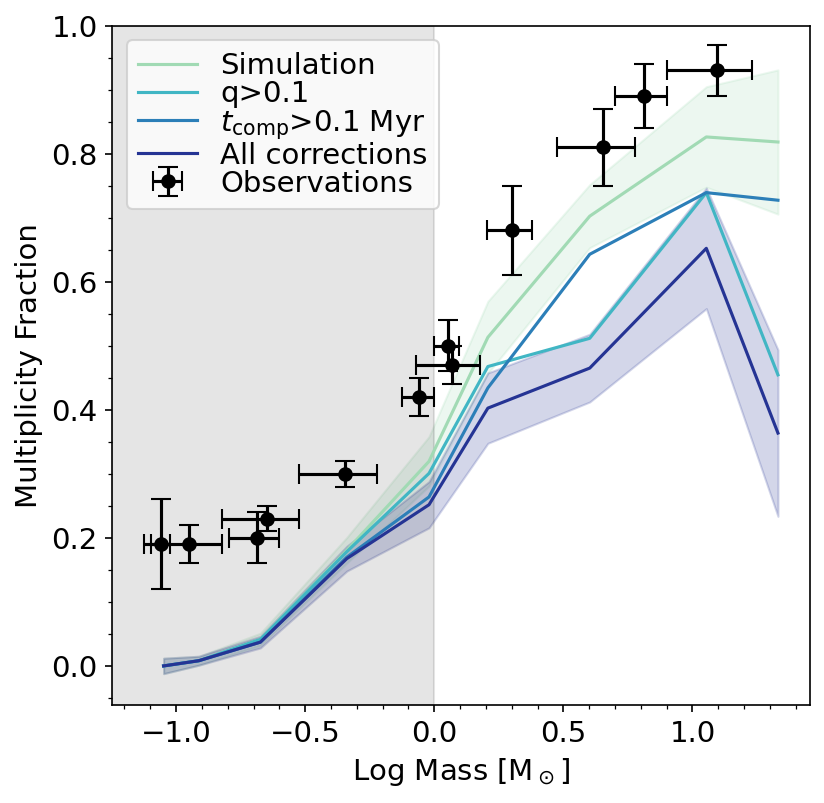}
\includegraphics[width=0.33\linewidth]{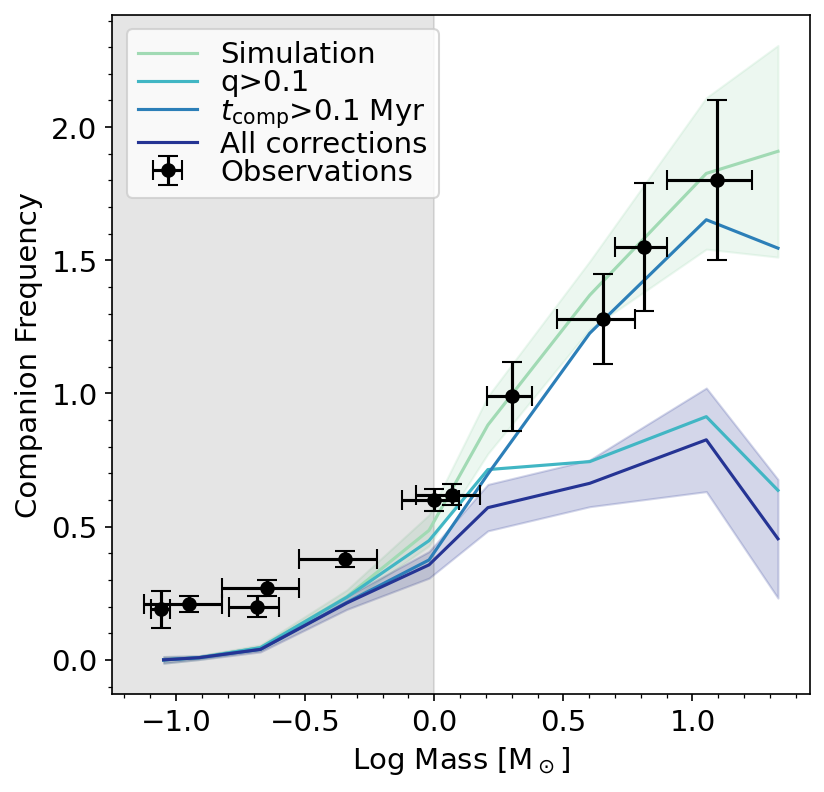}
\vspace{-0.4cm}
\caption{ Basic multiplicity properties at the end of the simulation in the fiducial run (\textbf{M2e4}) with a grey shaded region showing the mass range potentially affected by the $0.1\,\msun$ completeness limit of the simulations. \textit{Left}: Fraction of stars in different boundedness categories. \textit{Middle \& Right}: Multiplicity fraction ($\MF$) and companion frequency ($\CF$) as a function of primary stellar mass, showing the raw simulation values (solid) and the distributions after removing temporary companions, low mass-ratio companions and both. Shaded regions show the 1-$\sigma$ uncertainties, which are estimated using Eqs. \ref{eq:MF_sigma}-\ref{eq:CF_sigma} for $\MF$ and $\CF$ respectively. Observed values are taken from the review of \citet{offner_PPVII_multiplicity}. For an analysis of trends see \S\ref{sec:results_fiducial} in the main text.}
\label{fig:fiducial_CF_MF}
\vspace{-0.5cm}
\end {center}
\end{figure*}

The left panel of Figure \ref{fig:fiducial_CF_MF} shows the fraction of stars in different boundedness categories (see \S\ref{sec:metrics}) as a function of mass. As expected we find that massive stars ($>5\,\msun$) are more likely to be primary stars. We find that, up to companion masses of $\sim$~$10~M_{\odot}$, 20\% of stars are companions to more massive primary stars. These statistics derive from the fact that most high-mass stars are in multiple systems, while most low-mass stars are not. The middle and right panels of Figure \ref{fig:fiducial_CF_MF} show that the multiplicity fraction ($\MF$) and companion frequency ($\CF$) increase with the primary mass and are qualitatively similar to the observed values. Note for stars with masses below $1\,\msun$ both the $\MF$ and $\CF$ are affected by the $0.1\,\msun$ completeness limit of the simulation; the simulation does not resolve brown dwarfs. Removing short-lived companions has a mild effect on both $\MF$ and $\CF$. Applying an observationally-motivated $q>0.1$ cut-off, however, significantly reduces both the $\MF$ and $\CF$ for high-mass stars. This is because many high-mass stars in the simulations have Solar-type companions,
such that the system mass ratio falls just below the cut-off. Overall, we find that after corrections the simulations produce qualitatively similar but significantly lower values than those observed for both $\MF$ and $\CF$.

\subsection{Companion properties}\label{sec:results_fiducial_comp_properties}

\begin{figure}
\begin {center}
\includegraphics[width=\linewidth]{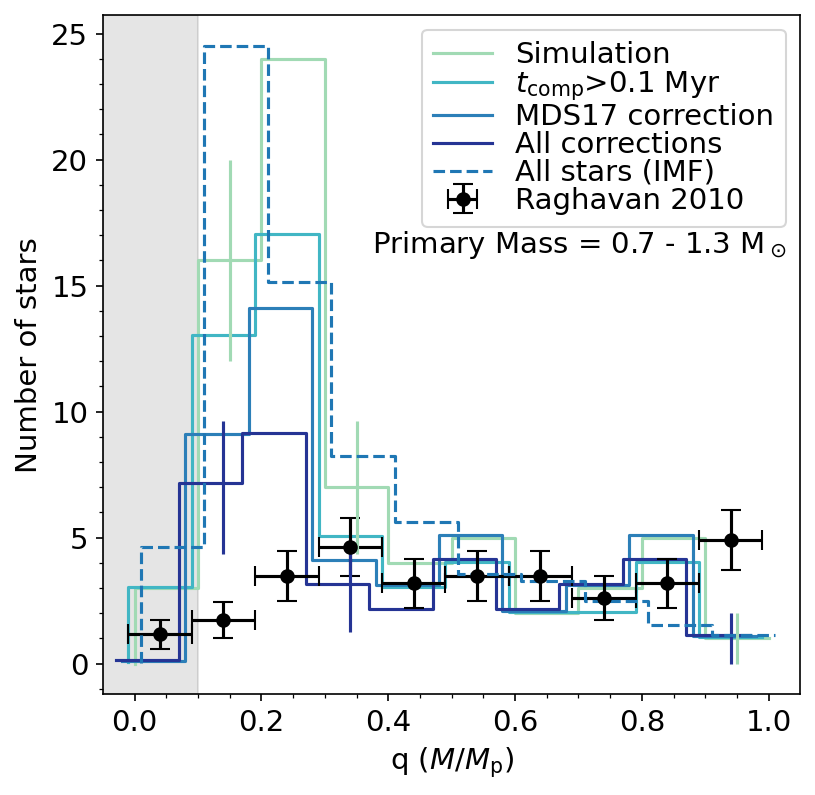}
\vspace{-0.4cm}
\caption{Distribution of the companion mass ratio $q$ for Solar-type stars in the fiducial run, using a similar notation to Figure \ref{fig:fiducial_CF_MF}, where we account for observational incompleteness reported in \citetalias{moe_distefano2016}. Bars show the Poisson error of the distributions in key bins. For a fixed primary mass each $q$ value corresponds to a stellar mass scale, so we plot the $<1\,\msun$ part of the stellar mass distribution (IMF, dashed line). We also show the observations for Solar-type binaries from \citet{raghavan2010}. A grey shaded region shows the mass ratio range potentially affected by the $0.1\,\msun$ completeness limit of the simulation. Overall the $q$-distribution before corrections is consistent with random sampling the IMF (dashed line) and only shows significant differences after removing low $q$ value companions. After all corrections are applied the distribution is much closer to the observed flat trend, with a significant absence of near-equal mass companions. }
\label{fig:fiducial_q_dist}
\vspace{-0.5cm}
\end {center}
\end{figure}

Observations find the distribution of the companion mass ratio $q$ is mostly flat for Solar-type stars, except for a peak at near-equal masses \citep{raghavan2010, offner_PPVII_multiplicity}. Figure \ref{fig:fiducial_q_dist} shows that in our simulation the distribution is not flat and exhibits a peak at $q\sim 0.2$. Comparing with the normalized stellar mass function of the simulation, i.e., the IMF, we find that the companion mass ratio distribution for Solar-type stars is consistent with random sampling from the IMF. After applying a correction for observational incompleteness of short period, low mass ratio companions (based on \citetalias{moe_distefano2016}) the distribution becomes flatter with a marginal peak at $q\sim 0.2$. Note that this marginal peak is dominated by low mass ratio, short-period companions so the significance of the peak strongly depends on the estimated observational completeness limit, which \citetalias{moe_distefano2016} estimated to be 25\%. After these corrections the distribution is qualitatively consistent with the observed trend, except for the lack of peak at unity mass ratio. 

\begin{figure*}
\begin {center}
\includegraphics[width=0.33\linewidth]{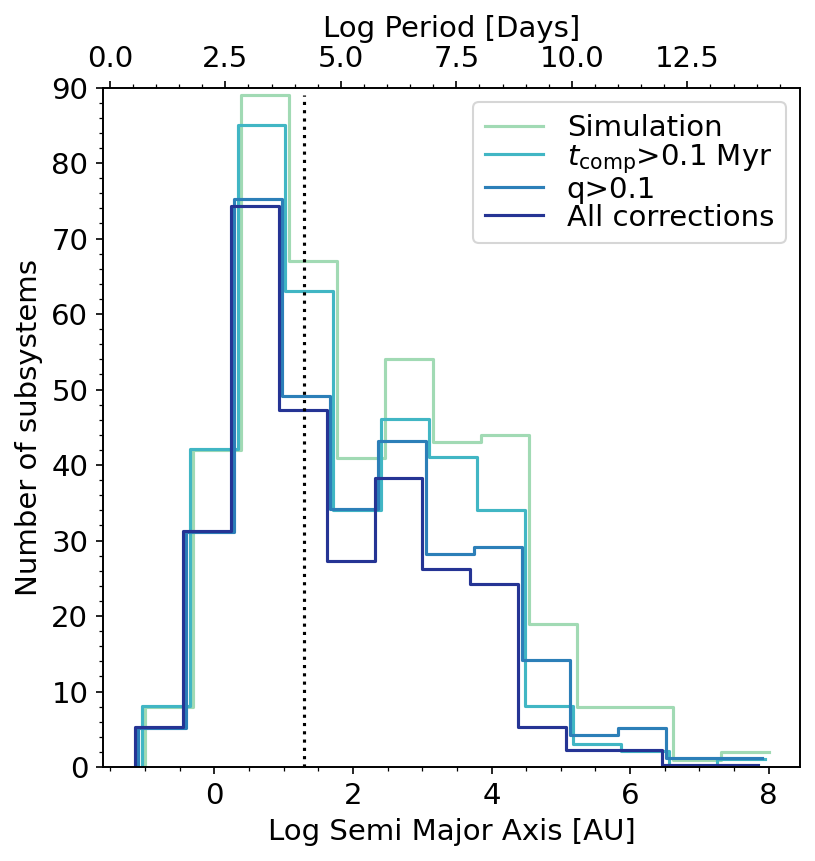}
\includegraphics[width=0.33\linewidth]{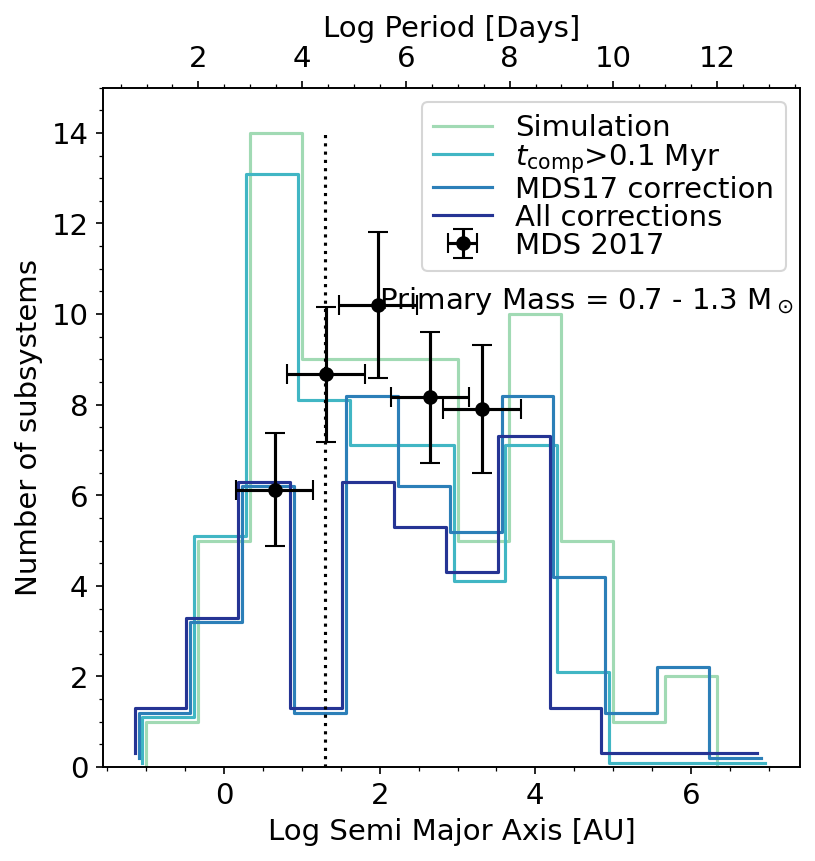}
\includegraphics[width=0.33\linewidth]{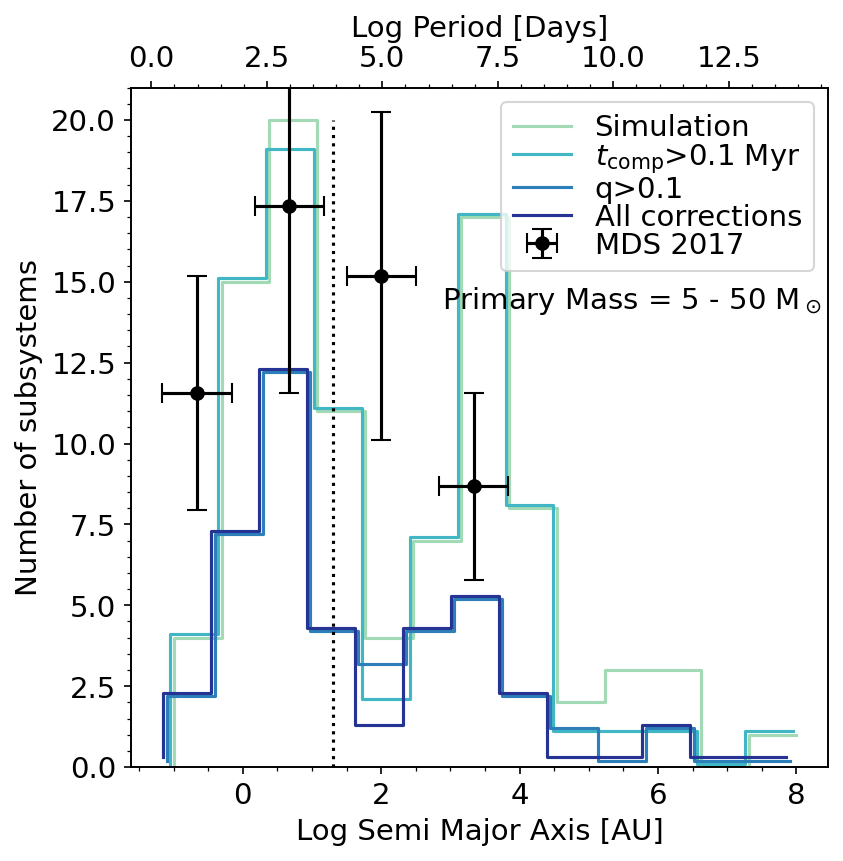}
\vspace{-0.4cm}
\caption{\textit{Left}: Semi-major axis/orbital period distribution for star systems in the fiducial run (\textbf{M2e4}), using the same notation as Figure \ref{fig:fiducial_CF_MF}. A vertical line represents the 20 AU gravitational softening length of the simulation. \textit{Middle \& Right}: Same but for Solar-type and massive ($>5\,\msun$) primaries only, also showing the corresponding observations from \citetalias{moe_distefano2016}. Note that for Solar-type stars we account for observational incompleteness based on \citetalias{moe_distefano2016}.}
\label{fig:fiducial_semimajor}
\vspace{-0.5cm}
\end {center}
\end{figure*}

Figure \ref{fig:fiducial_semimajor} shows the period/semi-major axis distribution for all stars, as well as for Solar-type and massive ($>5\,\msun$) stars only.  
In all cases we find a peak close to the gravitational softening length ($\sim 20\,\AU$), below which gravitational forces are artificially weakened. Above this value the number of companions declines with distance. Note that removing temporary companions significantly reduces the number of wide binaries, while removing low-mass ratio companions affects all scales. Note that in the case of Solar-type stars removing $q<0.1$ companions have little effect as we ignore all brown dwarfs in the simulation as they are below the completeness limit of the simulations presented in this work. However, observations of Solar-type and lower mass stars are incomplete for higher values of $q$ \citep{offner_PPVII_multiplicity}. Fig. \ref{fig:fiducial_semimajor} also shows the result if we also account for observational incompleteness of Solar-type stars for $q<0.5$ companions with $<30\,\AU$ separations. After these corrections, the simulation qualitatively agrees with observations from \citetalias{moe_distefano2016} for wide binaries, but the pile-up at the gravitational softening scale, which is likely numerical, prevents a detailed comparison (e.g., comparing the statistical significance of the apparent peak at 100 AU for Solar-type stars).

\begin{figure}
\begin {center}
\includegraphics[width=\linewidth]{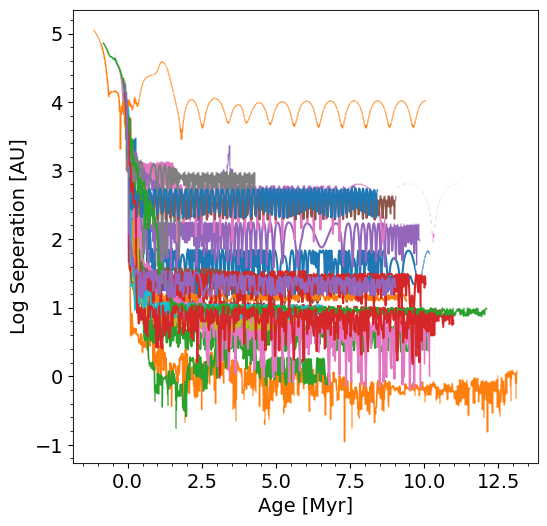}
\vspace{-0.4cm}
\caption{Separation of companions to their primary stars for a small sample of systems in the fiducial run as a function of the age of the bound system (note that the stars can form before the system becomes bound, hence the negative time values). Note that simulation snapshots are $\Delta t_\mathrm{snap}\approx 7\,\mathrm{kyr}$ apart, distorting the orbits of low separation companions. Overall we find that most companions form at larger distances ($\sim 10^4\,\AU$) and become bound almost immediately, then reach a stable orbit within $\sim 0.5\,\mathrm{Myr}$.}
\label{fig:fiducial_separation_evol_sample}
\vspace{-0.5cm}
\end {center}
\end{figure}

\begin{figure*}
\begin {center}
\includegraphics[width=0.49\linewidth]{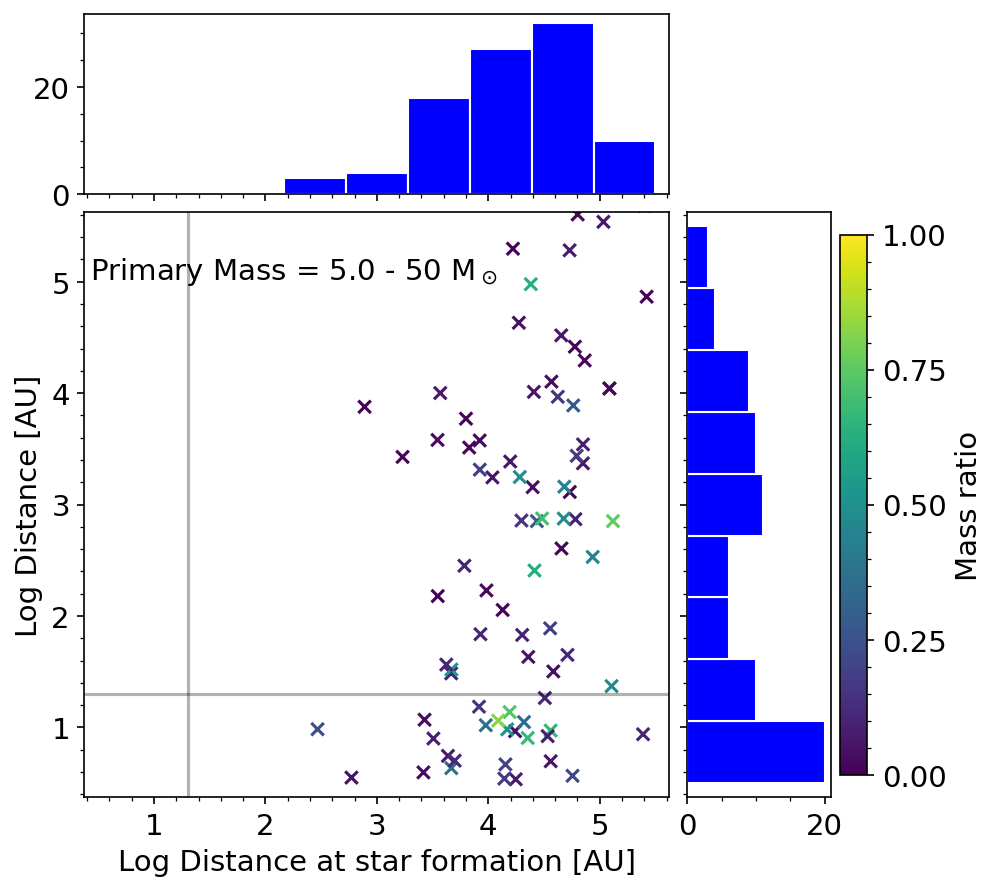}
\includegraphics[width=0.49\linewidth]{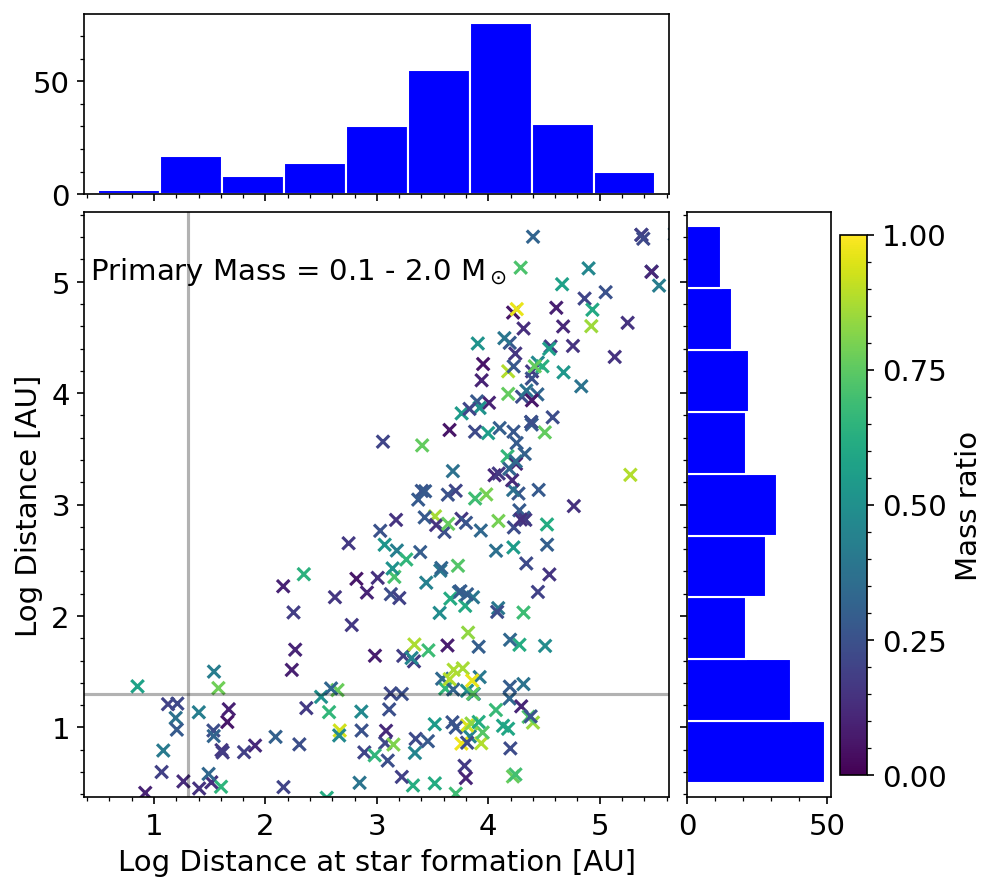}
\caption{Separation between primary and companions in the fiducial simulation at the snapshot just after the stars form and at the end of the simulation. Results are shown separately for high mass ($>5\,\msun$, left) and lower mass ($<2\,\msun$, right) primary stars. The symbols in the main scatter plot are colored according to the companion mass ratio $q$ in the system at the end of the simulation. The distributions of the individual metrics is shown above their respective axes. A horizontal line show the gravitational softening length of the simulation, which is also the exclusion radius of sink particles (i.e., no stars can form closer than this) denoted by a vertical line. There are still a few stars that appear to have formed at shorter distances, but this is just an artifact of us relying on discrete snapshots of the simulation ($\Delta t_\mathrm{snap}\approx 7\,\mathrm{kyr}$). Note that we are showing the data after having applied the corrections for both low mass ratio and temporary companions. }
\label{fig:fiducial_separation_dist}
\vspace{-0.5cm}
\end {center}
\end{figure*}

In addition to the semi-major axis distribution, it is instructive to see the separation distribution between primaries and companions, which we define as the instantaneous 3D distance between the pair positions. The stars, particularly just after formation, are not on stable orbits with a well-defined semi-major axis and pair separations and may evolve rapidly \citep{Offner_2010_turbulent_fragmentation_binary,Lee_2019_MHD_binary_separations} as shown in Figure \ref{fig:fiducial_separation_evol_sample}. We compare the separation at formation between primaries and their companions with their separations at the end of the simulation. The former quantity reflects the initial conditions and characteristics of the mechanism by which the multiples form. Figure \ref{fig:fiducial_separation_dist} shows the distributions for both separation metrics for massive ($>5\,\msun$) and lower mass ($<2\,\msun$) primaries. We find that most companions in the simulations formed between 1000 and 10000 AU from their primaries, which is the expectation for multiples formed via turbulent fragmentation \citep{Fisher_2004_binaries,Offner_2010_turbulent_fragmentation_binary,Offner_2016_jets,guszejnov_correlation}. 
As a result of dynamical interactions most of these companions end up with much closer separations than their initial birth separation (see Figure \ref{fig:fiducial_separation_evol_sample}). A significant fraction of companions migrate inwards until they reach scales at which gravitational softening impacts the dynamics, creating a peak in the distribution near the gravitational softening length. There is no clear trend in this behavior with regards to the companion mass ratio $q$: massive companions are as likely to \myquote{spiral in} as lower mass ones. Like prior numerical studies, we find that the separation evolution happens on a relatively short timescale of $<0.5\,\mathrm{Myr}$. Note that a few companions appear to have initial separations of $\lesssim 10^2\,\AU$. Due to the snapshot time increment ($\approx 7\,\mathrm{kyr}$), we don't have the separations at the moment of formation and, consequently, these short distances likely represent early rapid dynamical evolution, rather than formation on these scales. Note that if the simulation included the formation of multiples from unstable disks, we would expect to see a larger number of companions forming at such short separations.  

\begin{figure}
\begin {center}
\includegraphics[width=\linewidth]{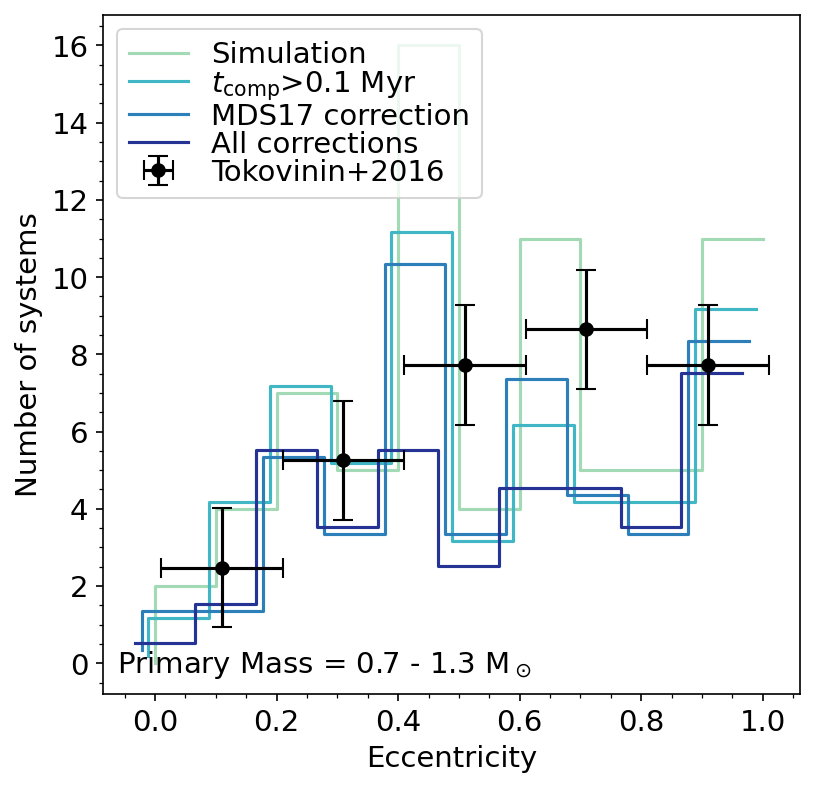}
\vspace{-0.4cm}
\caption{Distribution of the companion eccentricities for Solar-type stars in the fiducial run, using a similar notation to Figure \ref{fig:fiducial_CF_MF}, where we account for observational incompleteness reported in \citetalias{moe_distefano2016}. We also show the observations for Solar-type stars from \citet{Tokovinin_2016_solar_eccentricity_dist}. Overall the eccentricity distribution in the simulation is consistent with observations before corrections, but show a lack of companions at $e\sim 0.7$ after corrections.}
\label{fig:fiducial_eccentricity_dist}
\vspace{-0.5cm}
\end {center}
\end{figure}

In addition to the semi-major axis we calculate the eccentricity of each orbit. In Figure \ref{fig:fiducial_eccentricity_dist} we compare our results with the observations of \citet{Tokovinin_2016_solar_eccentricity_dist}, who examined companions of Solar-type stars with $>50\,\AU$ separation, which is above the $\sim20\,\AU$ gravitational softening length of the simulation. We find good agreement for all eccentricity values with the raw data. However applying all corrections leads to a deficit of companions in the $0.4<e<0.8$ range.

\begin{figure*}
\begin {center}
\includegraphics[width=0.45\linewidth]{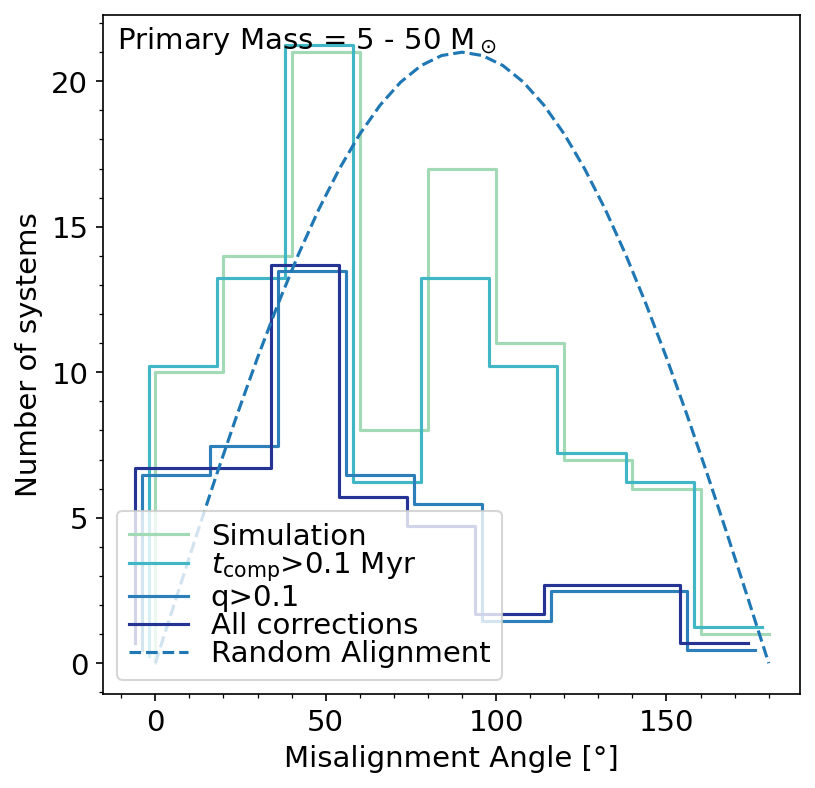}
\includegraphics[width=0.45\linewidth]{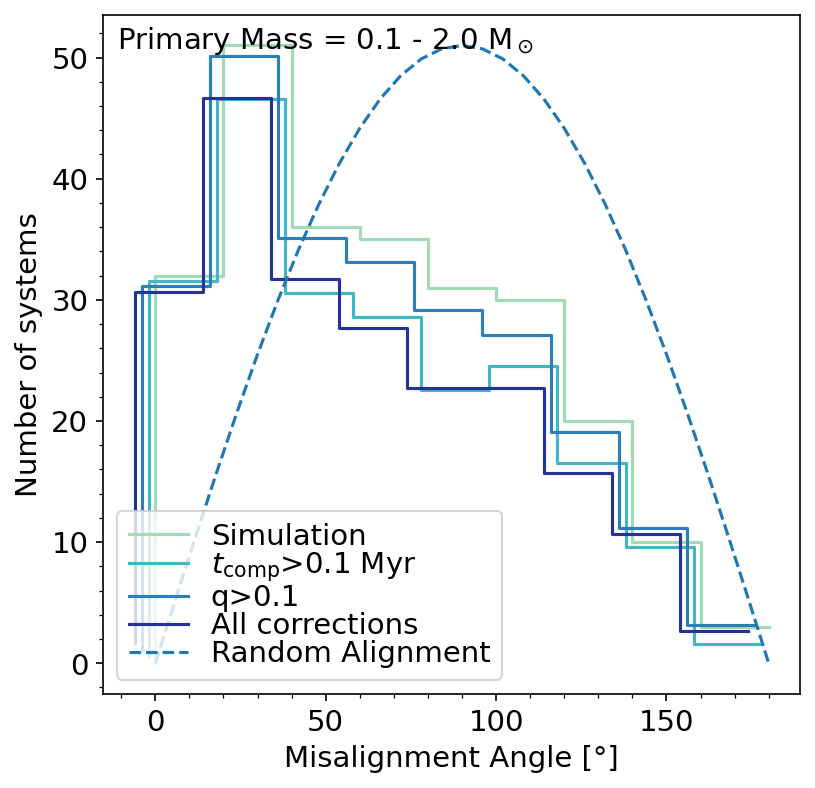}
\vspace{-0.4cm}
\caption{Distribution of the misalignment angle between primaries and companions for high mass ($>5\,\msun$, left) and lower mass ($<2\,\msun$, right) primary stars, using the same notation as Figure \ref{fig:fiducial_CF_MF}. We also show the distribution expected from randomly aligned companions. Overall, companions tend to be more aligned with their primaries compared to a random distribution, while massive stars exhibit slightly more misalignment.}
\label{fig:fiducial_angle_dist}
\vspace{-0.5cm}
\end {center}
\end{figure*}

The simulation tracks the angular momentum accreted by sink particles (stars), allowing us to analyse the spin alignment between stars and their companions. Note that the simulation does not allow stars to lose angular momentum via outflows or magnetic braking, so this total accreted angular momentum is significantly higher than the angular momentum of stars. Additionally, sink particles accrete angular momentum material from larger spatial scales than what may be actually accreted at the stellar surface, potentially leading to an overestimation of the accreted angular momentum. Nevertheless, the direction of the accreted angular momentum is a reasonable proxy for the direction of the stellar spin. Figure \ref{fig:fiducial_angle_dist} shows the distribution of the angle between the spins of the primary star and its companions 
for both high ($>5\,\msun$) and lower mass primaries ($<2\,\msun$). We find that in both cases companions are not randomly oriented, but instead are preferentially aligned with their primaries, a potential sign that these multiples formed via core fragmentation, however prior work found a significantly weaker preference for spin alignment \citep{Lee_2019_MHD_binary_separations}. 
However, massive primaries have a wider angle distribution, i.e., their spin is less likely to be aligned with that of their companions, which can be explained either by massive stars acquiring companions that formed in different regions or by the fact that massive stars accrete from a gas reservoir much larger than the initial core they form in. 


\begin{figure}
\begin {center}
\includegraphics[width=0.99\linewidth]{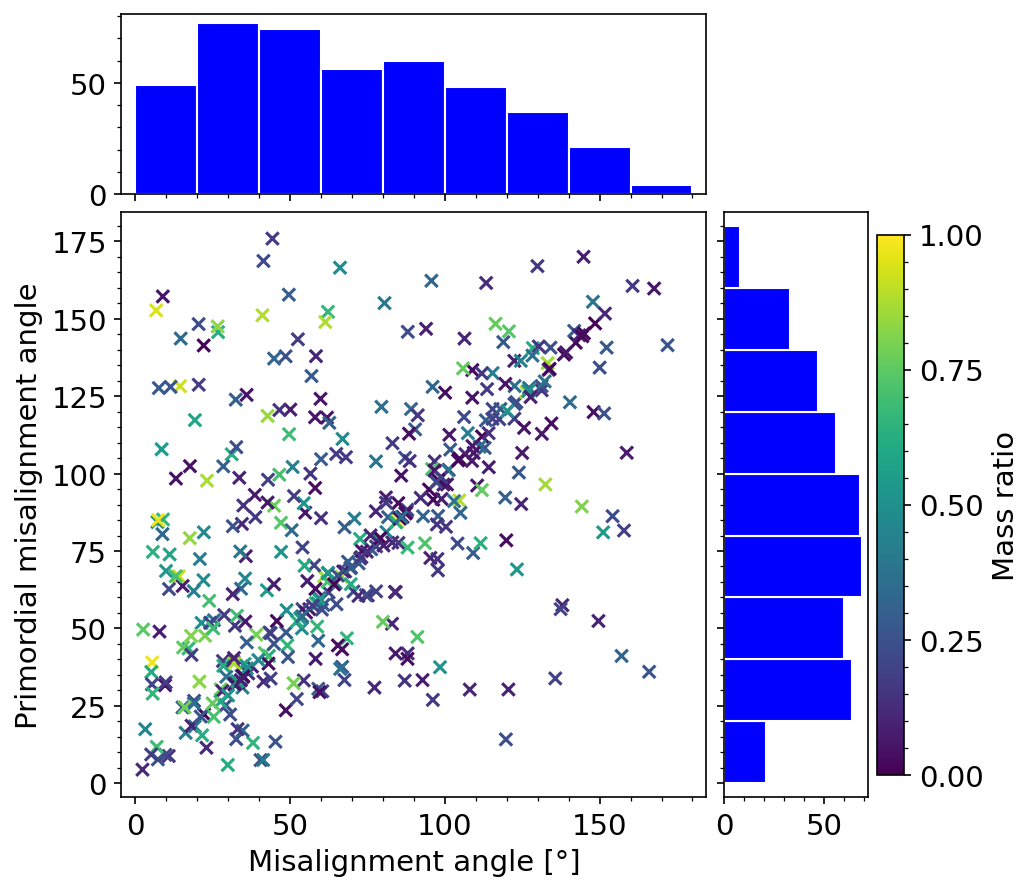}
\vspace{-0.4cm}
\caption{Misalignment angle between primaries and companions in the fiducial simulation at the snapshot just after the stars form and at the end of the simulation. The primordial misalignment angle is calculated at the time when the mass of the companion exceeds $0.08\,\msun$. The symbols and colorbar are set identical to those in Figure \ref{fig:fiducial_separation_dist}. Companions are preferentially aligned even in the early, primordial stage, and become even more aligned by the time star formation ends.}
\label{fig:primordial_angle_dist}
\vspace{-0.5cm}
\end {center}
\end{figure}

In Figure \ref{fig:primordial_angle_dist} we compare the primordial spin misalignment angle with the final value obtained at the end of the simulation. Taking the at-formation misalignment between companion and primary would yield similar results to random alignment as the spin direction of sink particles changes rapidly during the initial accretion. That is why we define primordial misalignment as the angle when the mass of the companion exceeds the mass scale of brown dwarfs ($0.08\,\msun$). Although companions tend to be aligned with their primary at both times, the primordial misalignment is closer to random. This is due to companions that accrete simultaneously with their primary star, bringing their spins closer to alignment.

\subsection{Time evolution of multiplicity properties}\label{sec:results_fiducial_evol}

\begin{figure}
\begin {center}
\includegraphics[width=\linewidth]{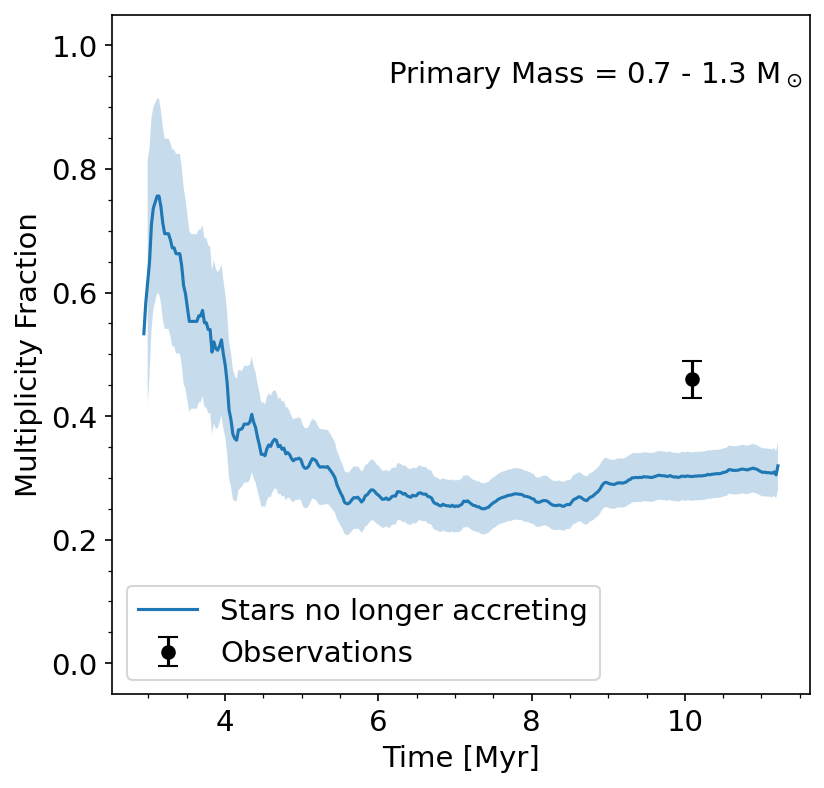}
\vspace{-0.4cm}
\caption{Multiplicity fraction of Solar-type stars that are no longer accreting (i.e., will remain Solar-type until the end of the simulation) as a function of time in the fiducial (\textbf{M2e4}) run. For reference we include the observed value for field stars from \citet{raghavan2010}.}
\label{fig:fiducial_MF_evol_solar}
\vspace{-0.5cm}
\end {center}
\end{figure}

\begin{figure*}
\begin {center}
\includegraphics[width=0.33\linewidth]{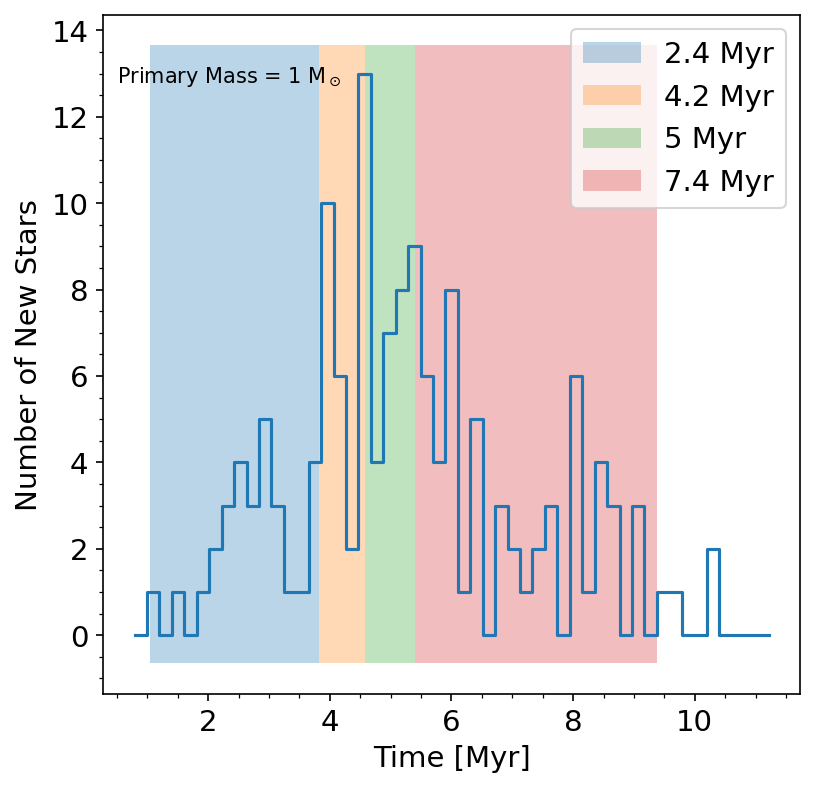}
\includegraphics[width=0.33\linewidth]{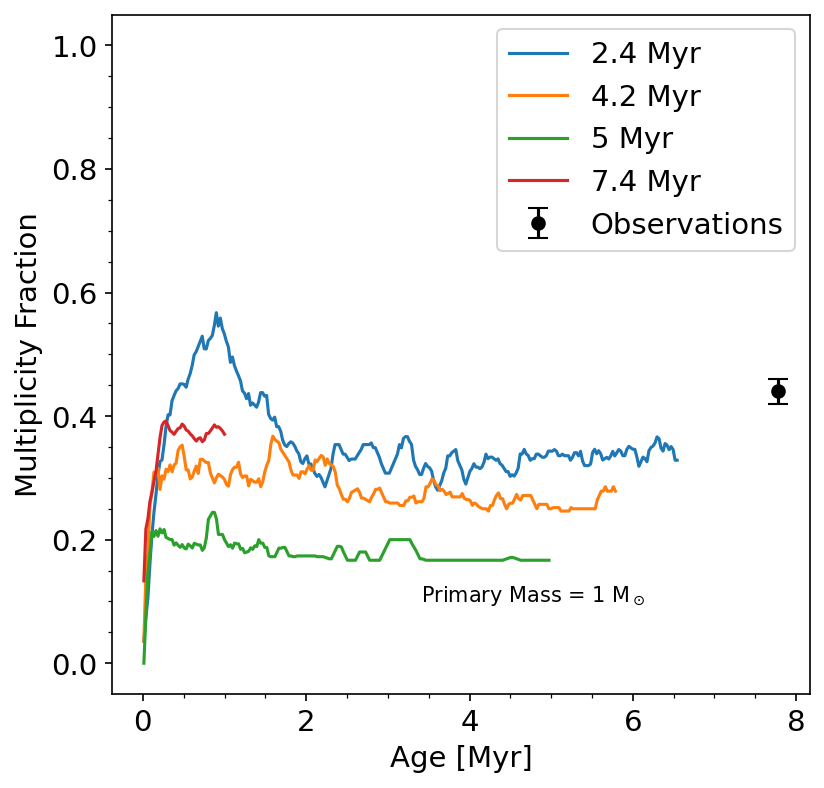}
\includegraphics[width=0.33\linewidth]{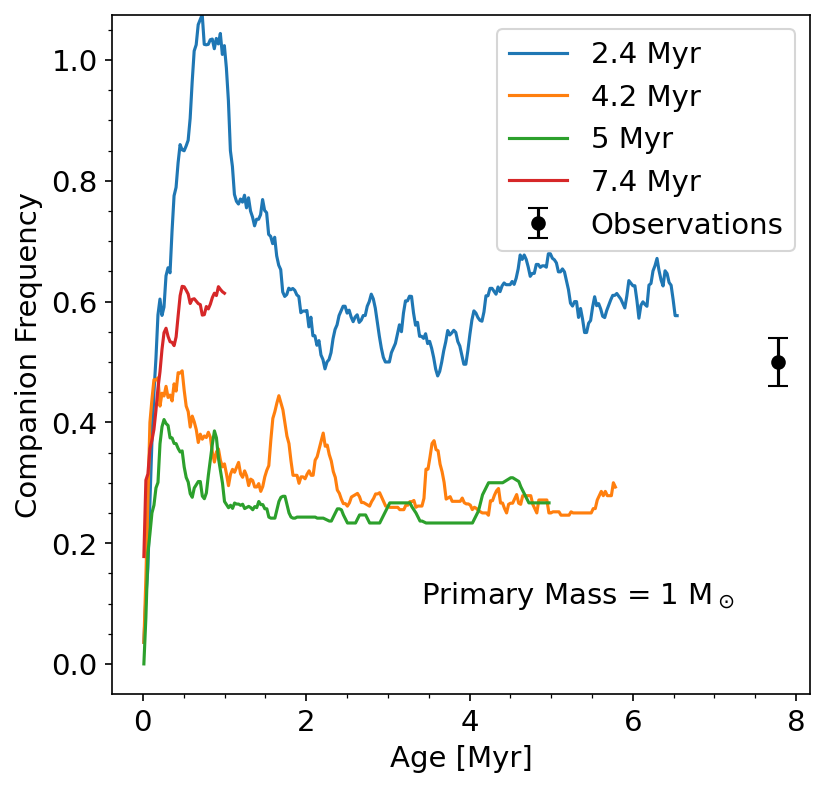}
\vspace{-0.4cm}
\caption{\textit{Left}: Distribution of formation times for Solar-type stars in the fiducial run. Colored regions show the formation time bins in which the multiplicity properties are calculated in the other panels. \textit{Middle \& Right}: The evolution of $\MF$ and $\CF$ as a function of age for Solar-type stars in various formation time bins. Note that any apparent discrepancy with Figure \ref{fig:fiducial_MF_evol_solar} is due to the different selection criteria since Figure \ref{fig:fiducial_MF_evol_solar} looks at stars that have already stopped accreting. For an analysis of trends see \S\ref{sec:results_fiducial} in the main text.}
\label{fig:fiducial_MF_CF_evol_for_timebins}
\vspace{-0.5cm}
\end {center}
\end{figure*}

Observations suggest that multiplicity evolves through dynamical interactions, such that over time systems lose members. Figure \ref{fig:fiducial_MF_evol_solar} shows the evolution of the $\MF$ for Solar-type stars in the fiducial run that have stopped accreting. We find a general decreasing trend, where Solar-type stars are much more likely to be primaries at the start of star formation than at later times. To produce this trend Solar-type stars must lose their companions as they age, or stars born at later times must have lower multiplicity. Figure \ref{fig:fiducial_MF_CF_evol_for_timebins} shows the formation rate of Solar-type stars and their multiplicity as a function of age. Note that Figure \ref{fig:fiducial_MF_evol_solar} and Figure \ref{fig:fiducial_MF_CF_evol_for_timebins} concern slightly different stellar populations: Figure \ref{fig:fiducial_MF_evol_solar} looks at Solar-type stars that have stopped accreting, while Figure \ref{fig:fiducial_MF_CF_evol_for_timebins} follows the multiplicity of stars throughout their lifetime. The $\MF$ remains roughly constant as the stars age, but the $\CF$ decreases. This means that trinary and quaternary systems containing Solar-type stars lose some companions over time but are unlikely to lose their last companion. This implies that most stars that form in a multiple system (e.g., not the earliest forming ones), \emph{stay} in a multiple system. This result is consistent with prior numerical studies \citep[e.g.,][]{Lee_2019_MHD_binary_separations} and will hold as long as the initial fraction of high-order systems is low and stellar densities are not too high. These changes, however, are relatively minor compared to the differences in both the $\MF$ and $\CF$ between stars born at different times. We find that among the first Solar-type stars that form about 40\% are primaries; among the last ones to form only 20\% are. 

\begin{figure*}
\begin {center}
\includegraphics[width=0.33\linewidth]{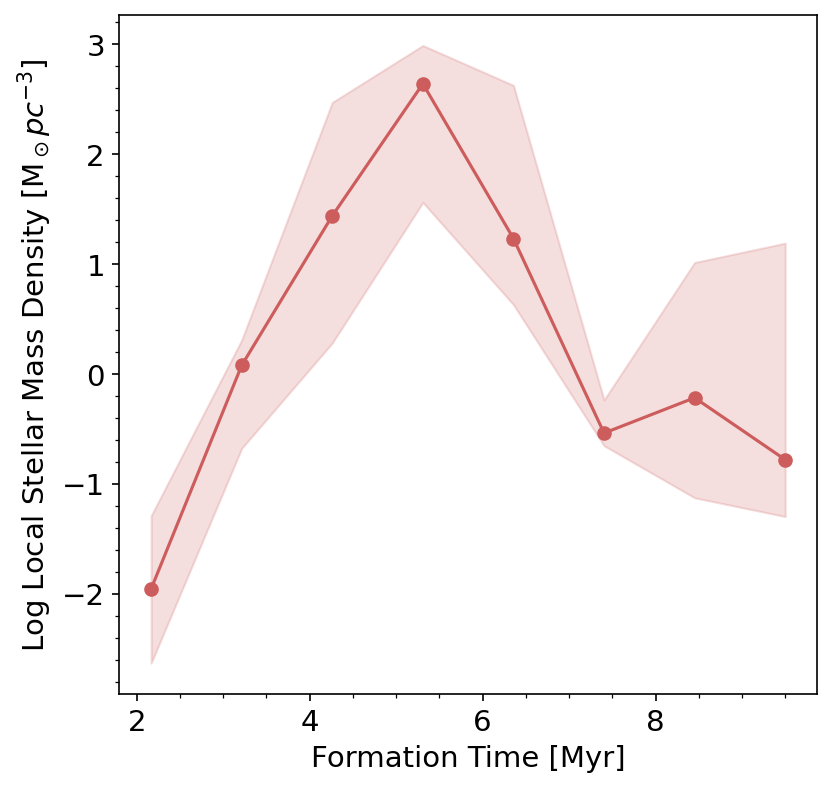}
\includegraphics[width=0.33\linewidth]{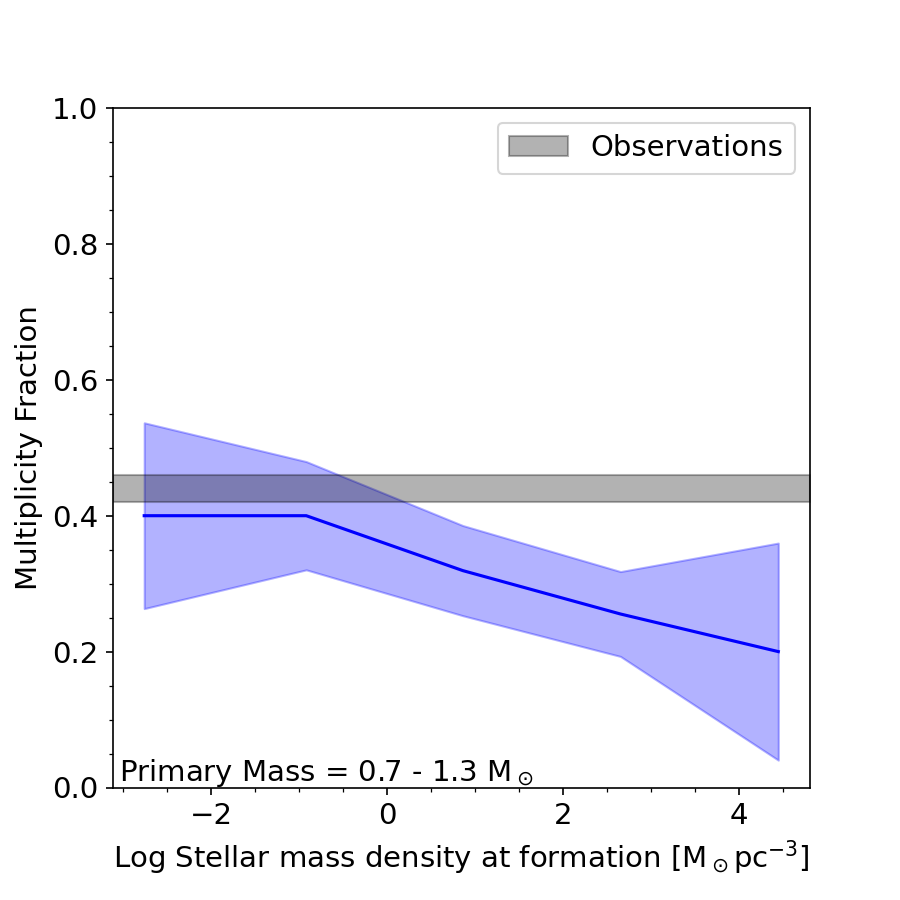}
\includegraphics[width=0.33\linewidth]{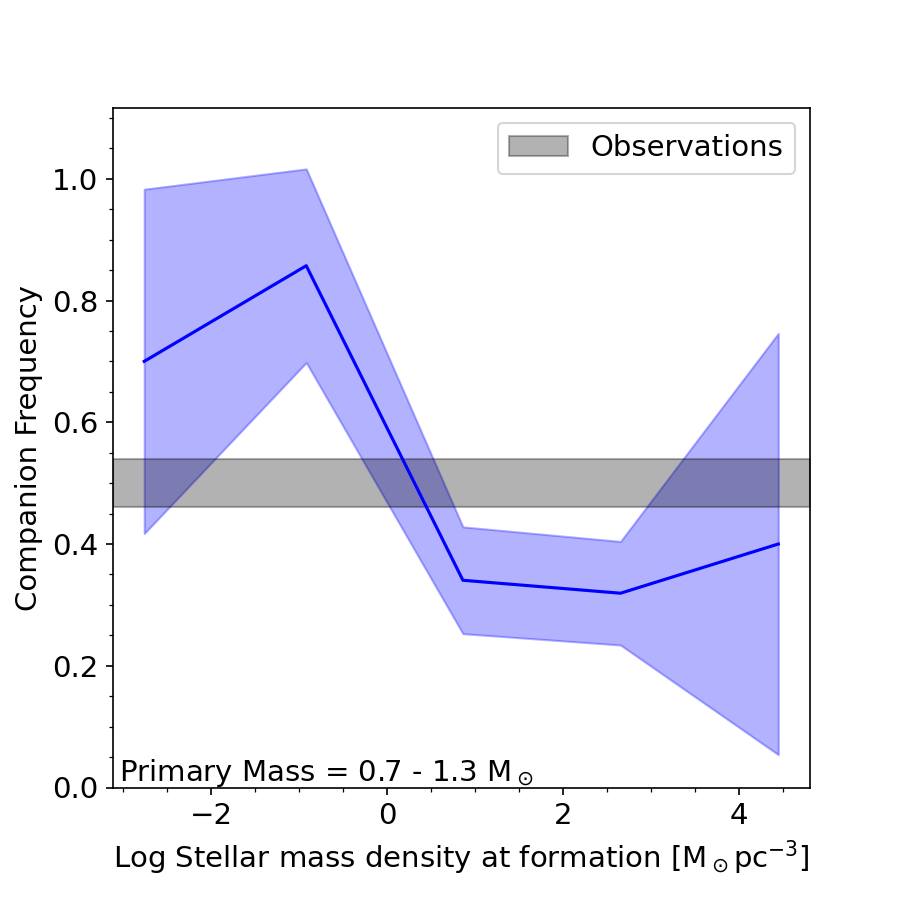}
\vspace{-0.4cm}
\caption{\textit{Left}: Median stellar mass density around newly formed stars in the fiducial run. The stellar mass density is calculated using the 32 nearest neighbors. Shaded regions show the 25th and 75th percentiles. \textit{Middle \& Right}: Multiplicity fraction and companion frequency in different stellar density bins. Shaded regions show the 1-$\sigma$ sampling uncertainties (see Appendix \ref{app:CF_MF_error} for details). The range of observed values for Solar-type field stars from \citet{raghavan2010} is also shown.}
\label{fig:fiducial_mass_density}
\vspace{-0.5cm}
\end {center}
\end{figure*}

A likely explanation for later forming stars having lower multiplicity is that they form in a different environment. Figure \ref{fig:M2e4_series} shows that the cloud undergoes global collapse and most star clusters merge to form one massive cluster surrounded by dense gas, until feedback from the stars expel the gas, weakening the gravitational potential well and leading to the expansion of the cluster (\citetalias{guszejnov_starforge_clusters}). The first stars form in relative isolation along filaments, while later stars form near existing star clusters. To examine how the \myquote{crowdedness} of the birth environment affects multiplicity we define the \emph{birth stellar density}, which we take to be the stellar mass density around the 32 nearest neighbors of a newly formed star. 
Figure \ref{fig:fiducial_mass_density} shows how this initial stellar mass density increases with time and starts to decline after 5 Myr when the cloud begins to disrupt and star formation is quenched in the central cluster (see Figure \ref{fig:M2e4_series}). The remaining gas-free clusters are gravitationally unbound and disperse (\citetalias{guszejnov_starforge_clusters}). The other panels of Figure \ref{fig:fiducial_mass_density} show that both the $\MF$ and $\CF$ for Solar-type stars decline with increasing stellar mass density at formation. 
This can be attributed to the higher likelihood of dynamical interactions (as there are more stars nearby), allowing for the newly formed star to be either captured by an existing star (increasing the multiplicity of earlier stars relative to later formed ones) or ejected from the gas reservoir. 

\subsection{Multiplicity of YSOs}\label{sec:results_fiducial_YSOs}

\begin{figure}
\begin {center}
\includegraphics[width=\linewidth]{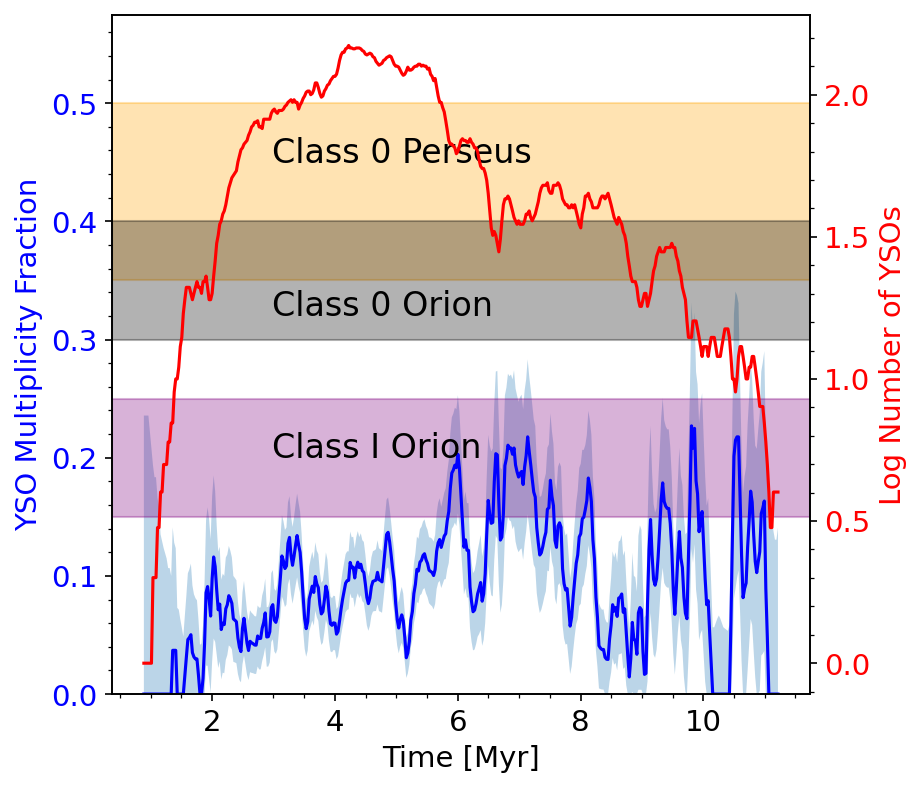}
\vspace{-0.4cm}
\caption{Properties of young stellar objects (YSOs) in the fiducial run, showing the number and multiplicity fraction of YSOs over time. Shaded rectangles show the observed multiplicity values by \citet{Tobin_2021_protostellar_multiplicity} for Class 0 and I protostars in Orion and Perseus, while the transparent blue shaded regions show the 1-$\sigma$ sampling uncertainty of the YSO multiplicity fraction. To make the plot easier to read we apply a 100 kyr rolling average.}
\label{fig:fiducial_YSO_evol}
\vspace{-0.5cm}
\end {center}
\end{figure}

Figure \ref{fig:fiducial_YSO_evol} shows the evolution of the YSO properties in the fiducial simulation. The number of YSOs, which we define to be stars younger than 0.5 Myr, essentially traces the star formation rate. As the cloud disrupts around 6 Myr star formation quenches and the YSO count decreases. 
To compare with the observations of \citet{Tobin_2016_protostellar_multiplicity_Perseus, Tobin_2021_protostellar_multiplicity} we calculate the multiplicity fraction of YSOs by taking only systems where \emph{all} members are YSOs and have a semi-major axis between 20 to $\mathrm{10^4}$ AU. We find that the YSO multiplicity in our simulation is comparable to that of Class I protostars in Orion. This is consistent with their expected ages of $\sim 0.1-0.5$Myr \citep{Dunham_2014_protostar_evol_review}.  

\section{Effects of initial condition variations on multiplicity}\label{sec:results_var}
  
In addition to our fiducial run we carry out a suite of simulations to explore the effects of initial conditions on multiplicity properties. We test for variations in the following initial parameters: the initial cloud surface density, virial parameter, magnetization, metallicity, as well as the interstellar radiation field (ISRF) and turbulent driving; see Table \ref{tab:var_guide} for specifics. These runs use the same turbulent initialization seed. In our analysis we also include two additional runs with the fiducial parameters but with different seeds, which provide a baseline of significance for variations between the runs.

To make the comparisons simpler all values shown in this section are the raw simulation results without corrections to remove short-lived or low-mass ratio companions.

\begin{table*}
    \setlength\tabcolsep{2.0pt} 
	\centering
	\begin{tabular}{ | c | c | c | c | }
	\hline
	Parameter & Default value & Tested variations  & Labels \\
	\hline
	Initial turbulence & $\alphaturb=2$ (Marginal boundedness) & x0.5, x2 & \textbf{M2e4\_a1}, \textbf{M2e4\_a4}  \\
	\hline
	Surface density & $\Sigma=63\,\msun/\pc^2$ (MW average) & x10, x0.1 & \textbf{M2e4\_R3}, \textbf{M2e4\_R30}  \\
	\hline
	Mass-to-flux ratio & $\mu=4.2$ (1\% relative magnetic energy) & x0.3, x0.1 & \textbf{M2e4\_mu1.3}, \textbf{M2e4\_mu0.4}  \\
	\hline
	Interstellar Radiation (ISRF) & Solar-circle values  \citep{habing1968,draine_1978_isrf}  & x10, x100 & \textbf{M2e4\_ISRF10}, \textbf{M2e4\_ISRF100}  \\
	\hline
	Metallicity & $Z=Z_\odot$ & x0.1, x0.01 & \textbf{M2e4\_Z01}, \textbf{M2e4\_Z001}  \\
	\hline
    \end{tabular}
        \vspace{-0.1cm}
 \caption{List of parameter variations investigated in \S\ref{sec:results_var} and the relevant IC labels from Table \ref{tab:IC_phys}}
 \label{tab:var_guide}\vspace{-0.5cm}
\end{table*}

\subsection{Initial level of turbulence}\label{sec:alpha}

\begin{figure*}
\begin {center}
\includegraphics[width=0.32\linewidth]{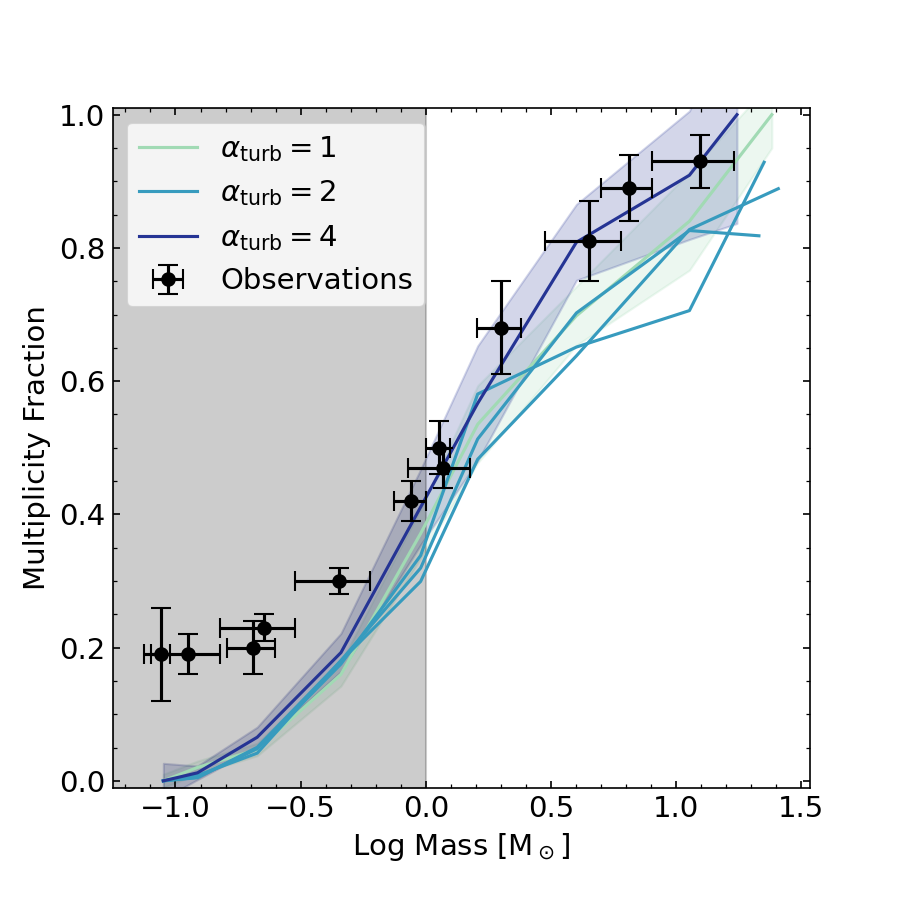}
\includegraphics[width=0.32\linewidth]{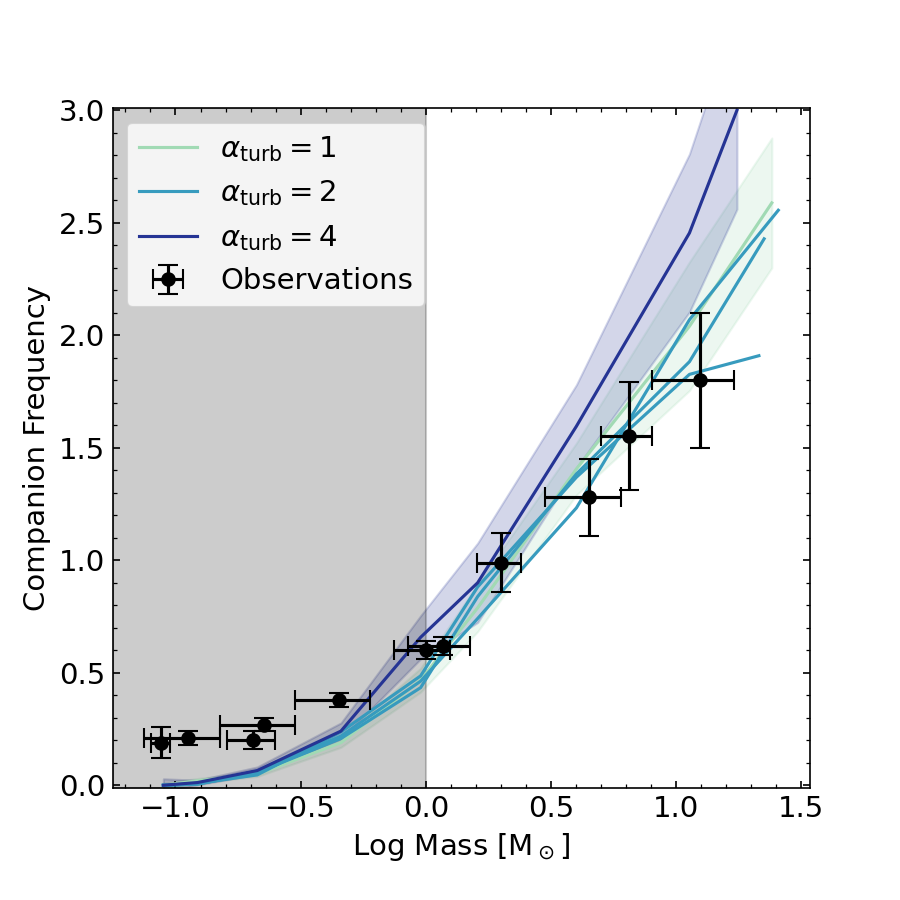}
\includegraphics[width=0.32\linewidth]{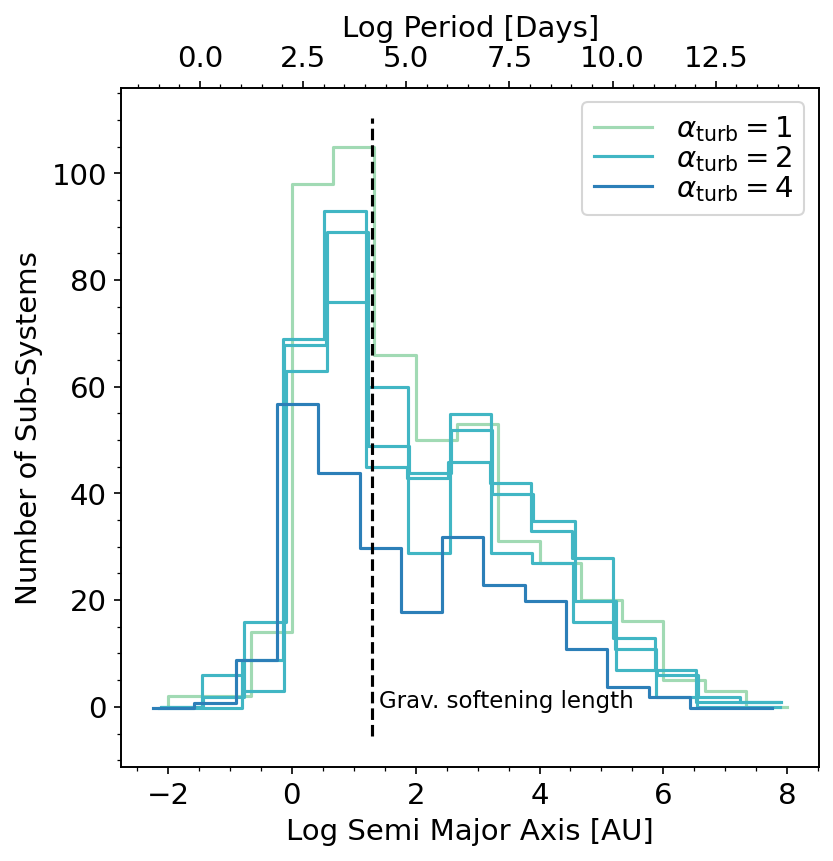}\\
\includegraphics[width=0.32\linewidth]{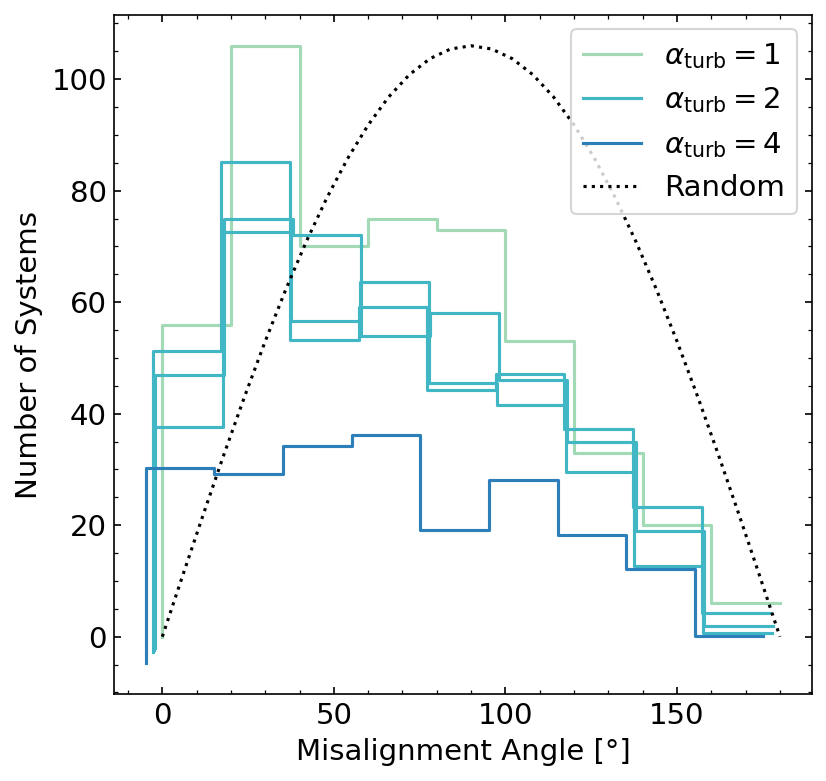}
\includegraphics[width=0.32\linewidth]{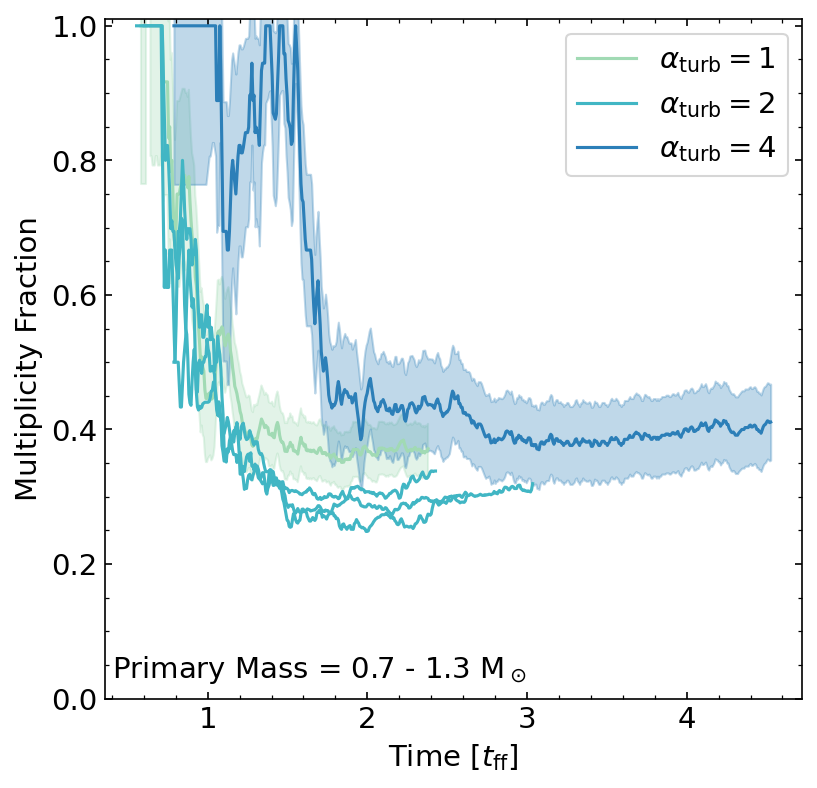}
\includegraphics[width=0.32\linewidth]{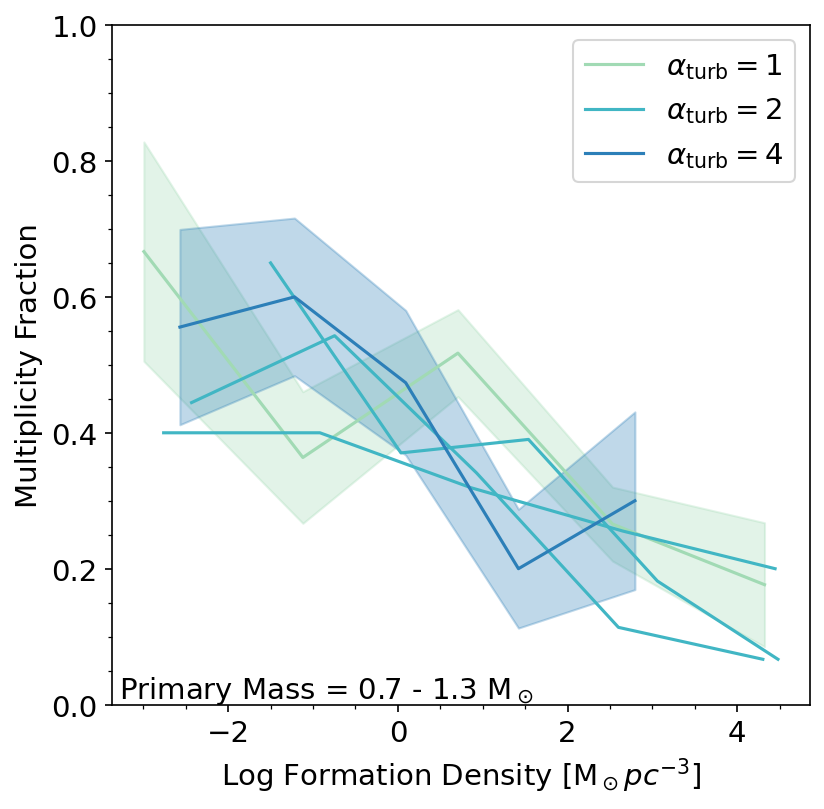}
\vspace{-0.3cm}
\caption{Multiplicity properties for different levels of initial turbulence (\textbf{M2e4\_a1}, \textbf{M2e4}, \textbf{M2e4\_a4}). The curves do not include corrections for short lived or low-mass ratio companions. For the fiducial \textbf{M2e4} IC we plot the results from three separate runs that have identical global parameters but different initial turbulent realizations. The \emph{top row} shows the multiplicity fraction $\MF$ (left), companion frequency $\CF$ (middle), and the distribution of the semi major axis for Solar-type stars (right). For $\MF$ and $\CF$ colored shaded regions show the 1-$\sigma$ sampling errors, which are not plotted for the fiducial \textbf{M2e4} runs. A grey shaded region shows the mass range potentially affected by the $0.1\,\msun$ completeness limit of the simulation. In the semi-major axis distribution the vertical line marks the gravitational softening length of the simulations. The \emph{bottom row} shows the misalignment angle distribution (left), the evolution of multiplicity for Solar-type stars that are no longer accreting (middle) and the multiplicity fraction for Solar-type stars as a function of birth stellar density (right). The multiplicity time evolution in the middle panel is normalized to the initial cloud freefall time to make comparisons between runs easier. In the left panel a dotted line shows the angle difference distribution resulting from a purely random draw of companion spins. Shaded regions show the 1-$\sigma$ sampling errors, similar to the top row.}
\label{fig:alpha_var}
\vspace{-0.5cm}
\end {center}
\end{figure*}

We compare three runs with different levels of turbulence as parameterized by the turbulent virial parameter $\alphaturb$. The runs all use the same initial turbulent seed, except for the fiducial run ($\alphaturb=2$), for which we show the results for two additional initial turbulent realizations. The change in velocity dispersion for the different $\alphaturb$ runs is achieved by scaling the initial velocity fields of the fiducial run (see Table \ref{tab:IC_phys}). Figure \ref{fig:alpha_var} shows that both the multiplicity fraction and companion frequency increases for $M>\msun$ stars with increasing turbulence, similar to the results of \citet{Cunningham_2018_feedback}, although the changes are comparable to the variations for different turbulent realizations. 

Except for a change in normalization (due to different star formation efficiencies among the clouds) the shape of the semi-major axis distribution is qualitatively similar. We find that increasing the level of turbulence shifts the peak of the misalignment angle distributions toward 90 degrees, similar to the distribution shape resulting from uncorrelated primary and companion spins. This suggests the higher global turbulence reduces the angular momentum correlation on smaller scales.

Increasing the initial turbulence delays star formation, but otherwise the multiplicity fraction of Solar-type stars follows a similar decreasing trend. All runs show decreasing multiplicity with birth stellar density, and the trends agree within 1-$\sigma$ error, but the highest achieved density decreases with the level of turbulence. This means that the increase in multiplicity with stronger turbulence could be explained by the more turbulent clouds having overall lower stellar densities. 

\subsection{Cloud surface density}\label{sec:sigma}

Cloud surface density is thought to be a key parameter of star formation \citep{krumholz08a, fkm2010, grudic_2020_cluster_formation} due to its influence on the dynamics of fragmentation and impact on stellar feedback. In addition to our fiducial cloud (\textbf{M2e4}), which has a surface density similar to the MW average ($\Sigma=63\,\msun/pc^2$) we 
run clouds with 10 times higher and lower values (\textbf{M2e4\_R3}, \textbf{M2e4\_R30}). Note that these runs have very different final star formation efficiencies (1\%, 9\% and 14\% in order of increasing surface density). Thus, the low surface density run (\textbf{M2e4\_R30}) has about a factor 10 fewer stars than the other runs, making its multiplicity metrics significantly more uncertain.

\begin{figure*}
\begin {center}
\includegraphics[width=0.32\linewidth]{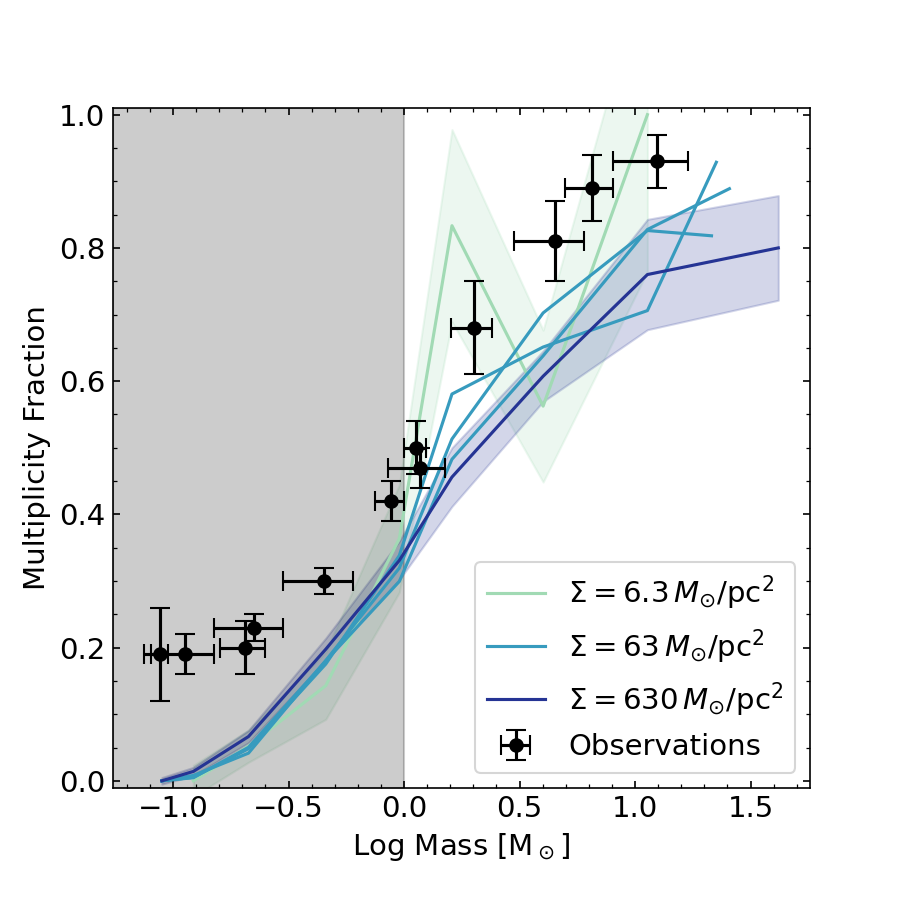}
\includegraphics[width=0.32\linewidth]{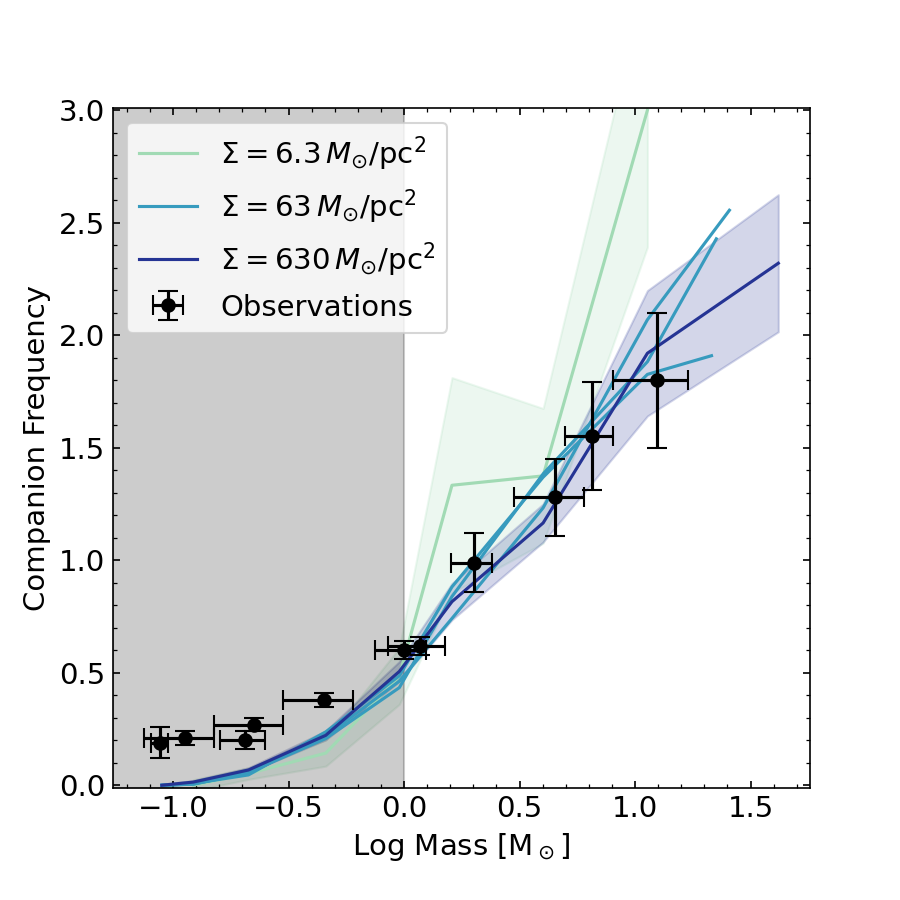}
\includegraphics[width=0.32\linewidth]{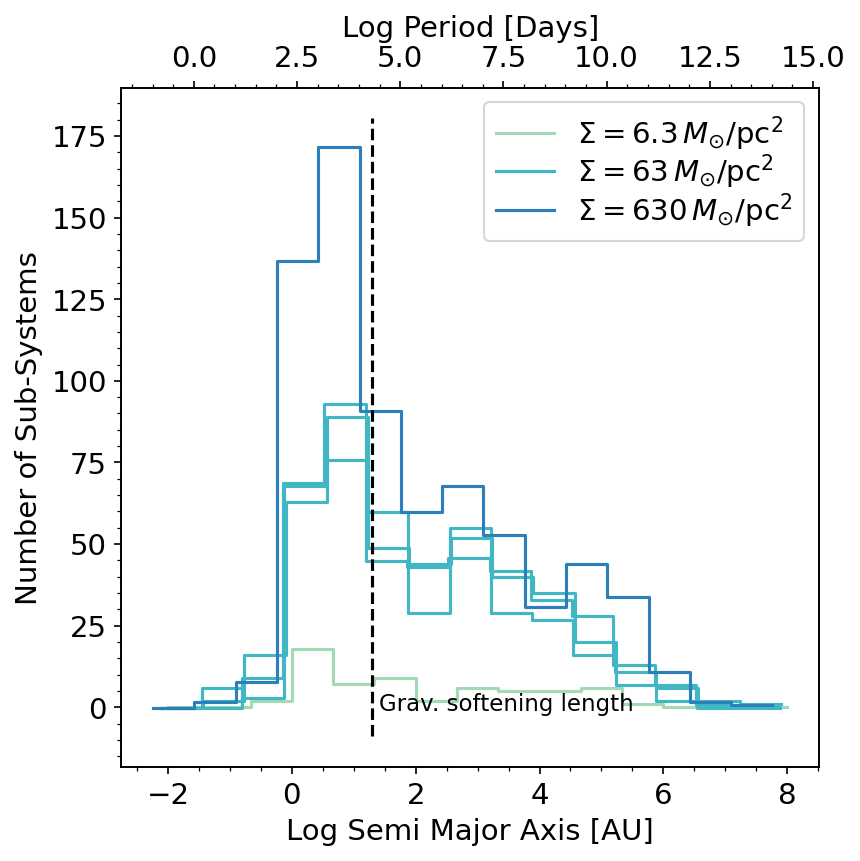}\\
\includegraphics[width=0.32\linewidth]{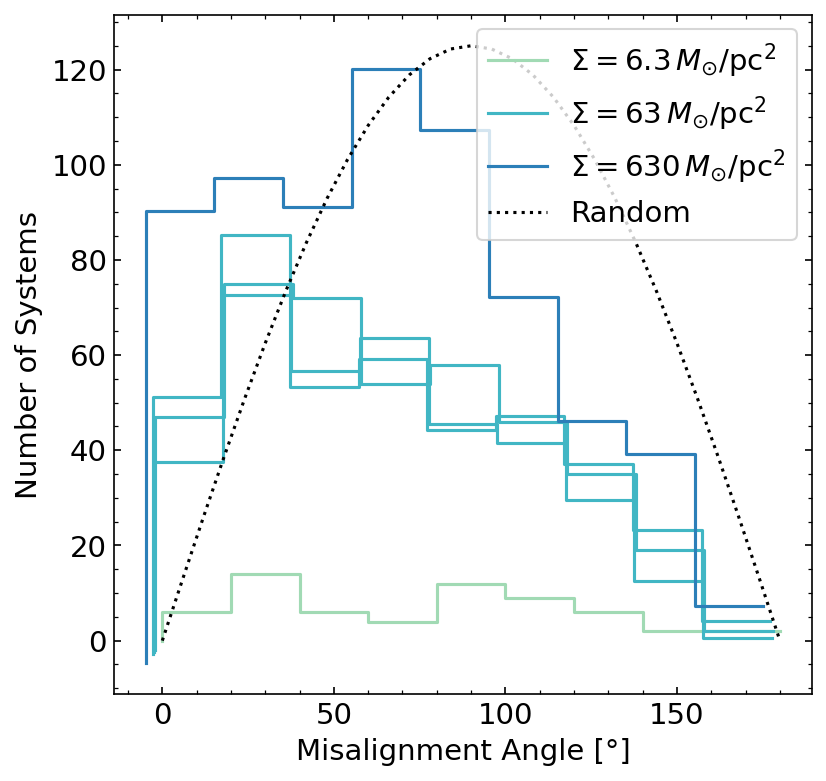}
\includegraphics[width=0.32\linewidth]{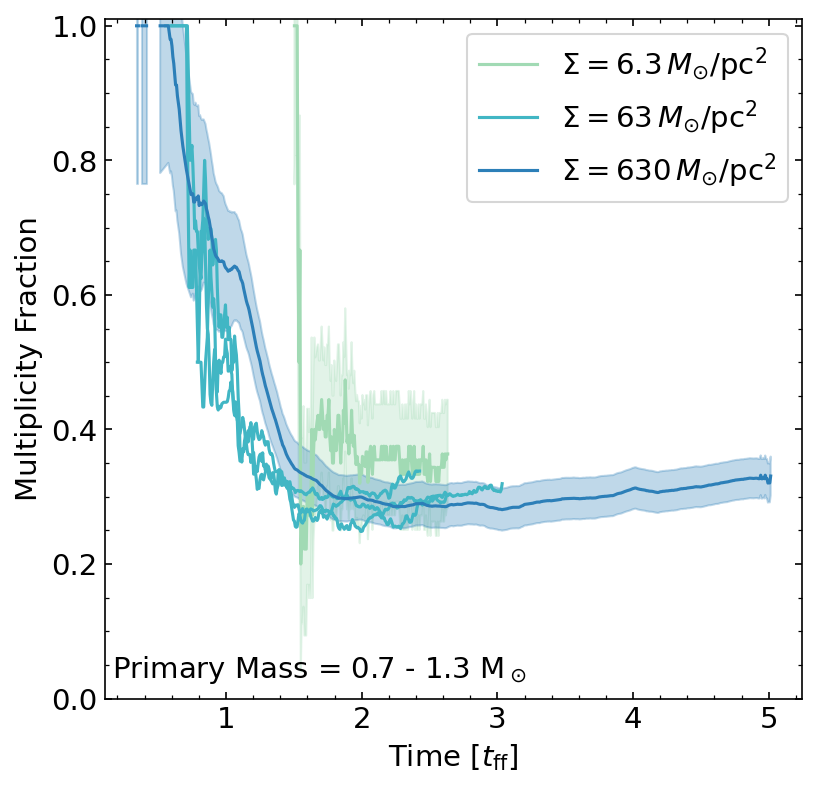}
\includegraphics[width=0.32\linewidth]{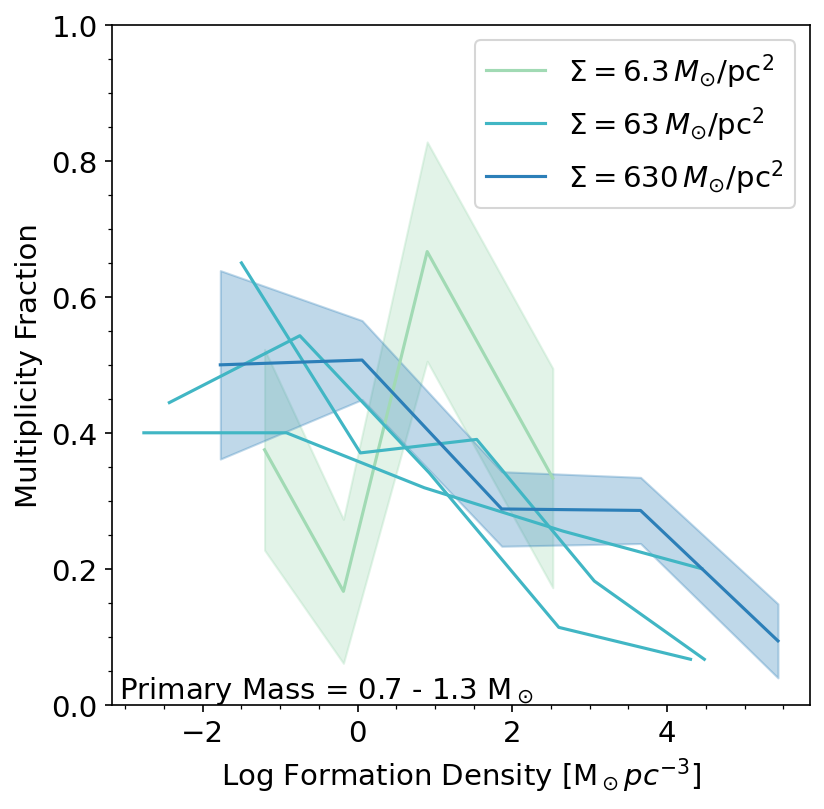}
\vspace{-0.3cm}
\caption{Same as Figure \ref{fig:alpha_var} but for different initial cloud surface densities (\textbf{M2e4\_R3}, \textbf{M2e4}, \textbf{M2e4\_R30}).}
\label{fig:sigma_var}
\vspace{-0.5cm}
\end {center}
\end{figure*}

Figure \ref{fig:sigma_var} compares the multiplicity properties across our runs with different initial surface density. We find that increasing surface density leads to lower multiplicity fractions and companion frequencies for higher mass stars as well as a much more pronounced peak in the semi-major axis distribution near the gravitational softening length ($\sim 20\,\AU$). Increasing the initial cloud surface density does not affect the spin alignment between primaries and their companions (bottom left panel of \ref{fig:sigma_var}), however the low surface density run shows an essentially flat distribution. As in all previously discussed runs, the multiplicity fraction of Solar-type stars decreases with time, which can be explained by the increasing stellar density around newly forming stars. We find that the relationship between the birth stellar density and the multiplicity fraction is similar between the runs, and their cut-off value increases with initial surface density. This also provides an explanation as to why both $\MF$ and $\CF$ decrease with increasing surface density: the denser the cloud, the higher the stellar density, leading to more dynamical interactions and thus lower multiplicity at the end of the simulations. 

\subsection{Cloud magnetization}\label{sec:mu}

Star formation efficiency is sensitive to the cloud mean magnetic field \citep[e.g.,][]{Padoan_2012_SF_law}, with efficiency decreasing with stronger fields (\citetalias{guszejnov_starforge_imf}). This result suggests multiplicity might also depend on the magnetic field.
In this section we present runs for clouds with increasing initial magnetic fields, corresponding to initial normalized mass-to-flux ratios $\mu$ of 4.2, 1.3 and 0.4 (\textbf{M2e4}, \textbf{M2e4\_mu1.3}, \textbf{M2e4\_mu0.4}, see Table \ref{tab:IC_phys}).

\begin{figure*}
\begin {center}
\includegraphics[width=0.32\linewidth]{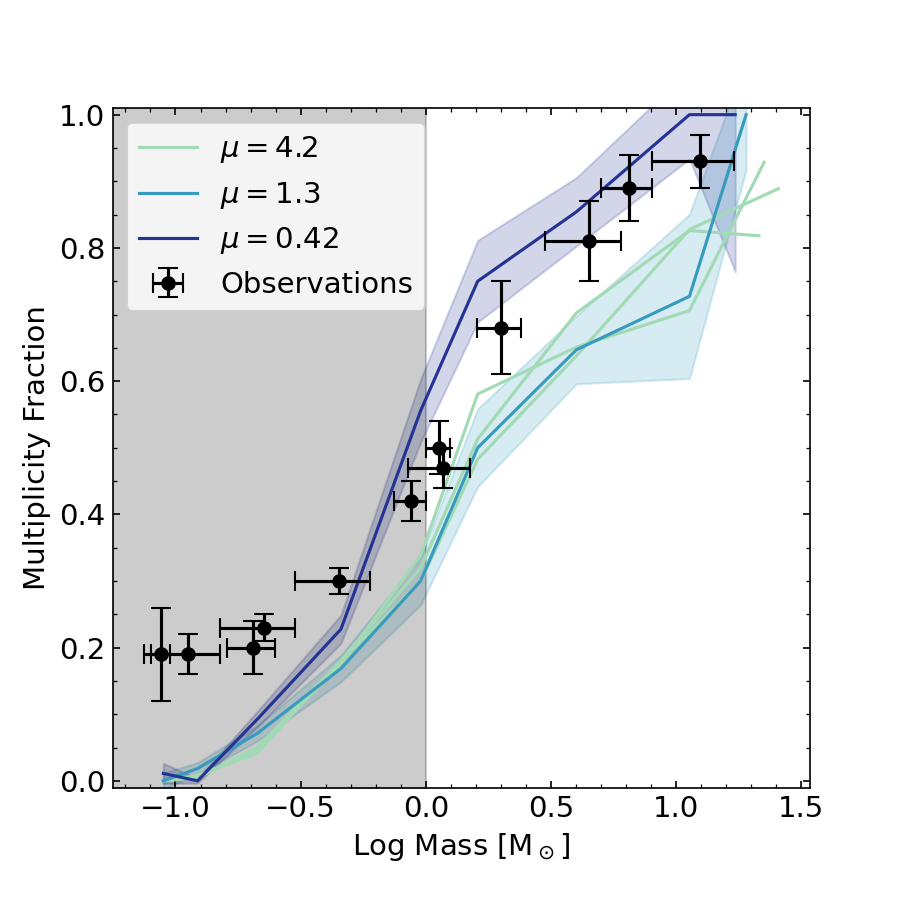}
\includegraphics[width=0.32\linewidth]{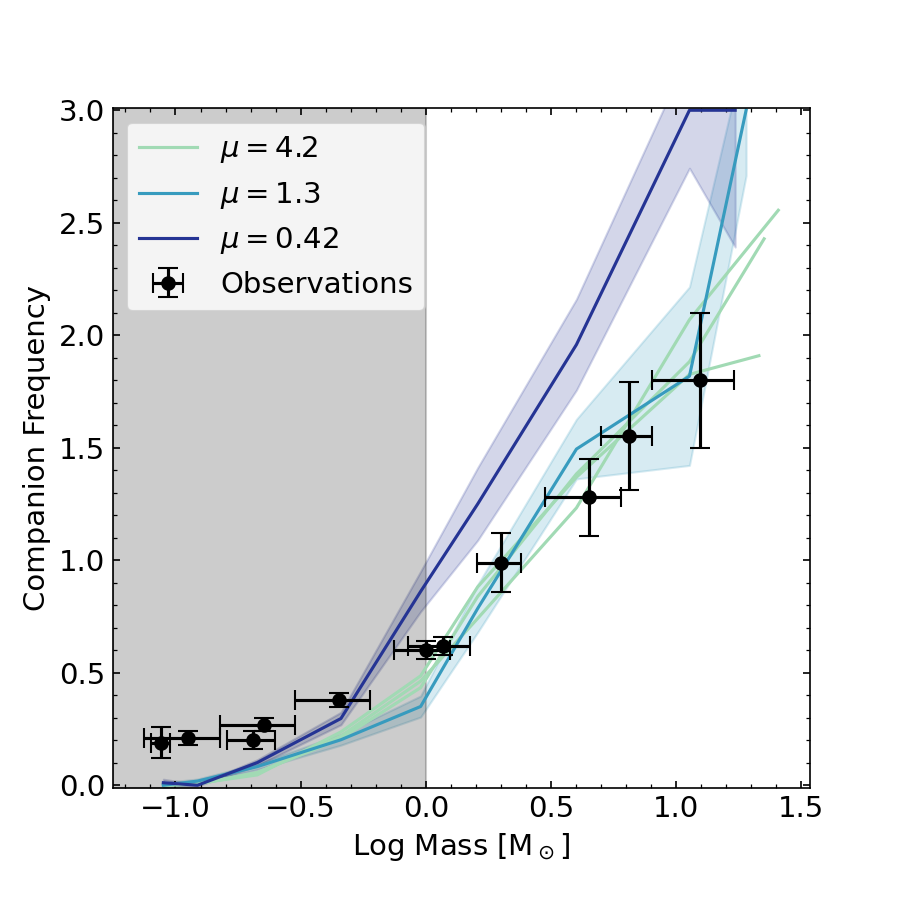}
\includegraphics[width=0.32\linewidth]{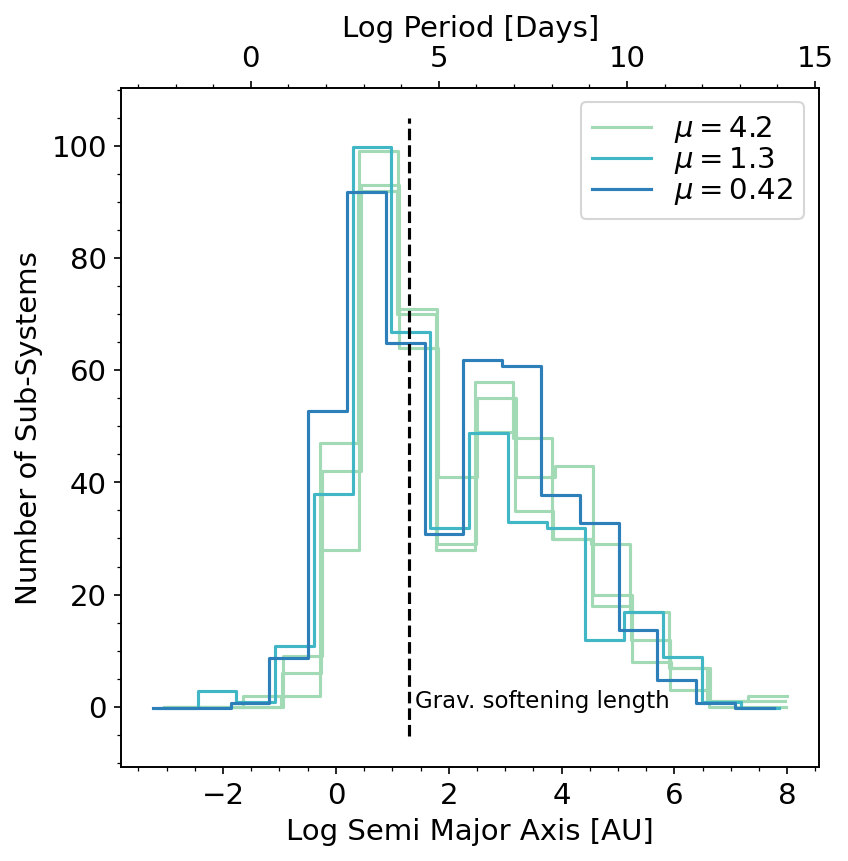}\\
\includegraphics[width=0.32\linewidth]{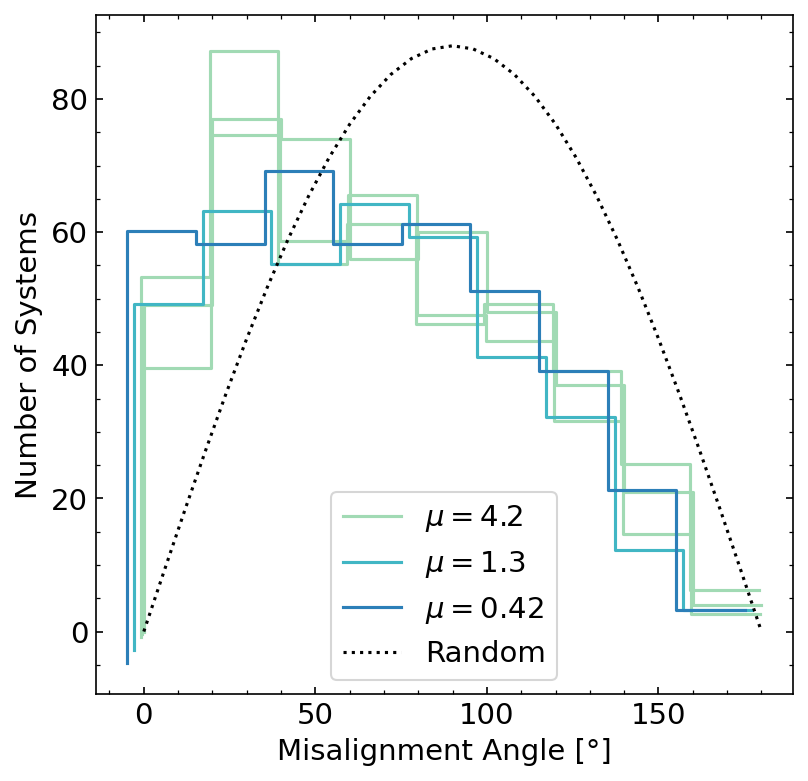}
\includegraphics[width=0.32\linewidth]{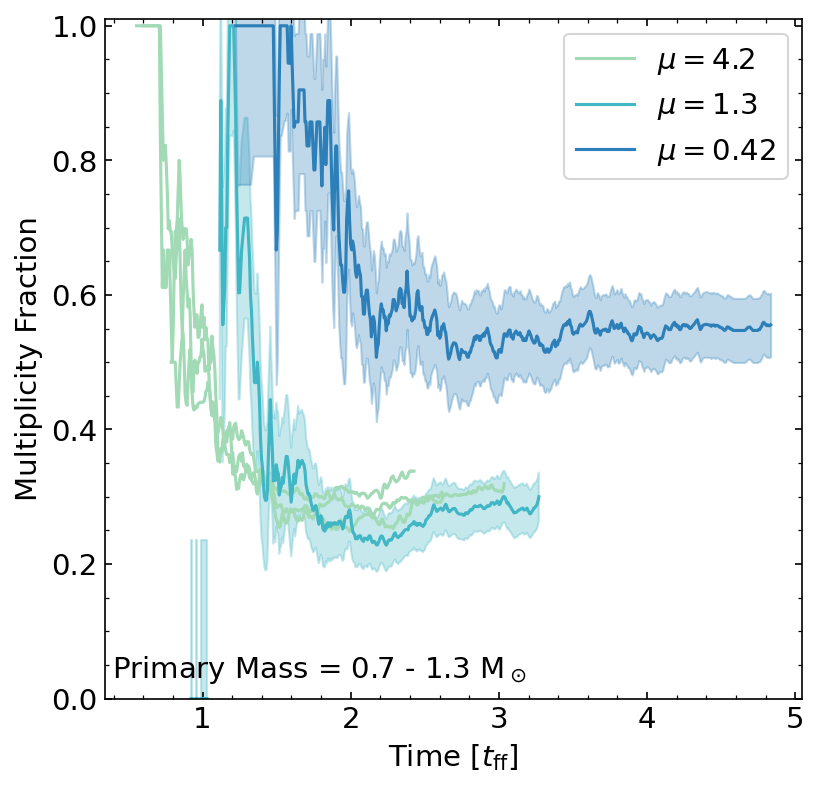}
\includegraphics[width=0.32\linewidth]{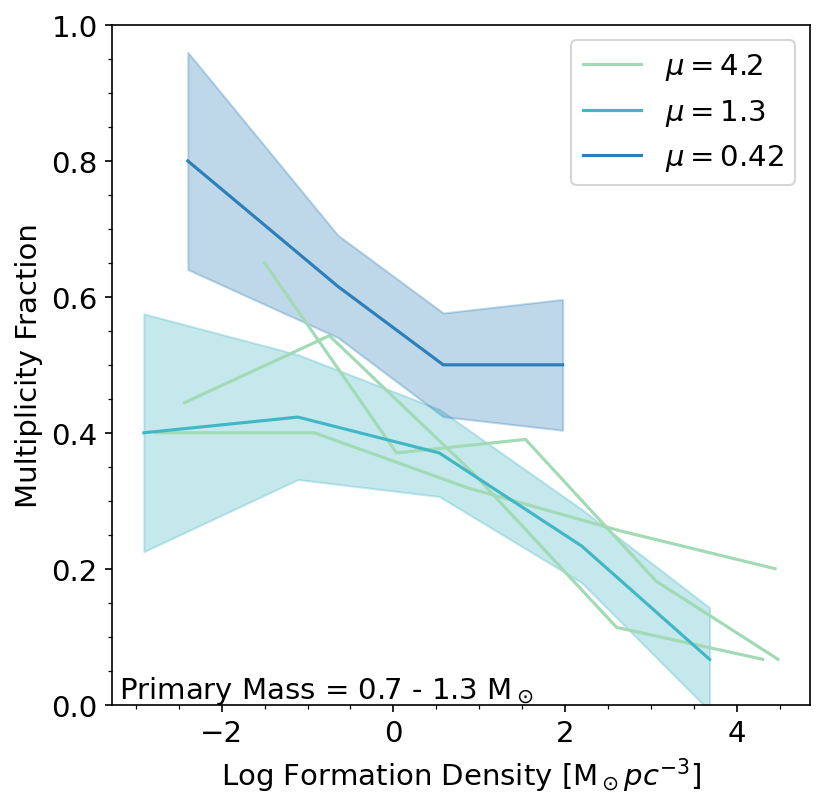}
\vspace{-0.3cm}
\caption{Same as Figure \ref{fig:alpha_var} but for different levels of initial magnetization (\textbf{M2e4}, \textbf{M2e4\_mu1.3}, \textbf{M2e4\_mu0.4}).}
\label{fig:mu_var}
\vspace{-0.5cm}
\end {center}
\end{figure*}

Figure \ref{fig:mu_var} shows that the strong field cloud has a significantly higher multiplicity fraction for Solar-type stars, an effect proposed by prior work \citep{Lee_2019_MHD_binary_separations}. However, there is essentially no change difference between the fiducial, weak ($\mu=4.2$) and intermediate ($\mu=1.3$) field runs.  For all three cases there are  no significant variations in either the semi-major axis distribution or the distribution of the misalignment angle. The increased magnetic fields provide significant support to the cloud against collapse, which delays star formation. Apart from this delay the multiplicity fraction of Solar-type stars follows a similar declining trend with time. The weak and intermediate field runs provide a similar relationship between the birth stellar density and the multiplicity fraction, while the highly magnetized run has significantly higher multiplicities at similar stellar densities. 
In \citet{Guszejnov_isoT_MHD} we show that regardless of the initial magnetic field strength, the magnetic energy density at high densities follow the same trend ($v_\mathrm{Alfv\'{e}n}\sim c_s$), due to the turbulent magnetic dynamo. This means that the effects of the global initial magnetic field do not propagate to densities higher than  $\rho_B>B_0^2/(\mu_0 c_s)$, where $B_0$, $c_s$ and $\mu_0$ are the initial magnetic field strength, the sound speed and the vacuum permeability respectively. So the initial magnetic field only influences multiplicity properties if $\rho_B$ is comparable to the densities of star forming cores. 

\subsection{Cloud metallicity}\label{sec:Z}

Metallicity is a key property of interstellar gas, which directly sets its thermodynamic behavior, so it is expected to have a major impact on star formation \citep{sf_big_problems}. In this section we present three runs with decreasing initial gas metallicities, corresponding to Solar, 10\% of Solar and 1\% of Solar values (\textbf{M2e4},\textbf{M2e4\_Z01} and \textbf{M2e4\_Z001} respectively, see Table \ref{tab:IC_phys}).

\begin{figure*}
\begin {center}
\includegraphics[width=0.32\linewidth]{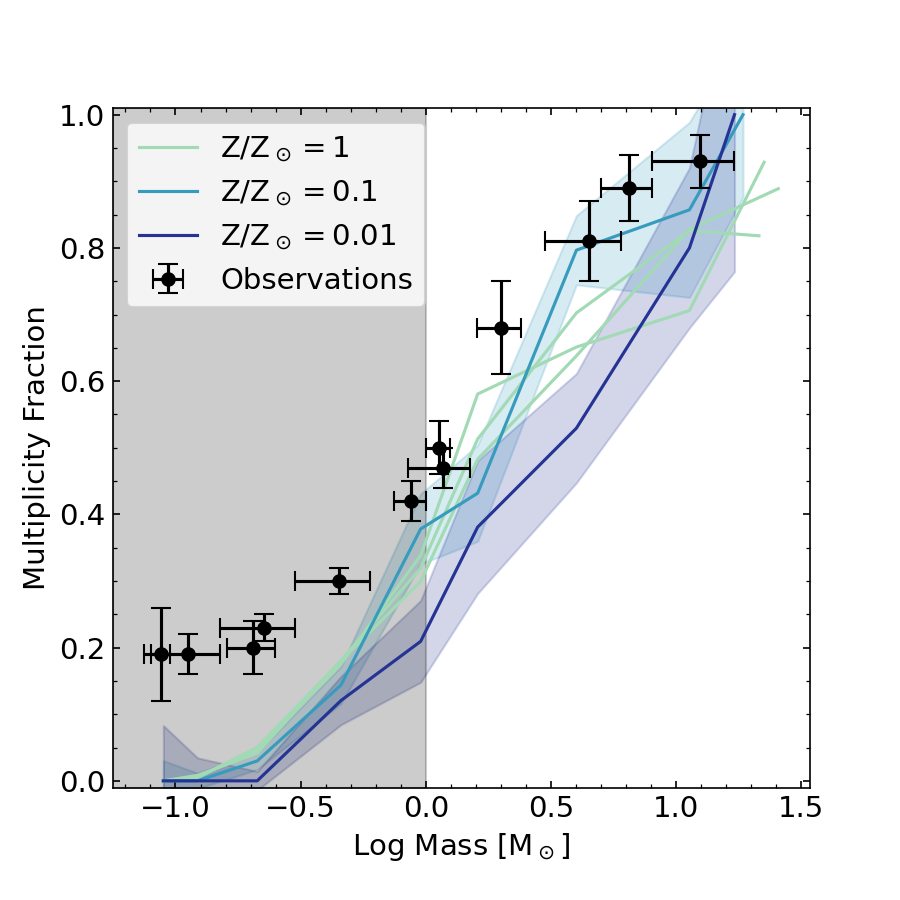}
\includegraphics[width=0.32\linewidth]{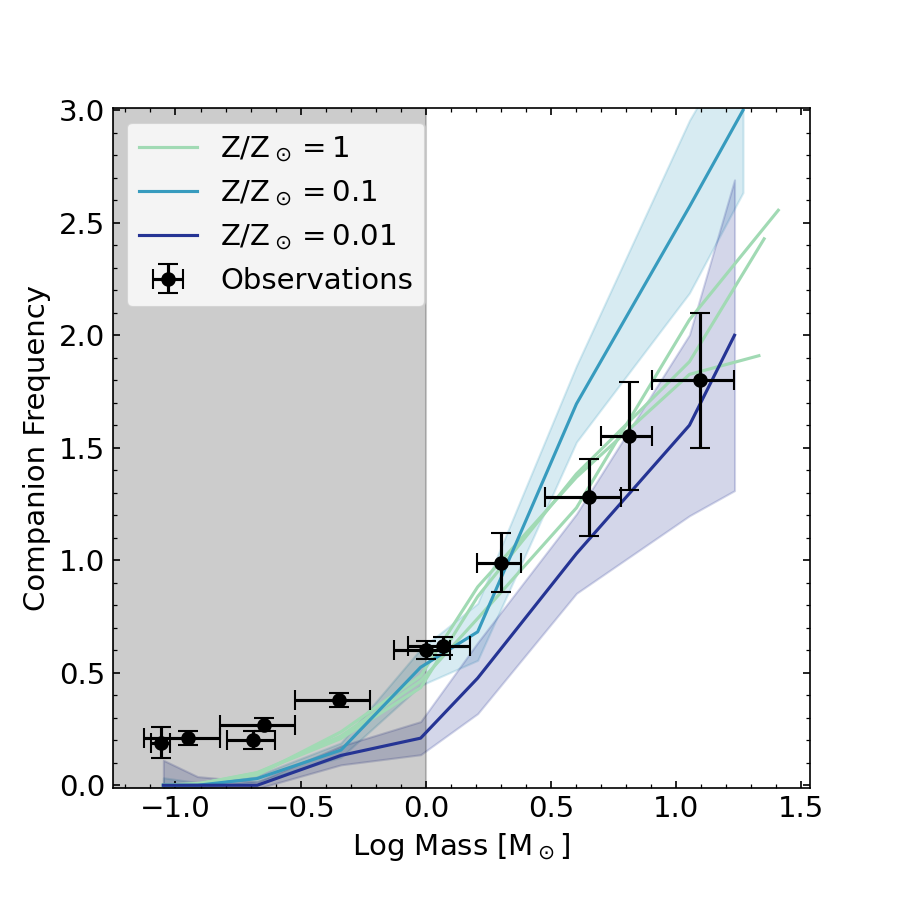}
\includegraphics[width=0.32\linewidth]{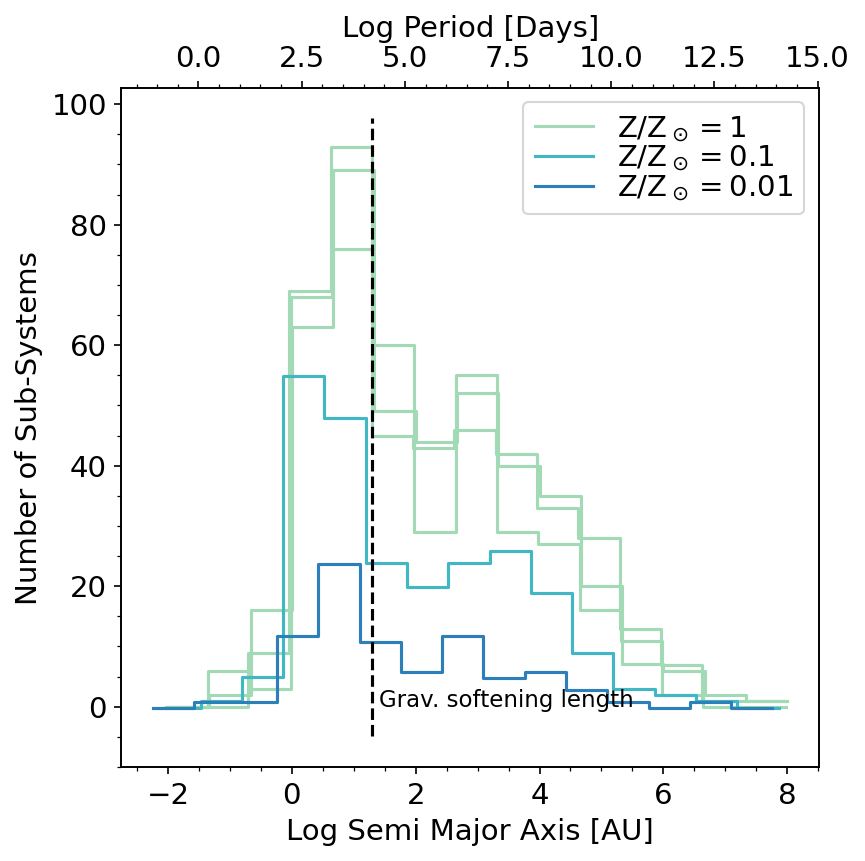}\\
\includegraphics[width=0.32\linewidth]{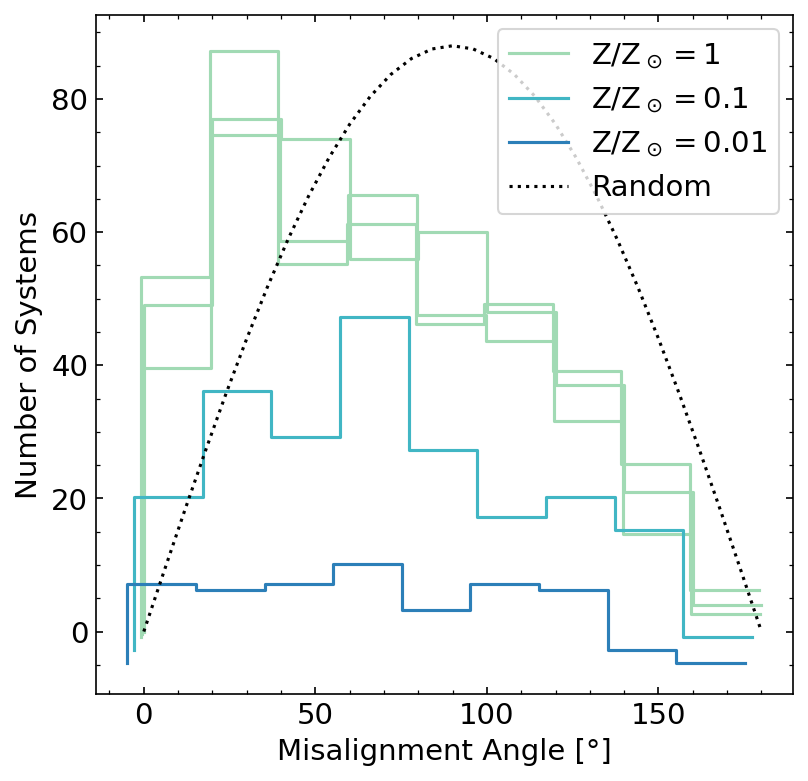}
\includegraphics[width=0.32\linewidth]{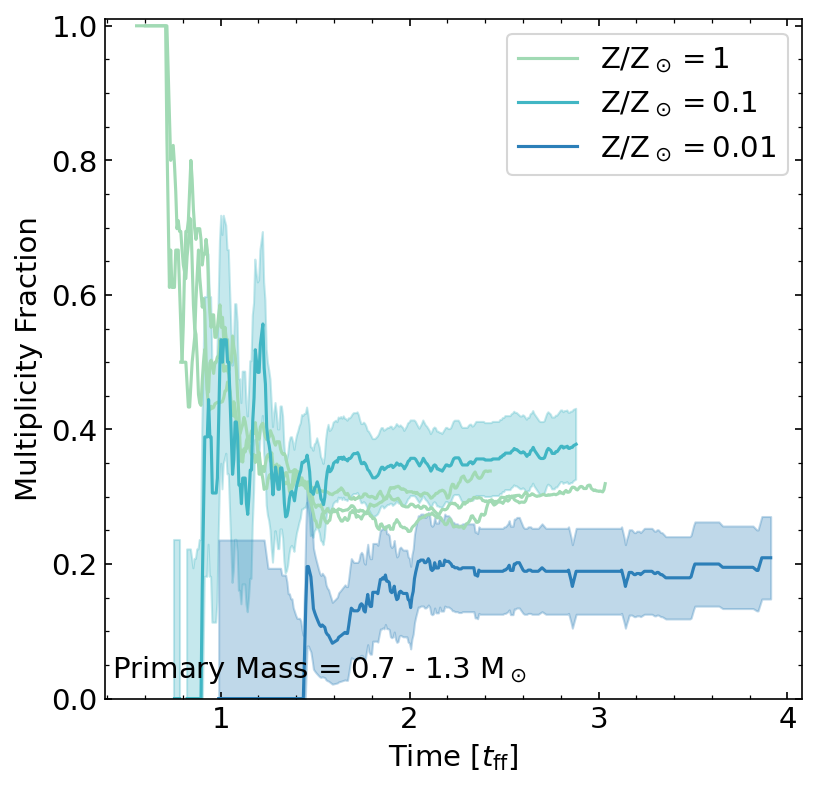}
\includegraphics[width=0.32\linewidth]{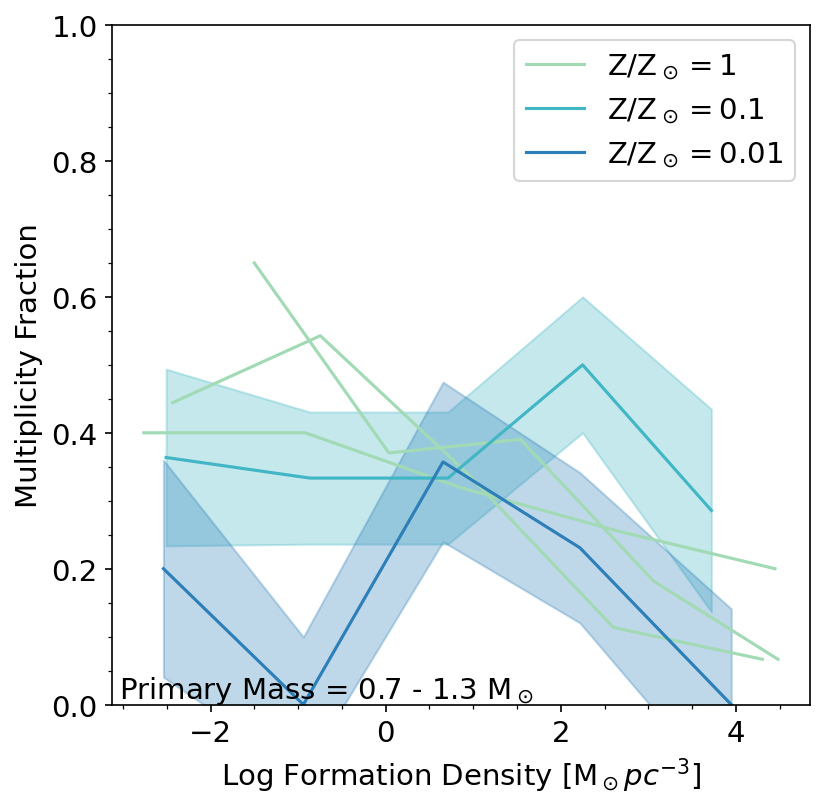}
\vspace{-0.3cm}
\caption{Same as Figure \ref{fig:alpha_var} but for different levels of initial gas metallicity (\textbf{M2e4}, \textbf{M2e4\_Z01}, \textbf{M2e4\_Z001}).}
\label{fig:Z_var}
\vspace{-0.5cm}
\end {center}
\end{figure*}

We find that metallicity significantly affects the star formation process, notably it shifts the IMF to significantly higher masses (see \citetalias{guszejnov_starforge_imf} for details). However, Figure \ref{fig:Z_var} shows that varying the metallicity of the gas has no clear effect on either the multiplicity fraction or the companion frequency, similar to the results of \citet{Bate_2019_Z_multiplicity}. While the normalization of the semi-major axis distribution is affected by the differences in the overall star formation efficiency, its shape appears to be similar between the three runs. Decreasing the metallicity mildly flattens the the misalignment angle distribution, i.e., makes anti-parallel companions slightly more likely. The evolution of the multiplicity fraction for Solar-type stars declines similarly for the fiducial and the 10\% Solar metallicity runs, but for the 1\% run we find significantly lower multiplicities and an increasing trend instead of a decreasing one. The relationship between the multiplicity fraction and the birth stellar density is also different between the runs; in the low metallicity clouds there is no clear relationship between the two quantities.

\subsection{Interstellar Radiation Field}\label{sec:ISRF}

The interstellar radiation field (ISRF) is set by the radiation of previously formed stars in the local galactic environment. The ISRF varies as a function of the galactocentric radius, so clouds located closer to the galactic center experience higher ISRFs. Thus, the radiative environment is expected to vary significantly between star-forming regions. We compare the multiplicity properties in three runs with progressively higher background radiation fields, starting from our fiducial run, which adopts the Solar-circle value of 1 Draine (\textbf{M2e4}), followed by runs with 10 times (\textbf{M2e4\_ISRF10})  and 100 times (\textbf{M2e4\_ISRF100}) higher radiation energy densities, see Table \ref{tab:IC_phys}. 

\begin{figure*}
\begin {center}
\includegraphics[width=0.32\linewidth]{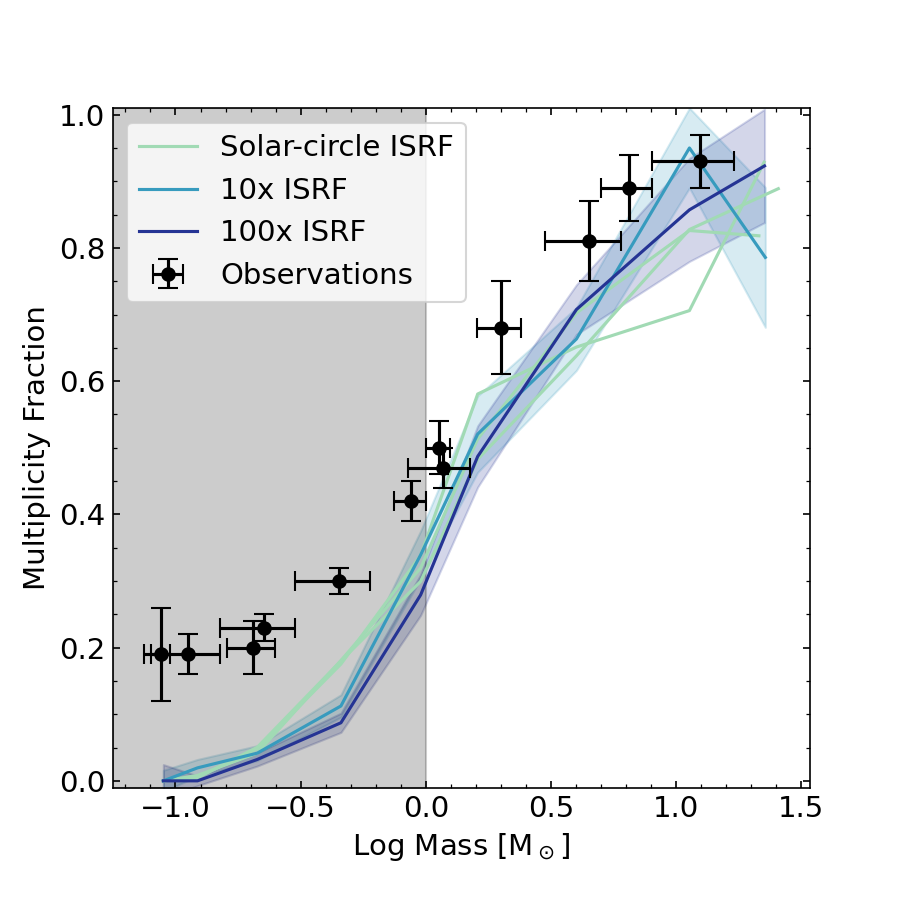}
\includegraphics[width=0.32\linewidth]{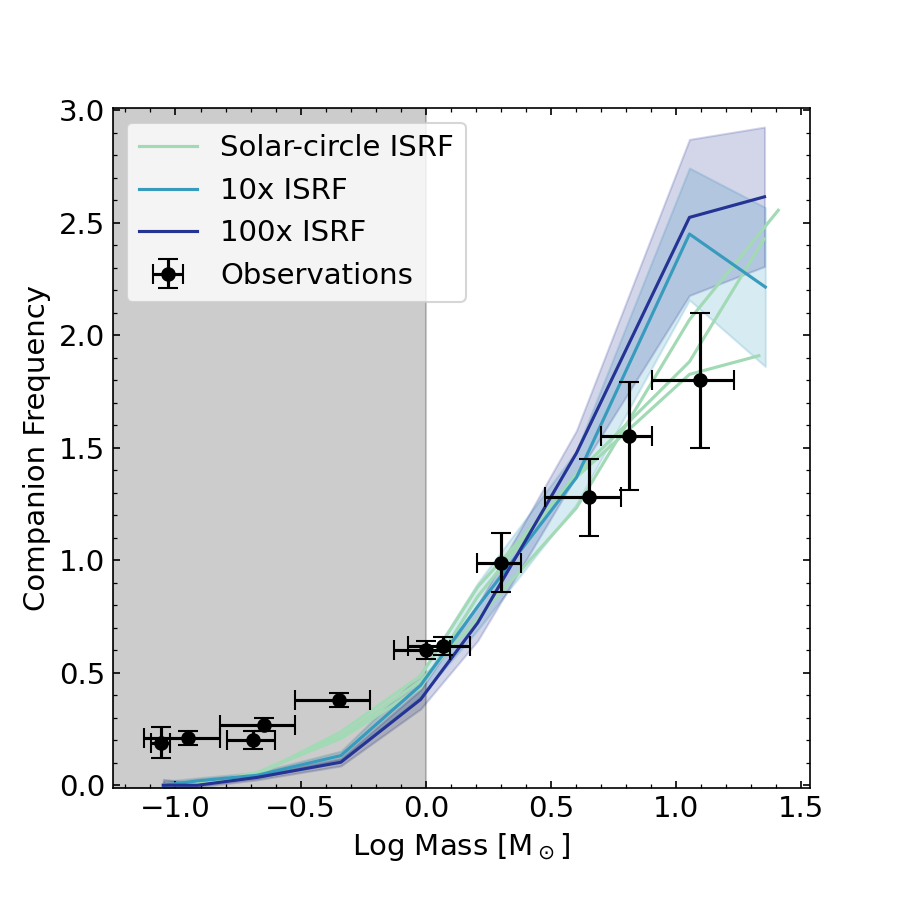}
\includegraphics[width=0.32\linewidth]{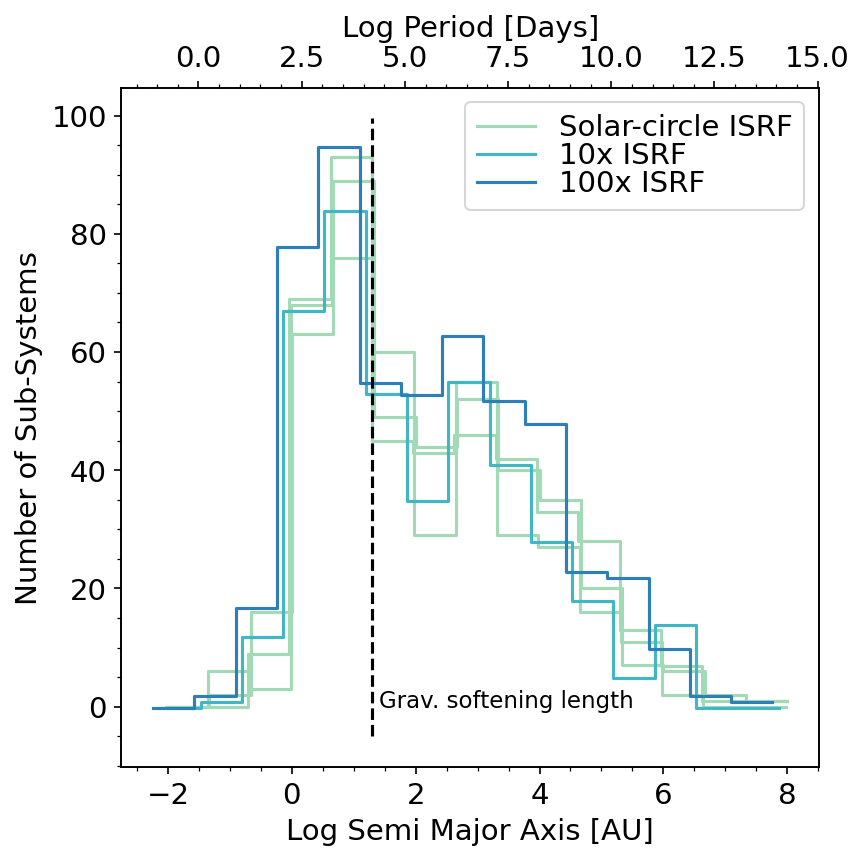}\\
\includegraphics[width=0.32\linewidth]{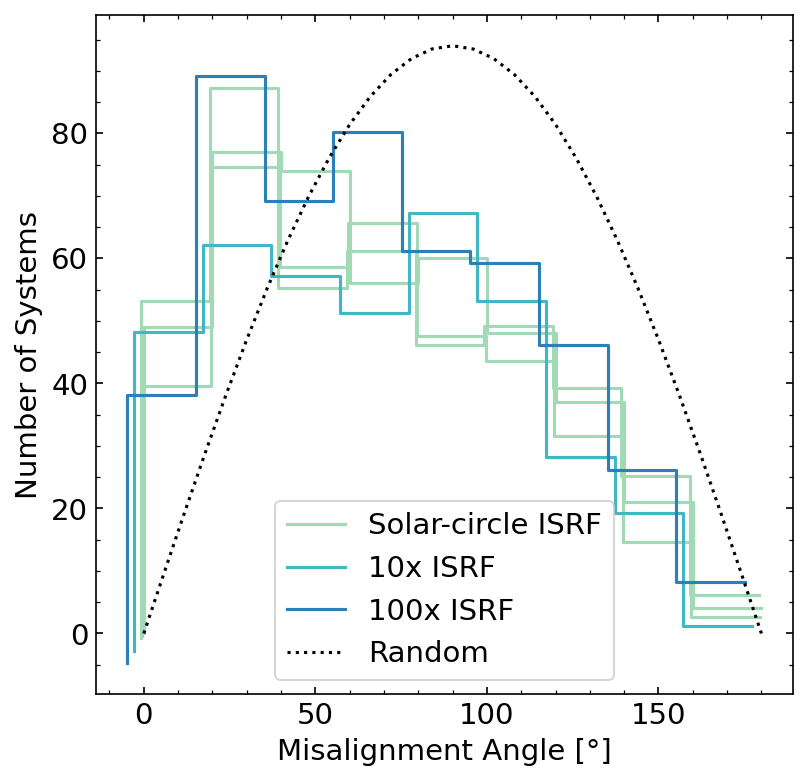}
\includegraphics[width=0.32\linewidth]{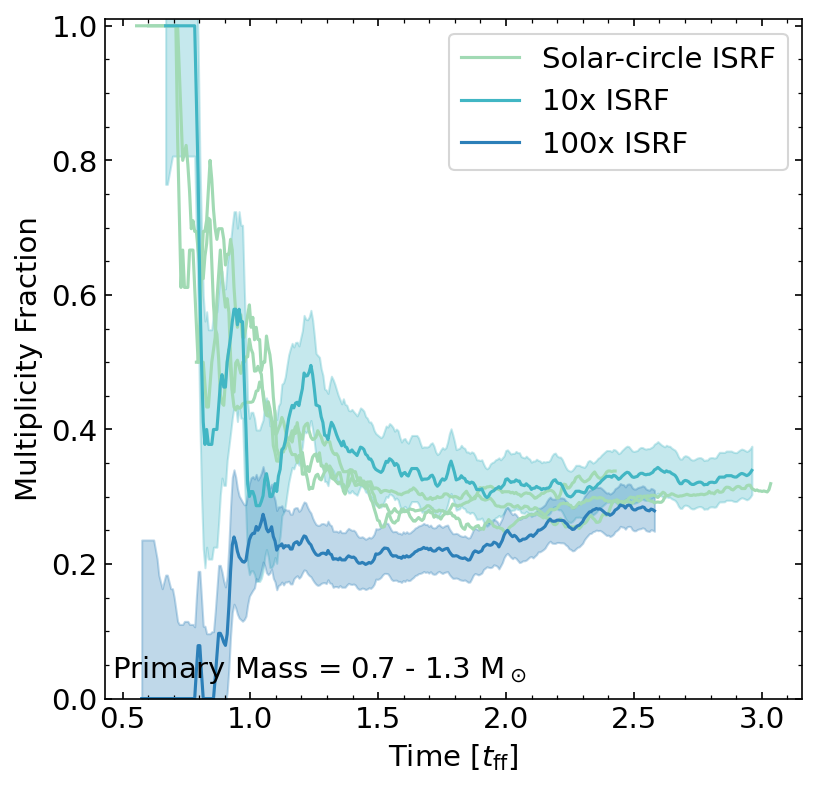}
\includegraphics[width=0.32\linewidth]{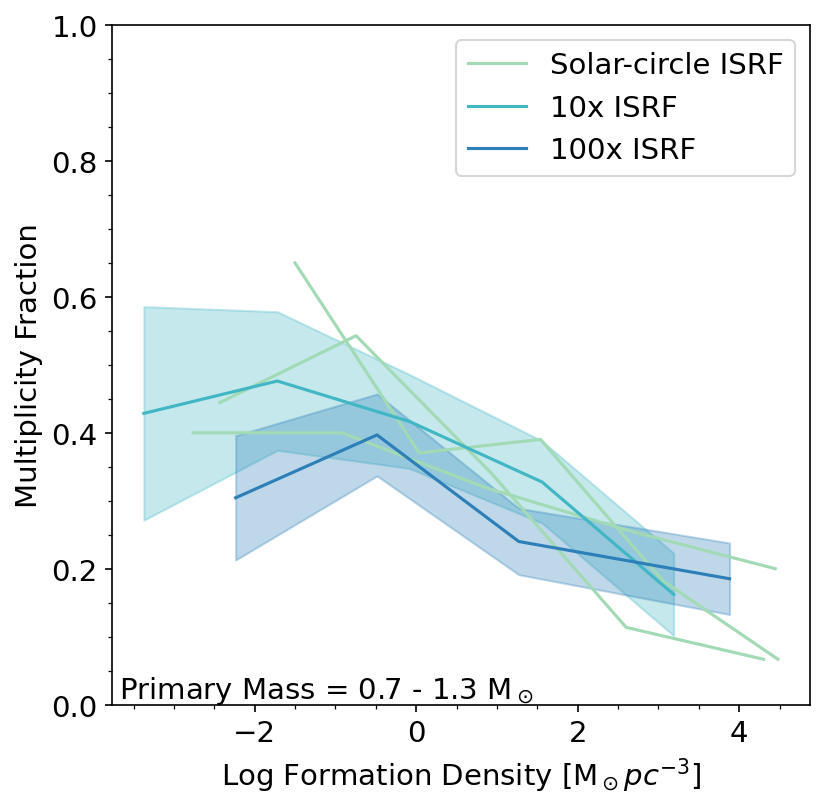}
\vspace{-0.3cm}
\caption{Same as Figure \ref{fig:alpha_var} but for different levels of interstellar radiation field (\textbf{M2e4}, \textbf{M2e4\_ISRF10}, \textbf{M2e4\_ISRF100}).}
\label{fig:ISRF_var}
\vspace{-0.5cm}
\end {center}
\end{figure*}

We find that increasing the ISRF increases the gas temperature and shifts the IMF to mildly higher masses (see \citetalias{guszejnov_starforge_imf} for details). However, figure \ref{fig:ISRF_var} shows a mild increase in both the multiplicity fraction and the companion frequency at high masses. The increased ISRF has little effect on the semi-major axis or the misalignment angle distributions. Similar to the case of metallicity variations, the mildly increased ISRF run shows similar $\MF$ evolution for Solar-type stars while \textbf{M2e4\_ISRF100} shows a qualitatively different evolution where $\MF$ increases with time. Nevertheless, the three simulations show a similar relationship between the stellar densities at formation and $\MF$, although the $\MF$ is consistently lower for \textbf{M2e4\_ISRF100}. Note that this is also the run with the most shift in the IMF towards higher masses, which likely affects the comparisons of Solar-type stars. 

\subsection{Cloud setup and turbulent driving (Box vs. Sphere)}\label{sec:box_vs_sphere}

We note in \S\ref{sec:initial_conditions} that there are several common assumptions in the literature for the geometry and boundary conditions of simulated star forming clouds. In this subsection we compare the results of a periodic Box configuration relative to our fiducial Sphere run. The Box runs differ from the fiducial run in two important aspects. First, periodic boundary conditions lead to both an order-of-magnitude shallower gravitational potential \citep{federrath_sim_2012} and prevent the escape of radiation and gas. Second, the Box setup starts from a self-consistent, pre-stirred state, and this external driving continues throughout the run, providing energy for turbulent modes on the box scale that cascade down to smaller scales. To disentangle the effects of these two factors, we compare three \textbf{M2e4} runs (Table \ref{tab:IC_phys}): 1) our fiducial Sphere run, 2) a Box run with continuous external driving and 3) a Box run where we turn off the driving after the initial \myquote{stirring} phase.

\begin{figure*}
\begin {center}
\includegraphics[width=0.32\linewidth]{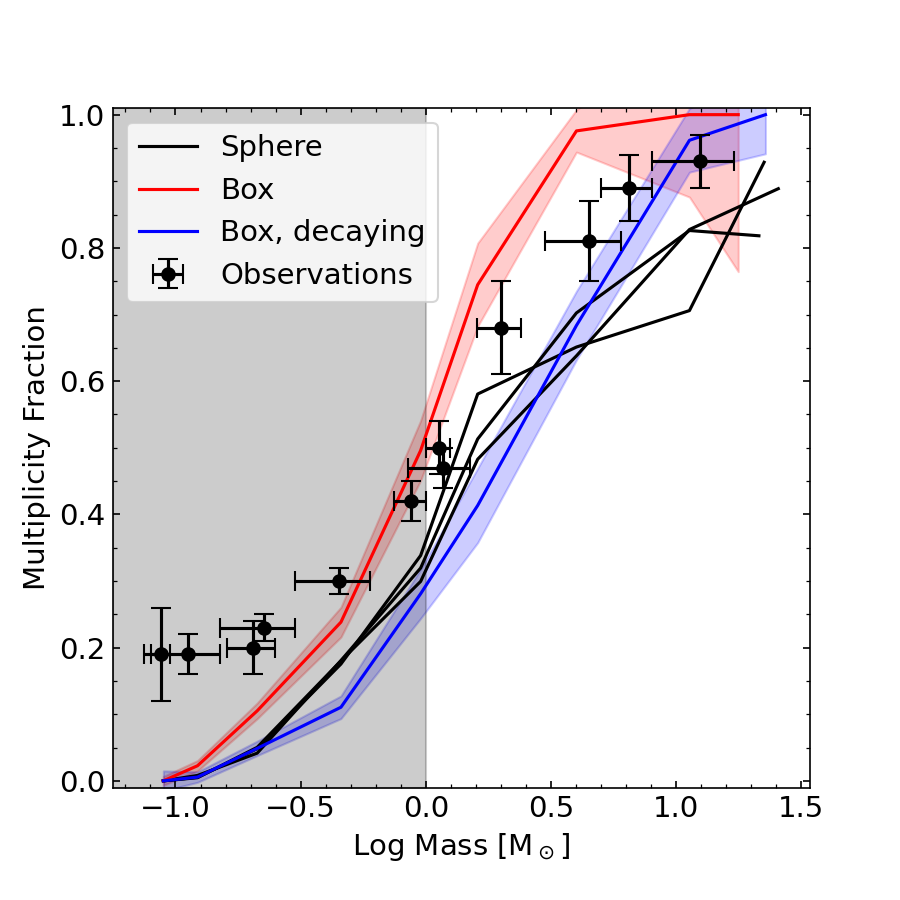}
\includegraphics[width=0.32\linewidth]{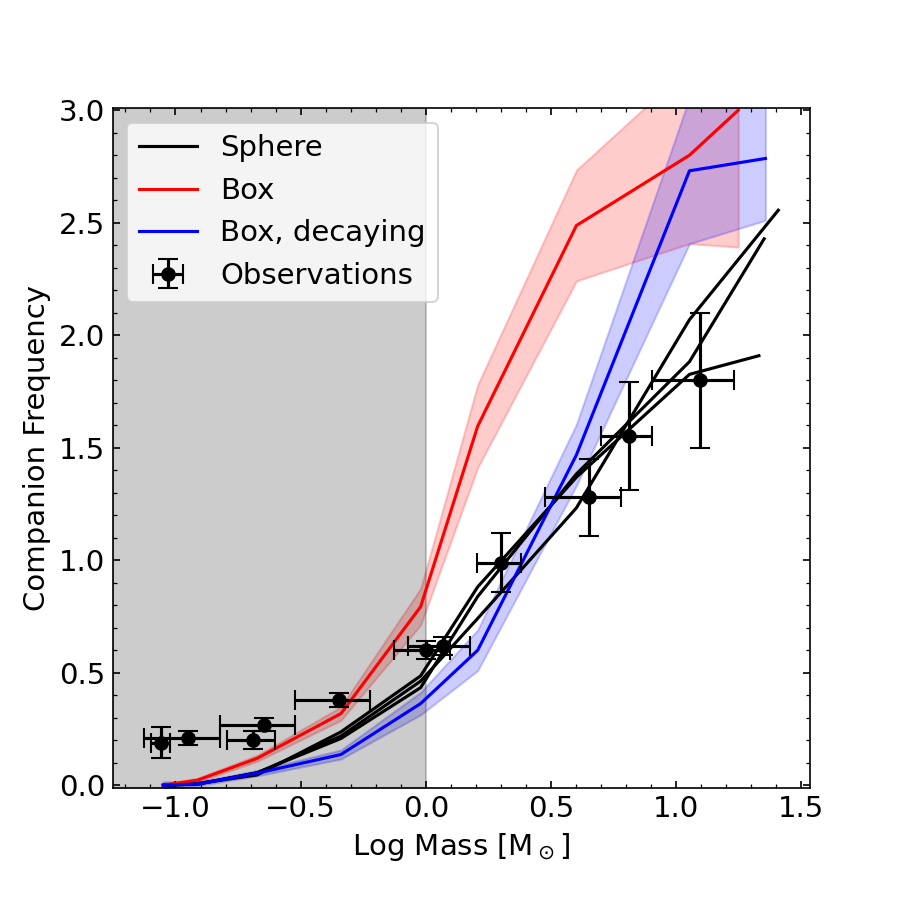}
\includegraphics[width=0.32\linewidth]{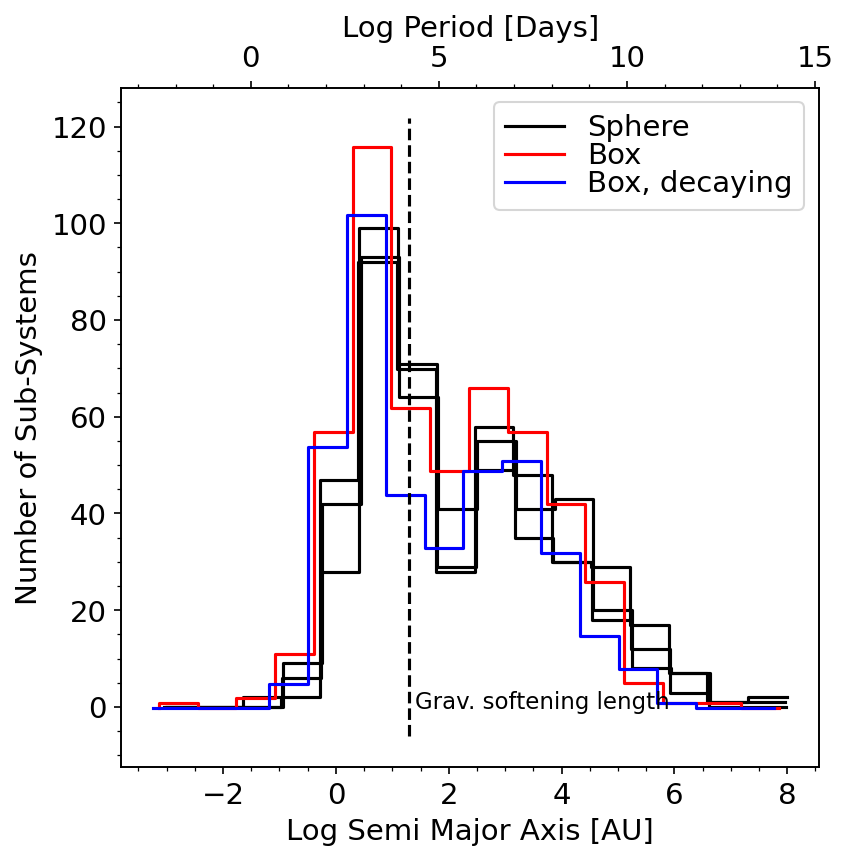}\\
\includegraphics[width=0.32\linewidth]{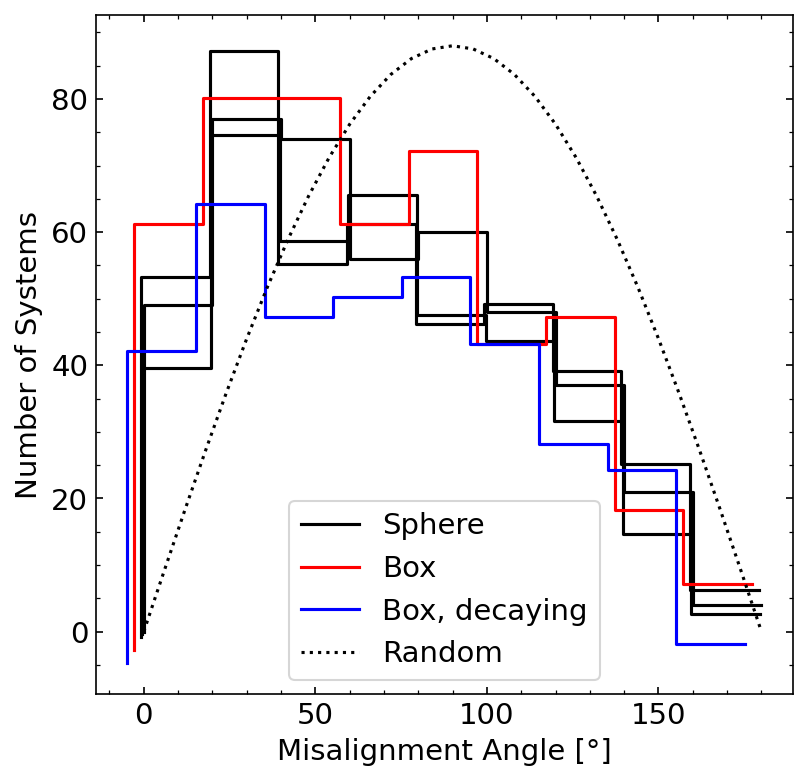}
\includegraphics[width=0.32\linewidth]{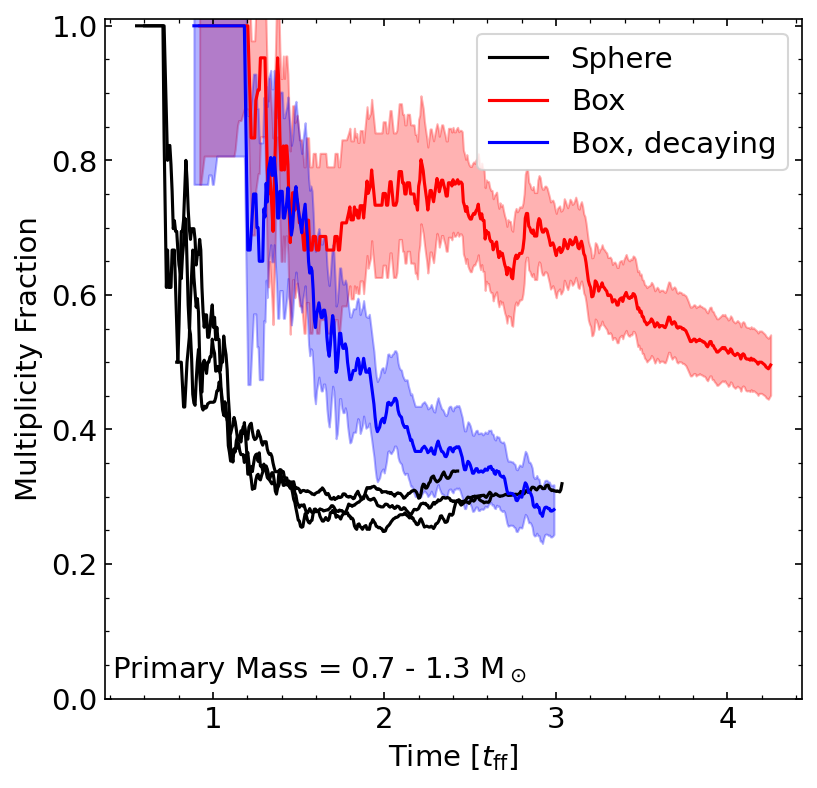}
\includegraphics[width=0.32\linewidth]{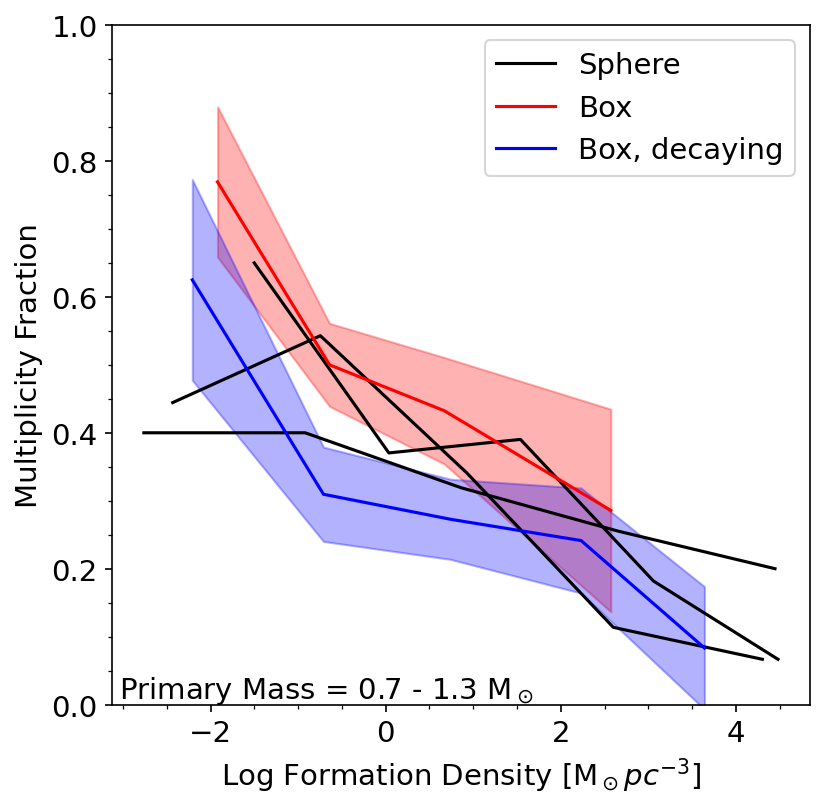}
\vspace{-0.3cm}
\caption{Same as Figure \ref{fig:alpha_var} for runs using fiducial M2e4 parameters but with different initial and boundary conditions (\myquote{Sphere}, \myquote{Box}, \myquote{Box, decaying}).}
\label{fig:Box_var}
\vspace{-0.5cm}
\end {center}
\end{figure*}

We find that the periodic boundary conditions have little effect on multiplicity properties when comparing the \myquote{Sphere} and \myquote{Box, decaying} runs, whose results agree within 1-$\sigma$ uncertainty for $\MF$, $\CF$, the semi-major axis and the misalignment angle distributions (see Figure \ref{fig:Box_var}). There is a difference in the length of the initial transient in the evolution of the multiplicity fraction of Solar-type stars. This delay is likely due to the stronger initial turbulent support in the Box run, since the periodic boundary conditions weaken the gravitational potential. As turbulence decays the \myquote{Box, decaying} run starts following the same trend as the fiducial Sphere run.

Turbulent driving, however, has a significant effect on the multiplicity fraction and companion frequency, leading to significantly higher values for both $\MF$ and $\CF$ on all mass scales. This is the only run in our parameter study that shows a change in sub-solar multiplicities. We attribute this difference to the turbulent driving, which weakens gravitational focusing and leads to lower stellar densities in star-forming regions. In other words, star formation is more distributed, which reduces the frequency of dynamical interactions thus leading to higher multiplicities on all mass scales.

We find that all three runs exhibit a similar relationship between the birth stellar density and the multiplicity fraction, but the Box run with turbulent driving has a lower maximum density, consistent with the higher multiplicity values we find for that run. 

\section{Discussion}\label{sec:discussion}

\subsection{Multiplicity in the fiducial simulated cloud}\label{sec:discussion_fiducial}

\begin{figure*}
\begin {center}
\includegraphics[width=0.33\linewidth]{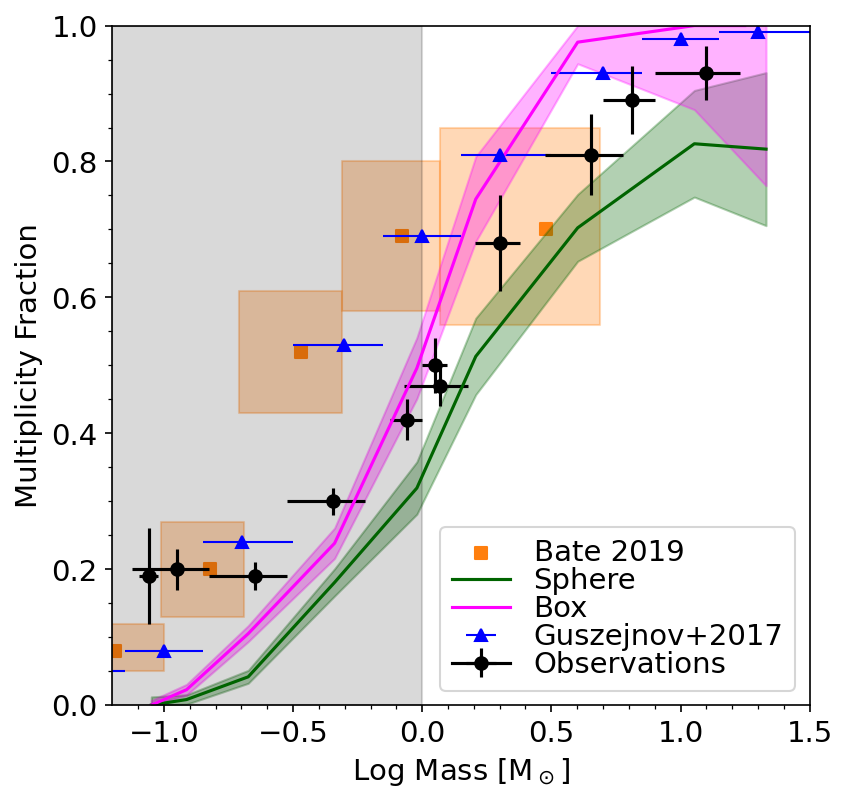}
\includegraphics[width=0.33\linewidth]{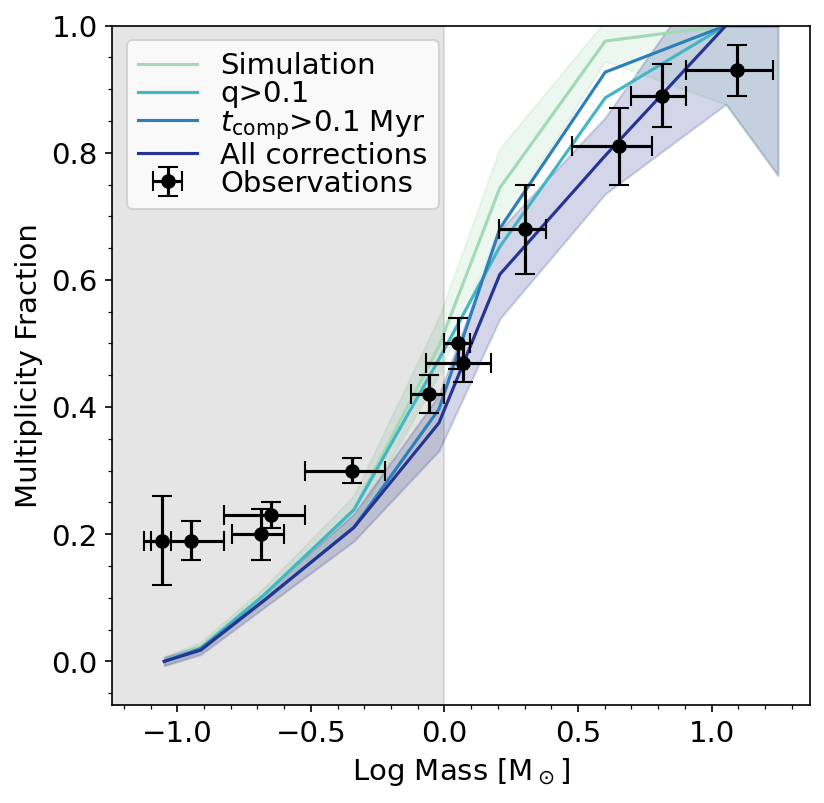}
\includegraphics[width=0.33\linewidth]{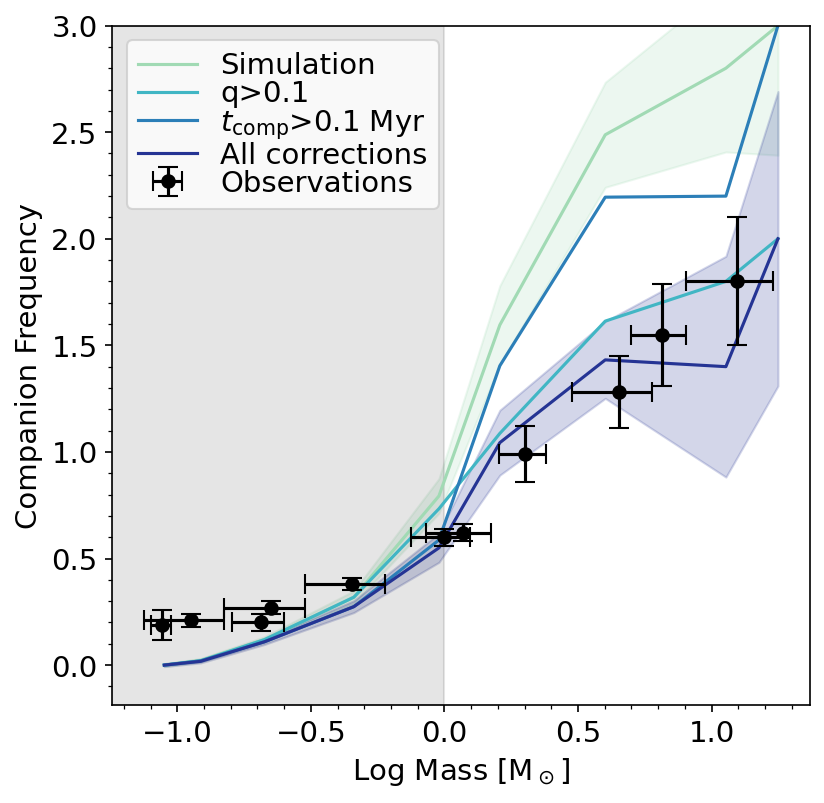}
\vspace{-0.4cm}
\caption{\textit{Left:} Multiplicity fraction as a function of primary mass in this work (both Sphere and driven Box runs, without corrections), in RHD simulations from \citet{Bate_2019_Z_multiplicity}, in semi-analytical predictions from \citet{guszejnov_correlation} and in observations \citep{offner_PPVII_multiplicity}. 1-$\sigma$ uncertainties are shown with either colored shaded regions or errorbars, while a grey shaded region shows the mass range potentially affected by the $0.1\,\msun$ completeness limit of the simulation. For high mass stars the Box results agree well with the semi-analytical core fragmentation model of \citet{guszejnov_correlation} and the RHD simulations of \citet{Bate_2019_Z_multiplicity}, while observations fall between the values in the Sphere and Box runs.\textit{ Middle \& Right}: Multiplicity fraction and companion frequency values for the driven Box run, using a similar notation as Figure \ref{fig:fiducial_CF_MF}. Unlike the Sphere run shown in Figure \ref{fig:fiducial_CF_MF}, the Box run results after corrections agree well with observations for both $\MF$ and $\CF$ (for stars with masses above $\msun$). Discrepancies at lower masses are likely due to the $0.1\,\msun$ completeness limit of the simulation.}
\label{fig:MF_comparison_plot}
\vspace{-0.5cm}
\end {center}
\end{figure*}

Similar to previous simulations in the literature \citep[i.e.,][]{bate_2009_rad_importance, bate12a,krumholz_2012_orion_sims,Mathew_Federrath_2021_IMF_multiplicity_2021} our simulations reproduce the rising trend with mass in both the multiplicity fraction and companion frequency. We find that the simulations match recent observations \citep{offner_PPVII_multiplicity} at all but the lowest mass scales (Figure \ref{fig:MF_comparison_plot}). The discrepancy at low masses can be explained by our choice of ignoring all brown dwarfs during the identification of multiples, motivated by the completeness limit of the simulation being at $\sim  0.1\,\msun$. 

We find that the multiplicity properties of stars depend on their formation time, i.e., early forming stars tend to have more companions than those that form near the end of the star formation process (Figure \ref{fig:fiducial_MF_CF_evol_for_timebins}). We find that the primary cause of this decrease is not stars losing companions, i.e., through dynamical interactions \citep{Heggie_1975_binary_dynamics}, but that later forming stars are born with fewer companions. We show that there is a correlation between multiplicity (i.e., $\MF$ and $\CF$) and the birth stellar density. This allows us to explain the decreasing trend with formation time, as locally collapsing regions merge 
and form a dominant, central, gas rich cluster, in which stars form at much higher stellar densities than in the early phase ($<4\,\mathrm{Myr}$) of cloud evolution when they formed along filaments (Figure \ref{fig:M2e4_series}). In this dense stellar environment dynamical interactions with other stars are much more likely, leading to newly formed stars being captured by existing ones, as well as companions being ejected. 


Similar numerical works in the literature mostly report only these \myquote{raw} values \citep[e.g.,][]{bate12a,Bate_2019_Z_multiplicity, Mathew_Federrath_2021_IMF_multiplicity_2021} without correcting for observational completeness limits and  chance alignments that the algorithm mistakenly identifies as a multiple star system. In this work we apply two simple corrections to account for these effects: we ignore companions with mass ratios below most observational completeness limits ($q<0.1$) and those that are not bound to their companion for at least 100 kyr and two full orbits. We find that the combined effects of these corrections dramatically reduces the number of companions for $>1\,\msun$ stars and consequently lower $\MF$ above a few $\msun$ (as these stars tend to have lower $q$ companions). Overall this means that our simulations under-predict both the $\MF$ and $\CF$ compared to observations. One possible explanation for this discrepancy is that stars in our simulations lose companions due to the inaccurate short-range gravitational forces in the simulation (i.e., having finite gravitational softening). We find this explanation to be unlikely as we find a pile-up of companions at the gravitational softening length (Figures \ref{fig:fiducial_semimajor}-\ref{fig:fiducial_separation_dist}), the net effect of the gravitational softening is likely to \emph{increase} the number of companions by trapping them at that length scale and preventing violent, short-range N-body interactions that could eject companions\footnote{This can be understood by noting that in the highly-softened limit $R << \epsilon$ for softening length $\epsilon$, the form of the gravitational force law becomes $g\left(R\right)=G M\left(<R\right)/R^2 \approx \left(4\uppi/3 \right) G \rho R $ for any softening kernel corresponding to a mass distribution with a flat central density $\rho$. Hence stars orbiting deep within the softening kernel behave as if connected by springs obeying Hooke's law, which has a stable solution expressible as normal eigenmodes for all $N$, in stark contrast to the chaotic Keplerian $N$-body problem.}. A more likely explanation is the apparent lack of stable protostellar disks in the simulations. This means that from three main pathways of binary formation (core fragmentation, disk fragmentation and capture, see \citealt{tohline2002, kouwenhoven_2010}) our simulation is missing disk fragmentation, which could conceivably make up for the missing companions relative to observations \citep{Kratter_2016}. Furthermore, the presence of protostellar disks would push companions mass ratios towards unity \citep[see e.g., ][]{kratter10a, Farris_2014,Duffel_2020_binary_disk_evol}. In addition, removal of angular momentum by magnetic breaking tends to drive accretion onto the more massive primary \citep{ZhaoLi2013}, thereby decreasing the mass ratio. Thus, the influence of disks and inclusion of non-ideal MHD together would likely shift currently small mass ratios above the $q=0.1$ limit, significantly increasing $\MF$ and $\CF$ at the high mass end after correcting for observational incompleteness (i.e., the difference between the \myquote{raw} and \myquote{corrected} results in Figure \ref{fig:fiducial_CF_MF} would be smaller). 
Finally, it is possible that our choice of initial conditions (i.e., geometry and turbulent driving) is the main cause of the discrepancy (see \S\ref{sec:discussion_vars}). This explanation is further supported by the fact that we find good agreement between the Box run and the semi-analytic core fragmentation model of \citet{guszejnov_correlation}, which follows only core fragmentation and has similar initial and boundary conditions (see Figure \ref{fig:MF_comparison_plot}). Note that \citet{guszejnov_correlation} ignores dynamical interactions, which is likely the explanation for the slightly higher multiplicity values it predicts relative to the Box run. After correcting for observational biases the Box results agree well with the observed $\MF$ and $\CF$ for $M>\msun$ stars, which is also the mass range unaffected by the $0.1\,\msun$ completeness limit of the simulation. 

For all primary masses we find that the mass ratio distribution in the fiducial run is consistent with randomly drawing companions from the initial mass function of the simulation (Figure \ref{fig:fiducial_q_dist}). For Solar-type stars, correcting for chance alignments leads to lower values at lower mass ratios ($q<0.2$). Observations find the q-distribution of Solar-type stars is flat \citep{raghavan2010} with an slight peak at $q\approx 1$ (see \citetalias{moe_distefano2016}). This uniform distribution is inconsistent with randomly drawing from the observed MW IMF, but it should be noted that the discrepancy is only significant at $q\approx 0.2$. Note that these observations are incomplete in this mass ratio range for short-period binaries, i.e., for companions with periods $\mathrm{Log} P/\mathrm{day}<4.5$ the observations are only complete for $q>0.5$ (see Figure 28 in \citetalias{moe_distefano2016}). After applying this correction to our results, we find a flatter distribution with a marginal peak around $q\approx 0.2$. This peak is dominated by low-$q$ companions at the softening length from their primary, so its significance strongly depends on the applied observational completeness limit. Similarly, we find that the agreement between the simulation and observations improves for both the semi-major axis and the eccentricity distributions after all corrections are applied. However, applying this correction significantly reduces the $\MF$ and $\CF$ for Solar-type stars, increasing the discrepancy with observations. It should be noted that the aforementioned \myquote{pile-up} of companions at the gravitational softening length ($\sim20\,\AU$) plays an out-sized role in this dramatic change (see Figure \ref{fig:fiducial_semimajor}). As previously noted we are unable to correct this pile-up as companions could spiral into shorter periods, be ejected or relax to longer orbits. Correcting for the observational bias for Solar-type stars (based on \citetalias{moe_distefano2016}) removes these companions, which implicitly assume that they either migrate to smaller scales or are ejected from the system. Furthermore, observations find a significant fraction of binaries have near-equal mass (twin) companions, which are missing in our simulations. 
This is likely due to the lack of long-lived protostellar disks, as disk fragmentation is more likely to produce near-equal mass companions \citep{kratter10a}, as disks allow companions to \myquote{steal} mass from the primary star. Disk accretion would also cause mass to be more equally distributed for secondaries that formed from turbulent fragmentation and migrated into or to close proximity of the primary's disk \citep{Duffel_2020_binary_disk_evol}. 

We track the angular momentum accreted by stars in the simulation and use its direction as a proxy for the direction of the spin of the star, as stars (i.e., sink particles) in the simulation can not lose angular momentum. Protostars are thought to inherit the angular momentum of their natal core, which would naturally lead to most binaries having similar spin alignments. Observations have found multiple protobinary systems where the protostellar outflows are misaligned \citep{Lee_2016}. We find that the distribution of the misalignment angle (i.e., angle between spins of the primary and its companions) is peaked towards lower values, i.e., companions tend to be aligned with their host stars (see Figure \ref{fig:fiducial_angle_dist}), exhibiting a less random distribution than prior results \cite{Offner_2016_jets,Lee_2019_MHD_binary_separations}. The distribution is fairly wide, and there is a significant number of companions with anti-parallel spin alignments. We find that the companions of more massive stars tend to be less aligned than companions of lower mass stars and that spin alignment increases over time. Figure \ref{fig:fiducial_separation_dist} shows that massive stars are slightly more likely to have companions that formed at large distances ($\sim 10^5\,\AU\sim 1 \,\pc$), making their spin directions more likely to be unrelated. Also, high-mass stars accrete from a significantly larger gas reservoir over a longer accretion time period rather than a more localized gas ``core" (\citetalias{grudic_starforge_m2e4}), and thus are less likely to have companions with aligned angular momentum vectors. Such misalignment has been found in recent observations of massive protostars \citep{Avison_2021_massive_protostar_accretion_outflow_obs}.  While multiple systems formed via turbulent fragmentation are less likely to have aligned spins compared to those formed by disk fragmentation \citep{Offner_2016_jets,Lee_2019_MHD_binary_separations}, systems that accrete from the same limited gas reservoir apparently still exhibit some spin correlation. 



\subsection{Connecting cloud properties and multiplicity}\label{sec:discussion_vars}

We analyze a suite of simulations where the initial properties of the cloud are varied (Table \ref{tab:var_guide}) and find that most multiplicity properties are insensitive to global cloud parameters. We find that multiplicity properties (i.e., $\MF$, $\CF$) can significantly vary between runs with identical global parameters but different turbulent realizations (see \S\ref{sec:results_var}), making it challenging to identify weaker trends. Note that observations are only able to constrain variations to changes in metallicity, as other properties of the natal cloud are not readily available once star formation ends. \citet{Moe_2019_solar_multiplicity_metallicity} showed that the multiplicity of Solar-type stars decreases with metallicity, due to a relative lack of close binaries.

These trends are summarized in Table \ref{tab:var_effect_summary}, note that the changes in the final star formation efficiency (SFE) and the shape of the IMF are investigated in detail in \citetalias{guszejnov_starforge_imf}, here we just state the results. Similar to \citet{Bate_2019_Z_multiplicity} we find that the initial cloud metallicity has no clear effect on multiplicity values, even though observations show a strong anti-correlation \citep{Moe_2019_solar_multiplicity_metallicity}. A possible explanation is that other cloud parameters (e.g., surface density) co-vary with metallicity for the observed multiples. Note that this trend was shown for close binaries only, which our simulation under-predict due to the lack of disk formation, which could also explain the discrepancy. 
We find that in most runs changes in the multiplicity fraction and companion frequency coincide with an opposite change in the stellar mass density around newly formed stars. This provides a potential explanation of these trends as an increasing stellar density means a higher chance for dynamical interactions, disrupting existing binaries and making it harder for newly formed stars to capture a companion. In our simulations an increase in the initial turbulence or continuous driving both weaken gravitational focusing in the cloud, leading to lower stellar densities. Starting from lower initial gas densities has a trivially similar effect. Overall, we find that multiplicity properties are sensitive to a different set of initial conditions than the IMF (see \citetalias{guszejnov_starforge_imf}).

We note that changing the initial surface density dramatically affected the fraction of companions at or below the gravitational softening length, implying that the surface density of the natal cloud likely influences the period distribution. Although we find no monotonic trend in either $\MF$ or $\CF$ with increasing initial magnetic field strength, we note that the run with the strongest field produces significantly higher multiplicity values, similar to the results of \citet{Lee_2019_MHD_binary_separations}.

The spins of companions in all our simulations are more aligned than random pairings with their primaries, however several initial parameters affect this distribution. Increased initial turbulence and reduced surface density both lead to more randomized spin alignments. The effects of surface density and turbulence can be potentially explained by the changes in how distributed star formation is within the cloud. Higher surface density or weaker turbulent support enhance the gravitational focusing of the parent cloud, leading to the formation of more massive and denser clusters \citep{guszejnov_starforge_clusters}. In a denser environment dynamical interactions are more common, so stars are more likely to both lose their original companions and capture new ones.  
Lowering the metallicity also leads to an increase in the randomness of spin alignment. This trend can potentially be explained by low $Z$ leading to higher gas temperatures, which makes protostellar cores larger, which leads (on average) to increased initial separation between companions that form through core fragmentation, making alignment less likely. 

\begin{table*}
    \setlength\tabcolsep{1.0pt} 
	\centering
	\begin{tabular}{ | c | c | c | c | c | }
	\hline
	Parameter & Final SFE & IMF change  & Stellar density  & Effect on multiplicity properties \\
	\hline
	Initial turbulence ($\alphaturb\uparrow$) & $\downarrow$ & Negligible & $\downarrow$ & $\MF\uparrow$, $\CF\uparrow$; spins more likely to be random  \\
	\hline
	Surface density ($\Sigma\uparrow$) & $\uparrow$ & Negligible  & $\uparrow$ & $\MF\downarrow$, $\CF\downarrow$; more companions at softening length  \\
	\hline
	Mass-to-flux ratio ($\mu\downarrow$) & $\downarrow$ & Steeper slope  & $\downarrow$ & $\MF\uparrow$, $\CF\uparrow$; variations only present in $\mu=0.42$ run \\
	\hline
	Metallicity ($Z\downarrow$) & $\downarrow$ & mild $\rightarrow$ shift  & No trend &  $\MF$, $\CF$ no longer correlated with stellar density; spins more likely to be random  \\
	\hline
	Interstellar Radiation (ISRF$\uparrow$) & mild $\uparrow$ & mild $\rightarrow$ shift  & $\uparrow$ &  Mild $\MF\uparrow$, $\CF\uparrow$ at high masses \\
	\hline
	Geometry (Box vs Sphere) & N/A & Negligible  & No trend &   Mild $\MF\uparrow$, $\CF\uparrow$ at high masses  \\
	\hline
	Turbulent driving &   N/A & Steeper slope & $\downarrow$ &  $\MF\uparrow$, $\CF\uparrow$ for all masses  \\
	\hline
    \end{tabular}
        \vspace{-0.1cm}
 \caption{Summary of results from \S\ref{sec:results_var}, showing the trends in the final star formation efficiency, the shape of the IMF, the average at formation stellar density and a general description on how multiplicity properties are affected (see Figures \ref{fig:alpha_var}-\ref{fig:Box_var} for details). Note that in case of the Box geometry there is no final SFE as the simulation is terminated when the periodic box is filled with an unphysical level of radiation. }
 \label{tab:var_effect_summary}\vspace{-0.5cm}
\end{table*}

\subsection{Caveats}\label{sec:caveats}

While the simulation presented here are the current state-of-the-art for simulating star-forming clouds, like other simulations in the literature STARFORGE employs a large number of significant approximations and assumptions to make the simulations computationally tractable (see the \citetalias{grudic_starforge_methods} for detailed discussions). 

In particular, the runs used here have a $\sim 30\,\mathrm{AU}$ Jeans-resolution, i.e. fragmentation on scales smaller than this are not resolved. This has dramatic effects on the formation of protostellar disks, causing the simulation to potentially miss close binaries that formed from disk fragmentation and overestimate stellar masses. Furthermore, the simulations have a $\sim 20\,\AU$ gravitational softening length that creates a \myquote{pile-up} of companions at this scale in the semi-major axis/period distribution (see Figure \ref{fig:fiducial_semimajor}). 

The simulations treat MHD in the ideal limit, assuming perfect coupling between the neutral gas and the magnetic fields. This approximation becomes invalid on the scale of protostellar disks, preventing the formation of long lived protostellar disks and the formation of binaries through disk fragmentation. Also, we show that the initial and boundary conditions of the cloud can affect multiplicity properties, so for a more predictive simulation a self-consistent connection to larger scales is required.

\section{Conclusions}\label{sec:conclusions}

In this work we analyze the stellar multiplicity properties in the STARFORGE radiation-magnetohydrodynamic simulations. These simulations follow the evolution of mid-sized molecular clouds ($M=20000\,\msun$) taking into account gravity, gas thermodynamics, turbulence, magnetic fields, and radiation as well as stellar feedback processes (jets, radiation, winds, SNe). The simulation suite consists of our fiducial cloud with MW average properties ($\Sigma=63\,\msun/\pc^2$, $\alphaturb=2$) and 12 clouds where we varied one of the initial conditions (see Table \ref{tab:IC_phys}).

We qualitatively reproduce the observed multiplicity fractions and companion frequencies for stellar masses significantly above the $0.1\,\msun$ completeness limit of the simulation. Previous works in the literature have drawn similar conclusions for simulations with less physics (i.e., \citealt{bate12a} does not include MHD or jets) and smaller cloud sizes \citep[i.e.,][]{Mathew_Federrath_2021_IMF_multiplicity_2021}. While the raw simulation results match well with observations, when we correct for observational incompleteness and chance alignments, we find that the fiducial simulation under-predicts both the multiplicity fraction and the companion frequency due to the significant fraction of low-mass-ratio ($q<0.1$) companions. This discrepancy can be explained by the simulation missing a key formation channel for binaries: disk fragmentation. 
Our simulations treat MHD in the ideal limit of perfect gas-field coupling, which leads to efficient magnetic breaking and greatly suppresses the formation of protostellar disks. This means that multiples in the simulation can only form either through the fragmentation of turbulent cores or the dynamical capture of a companion. Furthermore, disks have been shown to regulate the accretion of binaries and drive the system towards higher mass ratios, which likely explains the large fraction of low $q$ companions. Note that the multiplicity is sensitive to the simulation setup, such that our periodic box simulations that include external turbulent driving can reproduce observed values after accounting for observational incompleteness. Overall we conclude that capturing both disk fragmentation and having a realistic model for external driving are necessary for future simulations that aim to study stellar multiplicity.

We show that the multiplicity properties evolve over time. The primary reason for the evolution is not stars losing their companions, but that early-forming stars have significantly higher multiplicities than those that form near the end of the simulation. We find an inverse correlation between the stellar density around newly formed stars and their future multiplicity. This relationship can explain the trend in the $\MF$ and $\CF$ with several initial parameters. Specifically higher initial turbulence and lower cloud surface density both lead to lower stellar densities, and we find that these runs have higher multiplicity values for all masses. Also, replenishing turbulence (i.e., externally driving the turbulence in the cloud) significantly increases multiplicity values and lowers stellar densities. Despite having significant effects on the IMF, 
varying the metallicity or the interstellar radiation field showed no clear trend in either the $\MF$ or $\CF$.

We find that most companions form at $1000-10000\,\AU$ from their primaries, then \myquote{spiral in} within <1 Myr and settle at a much shorter orbital separation. A significant fraction of companions \myquote{pile-up} at the gravitational softening length, which prevents any further hardening of these binaries. We find that the fraction of companions at these length scales increases for higher initial surface densities, i.e., the average companion separation is smaller in higher density clouds.

The mass distribution of companions in the simulation agrees with random sampling from the IMF for both low and high mass stars. This appears to be in contradiction to observations, which find a flat distribution for Solar-type stars \citep{raghavan2010}. However, applying corrections for observational incompleteness dramatically flattens the distribution. This change is due to the high number of low-mass, short-period companions close to the gravitational softening length.

The spins of companions tend to be aligned with their primaries in the simulation, although the distribution is wide. Increasing turbulence or decreasing metallicity shifts the distribution towards random alignment.

Overall, our simulations allow us to predict the multiplicity statistics arising from either common core fragmentation or dynamical capture, with significantly better statistics than any previous work. In future work we will run simulations that account for all three channels of multiple formation by including non-ideal MHD effects and having significantly lower gravitational softening lengths ($\sim \AU$). A combined analysis of those results with the ones presented in this paper will give a detailed picture of the roles each formation channel plays in the formation of multiples.
  
 

\section*{Acknowledgements}
DG is supported by the Harlan J. Smith McDonald Observatory Postdoctoral Fellowship and the Cottrell Fellowships Award (\#27982) from the Research Corporation for Science Advancement. 
Support for MYG was provided by NASA through the NASA Hubble Fellowship grant \#HST-HF2-51479 awarded  by  the  Space  Telescope  Science  Institute,  which  is  operated  by  the   Association  of  Universities  for  Research  in  Astronomy,  Inc.,  for  NASA,  under  contract NAS5-26555.
Support for PFH was provided by NSF Collaborative Research Grants 1715847 \&\ 1911233, NSF CAREER grant 1455342, and NASA grants 80NSSC18K0562 \&\ JPL 1589742.
SSRO and ANR are supported by NSF Career Award AST-1748571 and by a Cottrell Scholar Award from the Research Corporation for Science Advancement. 
CAFG was supported by NSF through grants AST-1715216, AST-2108230,  and CAREER award AST-1652522; by NASA through grant 17-ATP17-0067; by STScI through grant HST-AR-16124.001-A; and by the Research Corporation for Science Advancement through a Cottrell Scholar Award.
ALR  acknowledges support from Harvard University through the ITC Post-doctoral Fellowship.
This work used computational resources provided by XSEDE allocation AST-190018, the Frontera allocation AST-20019, and additional resources provided by the University of Texas at Austin and the Texas Advanced Computing Center (TACC; http://www.tacc.utexas.edu).

\section*{Data availability}
The data supporting the plots within this article are available on reasonable request to the corresponding authors. A public version of the {\small GIZMO} code is available at \url{http://www.tapir.caltech.edu/~phopkins/Site/GIZMO.html}.

 


 \bibliographystyle{mnras}
 \bibliography{bibliography} 



\appendix

\section{Estimating multiplicity errors}\label{app:CF_MF_error} 

Although the STARFORGE simulation are (to date) the largest full physics star formation simulations that follow individual stars, they still represent relatively small molecular clouds, with our fiducial run having $M_0=2\times10^4\,\msun$, similar to the small GMCs in the Solar-neighborhood (e.g., Taurus). Thus they only form a small number of massive stars, which naturally leads to high sampling errors. In this Appendix we present a simple Bayesian model to estimate this error.

\subsection{Multiplicity Fraction}

Let's assume that in a certain mass bin we have $N$ primaries and we find $k$ of them to be in multiples. This naturally leads to the estimate that the multiplicity fraction for that mass bin is $\MF=k/N$. One would be tempted to estimate the uncertainty in $\MF$ by simply calculating the standard variations for a Poisson($k/N$) or binomial($N$, $k/N$) distribution. These, however, both fail in the $k\rightarrow 0$ limit. Instead, we use Bayes theorem to calculate the conditional probability density function $f(p|N,k)$:
\be
f(p|N,k) \dderiv p = P(\MF\in[p,p+dp]|N,k),
\ee
where $P(...|...)$ denotes conditional probability. Let us assume that $k$ is chosen from a binomial($N$, $p$) distribution (i.e., of $N$ systems each has $p$ chance of being a multiple) and use a uniform prior, i.e., $P(\MF\in[p,p+dp])=\dderiv p$. From  Bayes theorem it follows that
\be
f(p|N,k)  = \frac{(N+1)!}{(N-k)!k!}p^k (1-p)^{N-k},
\ee
For our estimate of $\MF$ we take the most likely value, which is simply $k/N$ (alternatively one could also use the mean value, which is $(k+1)/(N+2)$). For the error we take the standard variation, which is
\begin{eqnarray}
\sigma_{MF}^2(N,k) = \int_0^1{p^2 f(p|N,k)} - \left(\int_0^1{p f(p|N,k)}\right)^2 = \nonumber \\ \frac{(k+2)(k+1)}{(N+3)(N+2)} - \left(\frac{k+1}{N+2}\right)^2 = \frac{(N-k+1)(k+1)}{(N+3)(N+2)^2}. \label{eq:MF_sigma}
\end{eqnarray}
Note that for $N\gg 1$ and $k\gg 1$ the above equation simplifies to $\sigma_{MF}^2(N,k)\approx k(N-k)/N^3$, equal to what the naive binomial($N$, $k/N$) assumption would give.


\subsection{Companion Frequency}

We estimate the error of the companion frequency similarly to the approach we used to compute the error of the multiplicity fraction, but we instead assume that the number of companions follows a Poisson distribution with mean value $\lambda$. For $\lambda$ we adopt a uniform prior on [0,3] as we don't have any stars with more than 3 companions. Since the sum of similar Poisson variables also follows a Poisson distribution, we can easily construct the conditional probability density function $g(\lambda|N,k)$ for $N$ systems with $k$ companions in total, which yields
\be
g(\lambda|N,k) = \frac{N k!}{\gamma(k+1,3 N)}\frac{\lambda^k N^ke^{-\lambda N}}{k!},
\ee
where $\gamma(x,y)= \int_0^x t^{y-1}e^{-t}\dderiv t$ is the lower incomplete gamma function. As with $\MF$ we take the most likely value as our estimate for the companion frequency, so $\CF=k/N$. Using  $g(\lambda|N,k)$ we estimate the error with the standard variation, which yields
\begin{eqnarray}
\sigma_{CF}^2(N,k) = \int_0^3{\lambda^2 g(\lambda|N,k)} - \left(\int_0^3{\lambda g(\lambda|N,k)}\right)^2 = \nonumber \\ \frac{\gamma(k+3, 3 N)}{N^2 \gamma(k+1,3 N)} - \left(\frac{\gamma(k+2, 3 N)}{N \gamma(k+1, 3 N)}\right)^2. \label{eq:CF_sigma}
\end{eqnarray}


\bsp	
\label{lastpage}
\end{document}